\documentclass[prx,nobalancelastpage,twocolumn,nofootinbib,superscriptaddress]{revtex4-2}

\usepackage{xcolor,amsthm,amsmath,amsxtra,amsfonts,dsfont,graphicx,bm}

\usepackage{titlesec}

\def\Id{{\openone}}
\newcommand{\be}{\begin{equation}}
\newcommand{\ee}{\end{equation}}
\newcommand{\bea}{\begin{eqnarray}}
\newcommand{\eea}{\end{eqnarray}}
\newcommand{\bse}{\begin{subequations}}
\newcommand{\ese}{\end{subequations}}

\newcommand{\ket}[1]{\vert#1\rangle}

\newcommand{\tr}{\mathrm{tr}}

\usepackage[colorlinks,colorlinks,citecolor=blue,linkcolor=magenta,urlcolor=blue,breaklinks=true]{hyperref}

\usepackage{tcolorbox}
\tcbuselibrary{skins,xparse} 

\begin{document}

\title{Algorithms for quantum simulation at finite energies}

\author{Sirui Lu}
\author{Mari Carmen Ba\~nuls}
\author{J. Ignacio Cirac}
\affiliation{Max-Planck-Institut f{\"{u}}r Quantenoptik,
Hans-Kopfermann-Str.\ 1, 85748 Garching, Germany}
\affiliation{Munich Center for Quantum Science and Technology (MCQST), Schellingstr. 4, D-80799 M\"unchen, Germany}

\begin{abstract}
We introduce two kinds of quantum algorithms to explore microcanonical and canonical properties of  many-body systems. 
The first one is a hybrid quantum algorithm that, given an efficiently preparable state, computes expectation values in a finite energy interval around its mean energy. This algorithm is based on a filtering operator, similar to quantum phase estimation, which projects out energies outside the desired energy interval. However, instead of performing this operation on a physical state, it recovers the physical values by performing interferometric measurements without the need to prepare the filtered state. We show that the computational time scales polynomially with the number of qubits, the inverse of the prescribed variance, and the inverse error. In practice, the algorithm does not require the evolution for long times, but instead a significant number of measurements in order to obtain sensible results. 
Our second algorithm is a quantum-assisted Monte Carlo sampling method to compute other quantities which approach the expectation values for the microcanonical and canonical ensembles.
Using classical Monte Carlo techniques and  the quantum computer as a resource, this method circumvents the sign problem that is plaguing classical Quantum Monte Carlo simulations, as long as one can prepare states with suitable energies. All algorithms can be used with small quantum computers and analog quantum simulators, as long as they can perform the interferometric measurements. We also show that this last task can be greatly simplified at the expense of performing more measurements.
\end{abstract}

\date{\today}

\maketitle

\section{Introduction}
\label{introduction}

The advent of quantum simulators \cite{feynman1982simulating,Lloyd96}
opens many exciting opportunities to probe and understand fundamental
problems in physics, ranging from condensed matter to high energy
physics and quantum chemistry
\cite{CiracZollerPhysToday,Bauer2001.03685,GeorgescuRMP2014}.
Feynman's original proposal in 1982 was to build a universal digital
quantum computer that can imitate any physical systems. Although
tremendous progress has been made, building a universal quantum
computer that will fulfill Feynman's vision is still a long-term
task. However, both near term (noisy) quantum computers and analog
quantum simulators can already help us to address some of those
problems. The latter, where the interaction is engineered directly
according to the physical Hamiltonian under investigation, are
particularly advanced in different platforms, like cold atoms in
optical lattices \cite{RMP_BlochDalibard,Gross2017a}, trapped ions
\cite{BlattRossNatPhys2012}, Rydberg atoms \citep{Saffman2016b},
quantum dots \cite{VandersypenAnnalenderPhysik}, superconductors
\cite{LamataAdvancesinPhys2018}, photons
\cite{HartmannJofOptics2018}, etc. In particular, very controlled
experiments can be carried out with around 50 qubits
\cite{BlochManyBodyLocalin1D,BlochMBLin2D,Parsons2016,Chiu2019,Labuhn2016,Bernien2017a,Zhang2017,MonroeTimeCrystal,Arute2019,ZollerRoosBlattNature2019,BlattEE,Garttner2017,
DeLeseleuc2019,IBMGHZ18} and it is expected that this number will be
significantly increased in the coming years.

There are many questions that crave answers from quantum
simulators, especially physical properties of ground,
non-equilibrium, and finite temperature states. Most of the
theoretical work on quantum simulations has focused on the dynamics
of many-body quantum systems, as well as on their properties at zero
temperature. Since the first algorithm \cite{Lloyd96} that showed how
the dynamics could be efficiently simulated, large improvements have
been achieved leading to a very economic algorithm
\cite{HaahIEEE2018}. In practice, in analog quantum simulators the
dynamics are naturally implemented by letting the system evolve
according to the engineered Hamiltonian \cite{CiracZollerPhysToday}.
For ground state problems, the situation is quite different since
determining its properties is very demanding and, in general, it
requires exponential time in $N$, the number of qubits to be
simulated \cite{KitaevQMAHard}. A quantum simulator can still be of
big help since the corresponding classical simulator requires
exponential resources both in time and memory, whereas the quantum
one achieves a moderate speed-up albeit with polynomial memory. The
first algorithms
\cite{KitaevRussMathSurveys1997,AbramsLloydPRL97,AbramsLloydPRL99}
used quantum phase estimation to project onto an eigenstate of the
Hamiltonian, and have been successively improved
\cite{AspuruDutoiLoveScience2005,PoulinWocjanPRL09,Ge,Lin2020}. In
particular, in \cite{Ge} a cosine-filtering operator is used
to prepare a state close to the ground state with a very small
variance. It is similar in its conception to quantum phase estimation, 
but has a better scaling for that purpose.
This idea has also been used in the context of tensor
networks \cite{Banuls} to estimate the amount of entanglement required 
to achieve small energy variances along the whole spectrum of a
Hamiltonian. Although all these quantum algorithms were originally 
designed for scalable quantum computers, proposals to use them with
analog quantum simulators have recently been put forward
\cite{ZollerNatPhys2019}. This can be very convenient for small simulators, 
as the exponential scaling of the resources still limits their applicability 
to large systems. Other heuristic algorithms, like
adiabatic \cite{farhi2000quantum,aharonov2008adiabatic,Aharonov2003}
and variational \cite{peruzzo2014variational,farhi2014quantum} strategies, 
can be very useful and overcome the exponential scaling in certain cases
\cite{Bauer2001.03685}.

Quantum algorithms for excited states or finite temperature are more
scarce. In \cite{Chowdhury2017} it is shown how to realize the
imaginary time evolution operator to produce a Gibbs state, whereas
other algorithms propose sampling techniques
\cite{Temme2011b,Motta2019,Cohn1812.03607}. Phase estimation can also be directly used to prepare states at different energies, and thus
address quantum statistical questions in the microcanonical ensemble. All those algorithms may work well in practice. However, as for classical ones, they require an exponential time in $N$, although only polynomial memory resources.
Additionally, it remains challenging to implement most of them with the existing small quantum computers or analog quantum simulators.

\begin{figure*}[tbp]
\includegraphics{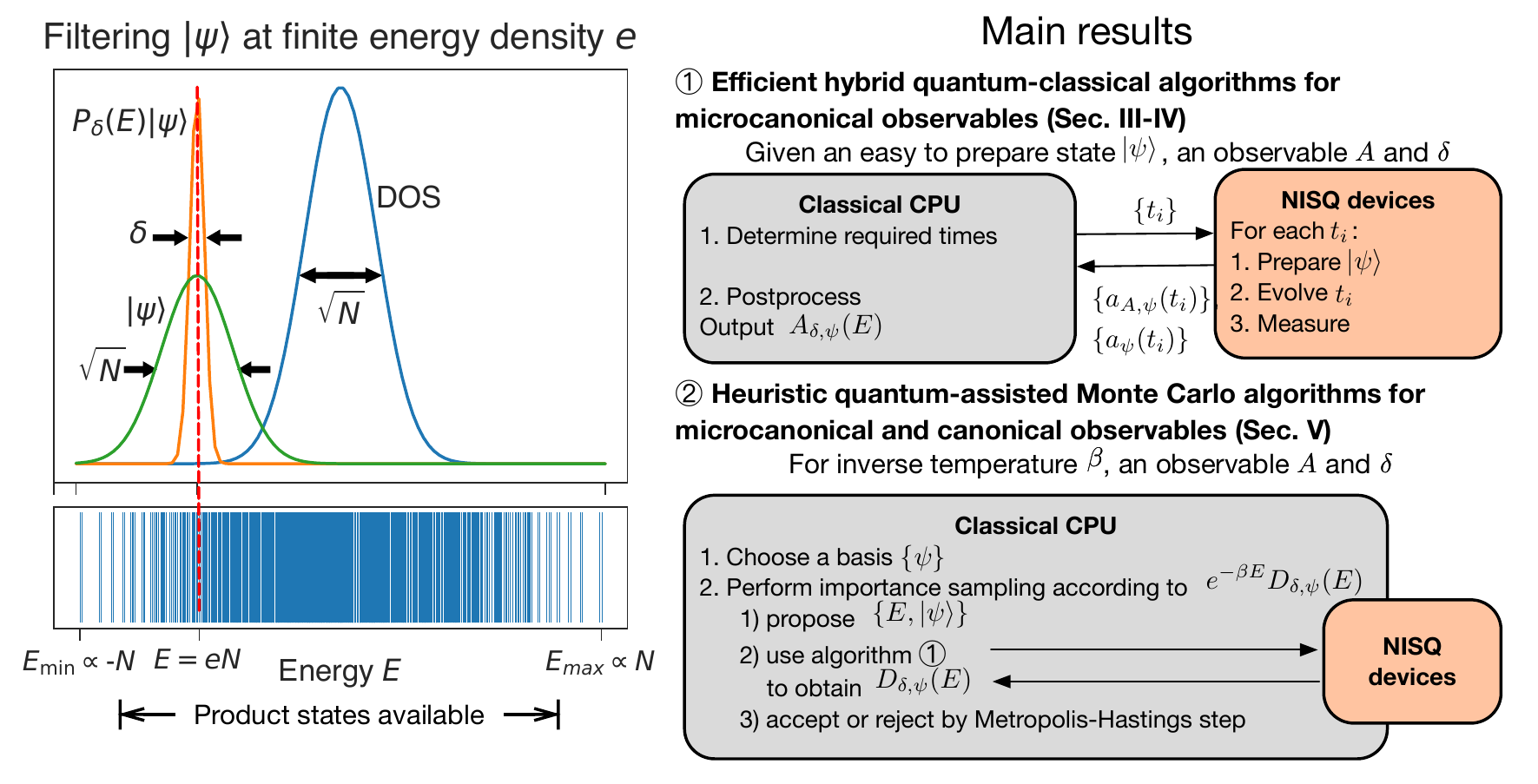}
\caption{Graphical summary of the concept and main results of the paper. \textbf{Left:} Our algorithms compute properties of a state with narrow energy variance (schematical spectral distribution shown in orange), which would result from applying a filter onto an easy to prepare state, e.g. a product state (green curve). For local Hamiltonians, for which the density of states in the thermodynamic limit approaches a Gaussian distribution of width proportional to $\sqrt{N}$ (blue curve), the accessible energy densities can lie on the tails of the spectrum. The box under the graph shows the actual energy spectrum for a $N=10$ Ising chain.
\textbf{Right:} The main contribution of our paper is the proposal of two sets of algorithms, schematically summarized in this figure. The first one (above) is a provably efficient hybrid quantum-classical algorithm for computing expectation values of observables in filtered states as illustrated on the left. It is based on repeated preparation, evolution and measurements on an easy to prepare state, run by a quantum device, and postprocessing by classical computation.
The second method is a set of quantum assisted sampling algorithms that can be used to compute microcanonical and canonical properties (summarized below for the canonical case). They perform a classical Monte Carlo importance sampling where the quantum device is used to determine the sampling probabilities efficiently. They require shorter coherence time than the first algorithm, at the price of more measurements. }
\label{fig:schematic}
\end{figure*}

In this paper we introduce and test two different types of quantum algorithms to determine physical properties in an energy interval or at finite temperature (see Fig~\ref{fig:schematic} for a graphical summary). The idea underlying our proposal  relies on the cosine-filter of \cite{Ge,Banuls} to target states with small energy variance at selected energies, in which the observables could be measured. Preparing those states may nevertheless be challenging in practice. 
We overcome this obstacle by showing that observations obtained after running quantum simulators for different stroboscopic evolution times are sufficient to determine the values of interesting quantities, without the need to prepare the filtered state at all. This is in the spirit of the time series approach introduced in~\cite{Somma2002} and later used for the related question of estimating the (binned averages of) Hamiltonian eigenvalues~\cite{Somma2019} (see also extensions based on Hamiltonian querying and Chebyshev expansions~\cite{aless2020spectral,rall2020quantum}). Our algorithms explicitly target finite energy properties, such as microcanonical expectation values, for which we can demonstrate efficiency. The quantities required in our method can be obtained with interferometric measurements, which involve the  conditional evolution depending on the state of a single qubit, and are specially suited for quantum simulators. 
More concretely: 
\begin{enumerate}
\item
Our first result is an efficient hybrid quantum-classical algorithm that targets the physical properties of states in an energy interval around that of any state that can be efficiently prepared (see Fig. \ref{fig:schematic} for an illustration). We prove that this algorithm can be carried out in time that is polynomial in $N$,
the inverse error, and the inverse width of the filtering operations. The later is related to the energy variance of the targeted state. Up to our knowledge, there is no classical algorithm achieving this polynomial scaling. 

\item 
Our second algorithm combines the quantum simulation with classical Monte Carlo methods, and provides practical methods to obtain both 
microcanonical and canonical expectation values of observables (see Fig. \ref{fig:schematic} for an illustration). These quantum assisted Monte Carlo methods use quantum simulators to compute the sampling probabilities that are required in Quantum Monte Carlo methods. Remarkably, they circumvent the sign problem \cite{WieseTroyer}, the main obstacle of applying such methods to many physics problems, so long as one can prepare (product or other kind of) states of suitable energies.
\end{enumerate}

Let us briefly mention that our proposal is different from other emerging families of quantum methods that make use of hybrid schemes, such as variational quantum eigensolvers~\cite{peruzzo2014variational,farhi2014quantum} and, in particular, 
hybrid ansatzes~\cite{VQAreview} or Quantum Subspace Diagonalization~\cite{QuKrylov} (see also \cite{mcclean2017hybrid,parrish2019quantum,Parrish2019d,Huggins2019a}). Different to our algorithms, such methods typically target ground state problems and, while some of them~\cite{QuKrylov} also use a superposition of a state evolved to different times, the coefficients of the superposition are variational parameters that need to be optimized, unlike in our proposal, where the coefficients are fixed by the filter. An even more significant difference is that these methods are not proven to be efficient.

Finally, the interferometric methods required for the algorithms presented
here have been used in the context of the Loschmidt echo \cite{Peres1984} in NMR \cite{Wisniacki2012}. They have also been proposed for ions
\cite{GardinerCiracZoller}, atoms \cite{Knap2013} and, more recently, to perform phase estimation with such systems \cite{ZollerNatPhys2019}.
We will give several alternative procedures to simplify that task, that can be applied in different situations. First, we will show that one can replace phase estimation by the ability of both preparing cat-like states \cite{GHZoriginalpaper} and having access to two additional
internal states. This capability already exists in different
platforms \cite{Laflamme1998NMRGHZ,Neumann2009GHZNV,Leibfried2005GHZ6,
Monz2011GHZ14,Dicarlo2010GHZ3,Song2017GHZ10,Wang2018GHZ18,
Friis2018GHZ20,LukinScienceSchCats,IBMGHZ18}. Then we will show that one can perform a similar procedure but without the requirement of the two additional states. Then, we will give a procedure that does not require the preparation of cat-like states, but proceeds with a sequence of measurements. Finally, we will give an even simpler method that works for Hamiltonians possessing certain symmetries, like XY or Hubbard models. The last two methods have an advantage with respect to the previous methods since no cat-like state needs to be evolved, and thus the method is more resilient against decoherence. However, they require a larger number of measurements.

The structure of this paper is as follows. In Section \ref{setup-and-cosine-filter} we introduce the models and the basic idea of the cosine-filter. In Section \ref{efficient-computation},
we present the first algorithm and show how one can use a quantum simulator to efficiently compute certain expectation values around fixed energies. In Section \ref{practical-computation} we give more practical methods for the same purpose and we test them numerically. In Section \ref{Sec:QAMC}, we present a family of quantum assisted Monte Carlo algorithms for microcanonical and canonical observables that combine classical Monte Carlo and quantum simulators. We explore numerically their performance 
and demonstrate that they are robust against certain noises.
In the appendices we present the methods to replace interferemotric measurements, describe the specific model we used in our numerical study, give details of the proof regarding the polynomial scaling of our algorithm, and investigate how much one has to decrease the variance in order to converge to the microcanonical and canonical results for a non-exactly solvable model.

\section{Setup and Cosine-Filtering}
\label{setup-and-cosine-filter}
\subsection{Setup}

We consider $N$ spins on a lattice and a Hamiltonian
 \be
 \label{Ham1}
 H = \sum_{n=1}^N h_n
 \ee
and denote by $E_{\rm min}$ and $E_{\rm max}$ the minimum and maximum eigenvalues of $H$. We will assume that $|E_{\rm min}|,|E_{\rm max}|<N/2$, so that the spectrum of $H$ lies within the interval $[-N/2,N/2]$. The Hamiltonian $H$ could be local in any spatial dimension, i.e. the term $h_n$ only acts on the $n$-th lattice site and its neighbors. However, we emphasize that the algorithms proposed here can also be applied to more general setups, where $h_n$ has long-range interactions or even with more complicated Hamiltonians, although some of our estimations will rely on the original form (\ref{Ham1}) or in its locality. The main requirement is that the evolution generated by $H$ can be efficiently implemented with the quantum simulator. More specifically, that given an initial state, $\psi$, and an observable $A$, one can efficiently determine
 \bse
 \label{ApsiA}
 \bea
 a_{A,\psi}(t) &=&\langle \psi |A  e^{-iHt}|\psi\rangle,\\
 a_\psi(t) &=& \langle \psi |e^{-iHt}|\psi\rangle
 \eea
 \ese
with a sufficiently small error.
Furthermore, if one can measure those quantities, one also has access to
 \be
 \label{Att}
 a_{A,\psi}(t_1,t_2)= \langle \psi |e^{iHt_1}A  e^{-iHt_2}|\psi\rangle
 \ee
since $a_{A,\psi}(t_1,t_2)=a_{A,\psi(t_1)}(t_2-t_1)$. 
The value of (\ref{Att}) can be readily measured using a quantum computer (see, e.g., Ref.~\cite{Ekert2002}). For analog quantum simulators, this may not be possible. In Appendix \ref{AppendixMeas} we give a series of alternatives to obtain such quantities. 
In general, by repeating the experiment $L$ times, the error will be additive and scale as $L^{-1/2}$.

\subsection{Initial states}
\label{initial-states}

In order for the simulation algorithm to be efficient, we will
choose as $\psi$ a state that can be prepared by the simulator. The simplest are product states
 \be
 |p\rangle = |p_1,p_2,\ldots,p_N\rangle,
 \ee
where $p_n$ are normalized states, e.g., for qubits,
 \be
 \label{product}
 |p_n\rangle= \cos(\theta_n)|0\rangle + e^{i\varphi_n} \sin(\theta_n)|1\rangle,
 \ee

In general, we denote the mean energy and variance of $H$ in the state $\psi$ by
 \bse
 \bea
 E_\psi &=& \langle \psi |H|\psi\rangle,  \\
 \sigma_{\psi}^2 &=& \langle \psi |(H-E_{\psi})^2|\psi\rangle.
 \eea
 \ese
In case $H$ is local and the state $\psi$ has finite correlation length, both $E_\psi$ and $\sigma_\psi^2$ will scale as $N$.

The energies of states that can be efficiently prepared will determine the range of energies that our algorithm can efficiently explore. If we restrict ourselves to product states, the mean energy $E_p$ does not cover the whole spectrum of $H$; there are energies $E_{p,{\rm min}}$ and $E_{p,{\rm max}}$ such that we can always choose a state $p$ with $E_p$ in the interval $[E_{p,{\rm min}},E_{p,{\rm max}}]$ but never outside (see Fig. \ref{fig:schematic} for an illustration). It has been shown that the range of the interval can be extensive in $N$ for local Hamiltonian (See Appendix~\ref{AppendixEfficient} for more details). Finding product states within this interval amounts to solving a mean-field problem, thus can be done efficiently on classical computers. In order to access energies outside this interval, one can consider other states, $\psi$, that are still easy to prepare but can cover a wider range of energies.
In particular, we could consider products of spin blocks, matrix product states \cite{Schuch2010a,Verstraete2008a,schollwock2011density}, or states obtained through adiabatic evolution or variational methods. Alternatively, the state $\psi$ could be prepared by starting from a product state and running the quantum simulator with a different Hamiltonian for some time.

\subsection{Cosine Filter}

Following \cite{Ge}, we define the cosine-filtering operator
 \be
 \label{Cosinefilt}
 P_\delta(E)= \left [\cos\left(\frac{H-E}{N}\right) \right]^{\lfloor N^2/\delta^2\rfloor_2},
 \ee
where we use $\lfloor...\rfloor_2$ to indicate the nearest even integer. Here $\delta$ can take arbitrary values, including decreasing (eg $\delta \sim 1/N$) or constant ($\delta\sim 1$) with $N$, that will be considered in the following sections. In order to interpret the action of this operator, it is useful to approximate \cite{Ge}
 \be
 \label{Gaussfiltapprox}
 P_\delta(E)\simeq e^{-(H-E)^2/2 \delta^2},
 \ee
as long as the spectrum of the operator that appears in the argument of the cosine lies in the interval $[-\pi/2 ,\pi/2]$ (in fact, this is also true in a bigger interval, see Appendix \ref{app:cos_gaussian}). Thus, it basically projects out the eigenstates of $H$ that have an energy $E'$ with $|E'-E|\gg\delta$, and thus acts as a filter around $E$~\cite{Schrodi2017}. By definition, $0<P_\delta(E)\le \Id$.

As in \cite{Ge}, we approximate
 \be
 \label{Cosfilt}
 \cos^M(X) \approx \sum_{m=-\lfloor x\sqrt{M}\rfloor}^{\lfloor x\sqrt{M}\rfloor} c_m e^{-i2m X}
 \ee
up to an error (in operator norm) bounded by ${2}e^{-x^2/2}$ for $\|X\|_\infty\le 1$, and
where
 \be
 \label{cm}
 c_m = \frac{1}{2^M} \left(\begin{array}{c} M \\ M/2-m \end{array} \right).
\ee

We can use this expansion to express the cosine-filter (\ref{Cosinefilt}) in terms of the evolution operator $e^{-iHt}$ for certain times $t$. For $|E|\le N/2$, we take $X=(H-E)/N$ so that
 \be
 \label{Expfilt}
 P_\delta(E) \simeq \sum_{m=-R}^{R} c_m e^{-i(H-E)t_m}
 \ee
where
 \be
 \label{Mt}
 R= x N/\delta, \quad t_m=2m/N.
 \ee
The idea will be, as in \cite{Banuls}, to apply (\ref{Cosfilt}) to certain states in order to filter them around some energy $E$, and then to obtain expectation values of observables with the resulting states. However, instead of preparing the state, we first express the desired values in terms of (\ref{ApsiA}), and then use the quantum simulator to measure those values separately. The number of measurements will be $2R$ times the number of repetitions required to obtain a prescribed accuracy. Each of the runs of the simulator will be for a time $t\le 2x/\delta$. In the end, we perform the multiplications and sum classically. 

\subsection{Ising Model}
\label{Model}

In order to benchmark our algorithms, we will use a rather trivial model for which we can obtain numerical results for values of $N\sim 100$ qubits, which should be attainable in present or planned quantum simulators. Let us take an even number, $N$, of fermionic modes and a Hamiltonian
 \be
 \label{Hamfermions}
  H = \frac{g}{2} \sum_{n=1}^N (a_n+a_{n}^\dagger) (a_{n+1}- a_{n+1}^\dagger) + h \sum_n (a_n^\dagger a_n-1/2),
 \ee
where $a_n$ are annihilation operators of the vacuum $|{\rm vac}\rangle$ and we have chosen periodic boundary conditions for the fermions: $a_{N+1}=a_1$. Through the Jordan-Wigner transformation, this corresponds to the Ising Hamiltonian
 \be
  H = \frac{g}{2} \sum_{n=1}^N \sigma_{n,x}\sigma_{n+1,x} + \frac{h}{2} \sum_{n=1}^N \sigma_{n,z},
 \ee
where $\sigma_{x,z}$ are Pauli operators, with appropriate boundary conditions.

Defining the operators in momentum representation
 \be
 b_k =\frac{1}{\sqrt{N}} \sum_{n=1}^N e^{i2\pi kn/N} a_n,
 \ee
where $k=-N/2+1,\ldots,N/2$, we will perform some computations with (Fock) states of the form
 \be
\label{kkk}
 |k\rangle = b_{k_1}^\dagger  \ldots  b_{k_\ell}^\dagger |{\rm vac}\rangle
 \ee
where $k_1<k_2\ldots<k_{\ell}$. 
They form an orthonormal basis and, even though they are not eigenstates of $H$, they are easy to deal with for large values of $N$. In appendix \ref{AppendixIsing} we provide the analytical formulas that allow us to obtain exact numerical results for this model. The minimal eigenvalue of $H$, $E_{\rm min}$, and the minimal energy attained by a state of the form (\ref{kkk}), $E_{p,{\rm min}}$ can be easily computed with those formulas.

Even though we are interested in energies above $E_{\rm min}$ and finite temperatures, let us notice that at zero temperature the Hamiltonian (\ref{Hamfermions}) features a phase transition at $g=h$. For $h\gg g$ the ground state is the vacuum, whereas for $h\ll g$ it is a superposition of the vacuum with states where excitations occur in pairs of momenta $\pm k$.

\section{Efficient quantum algorithm for observables at finite energy}
\label{efficient-computation}

Given a state $\psi$ and an observable $A$, we define
 \bse
 \label{Adeltapsi}
 \bea
 \label{Adeltapsi1}
 A_{\delta,\psi}(E) &=& \frac{\langle \psi|[A P_{\delta}(E) + P_{\delta}(E) A] |\psi\rangle}{2 \langle \psi| P_{\delta}(E) |\psi\rangle},\\
 \label{Adeltapsi2}
 A'_{\delta,\psi}(E) &=& \frac{\langle \psi| P_{\delta}(E)A P_{\delta}(E)|\psi\rangle}{\langle \psi| P_{\delta}(E)^2 |\psi\rangle}.
 \eea
 \ese
Both quantities are related to the microcanonical expectation value of $A$. In particular, if $\langle E|\psi\rangle\ne 0$, where $|E\rangle$ is the eigenstate of $H$ corresponding to the energy $E$,
\eqref{Adeltapsi2} converges to that value in the limit $\delta\to 0$. 

The numerator and denominator of (\ref{Adeltapsi}) can be expressed in terms of $a_{A, \psi}(t_n,t_m)$. Thus, the quantum algorithm uses the quantum simulator to determine those quantities up to the required precision, computes classically $c_m$ and $\exp(iEt_m)$, and then performs (classically) the required sums and multiplications.
We will show that both (\ref{Adeltapsi1},\ref{Adeltapsi2}) can be efficiently computed using a quantum simulator that has access to (\ref{ApsiA}). But for that, we have first to explain what we mean by "efficiently" and also formulate the problem more precisely.

We say that a state $\psi$ can be efficiently prepared if, for any prescribed error, $\epsilon>0$, we can obtain a state $\varphi$  with $\|\varphi-\psi\|^2 <\epsilon$ in a time
 \be
 \label{Polytime1}
 T = {\rm poly}(N,1/\epsilon).
 \ee
Furthermore, we say that the quantum simulator can efficiently measure $A$ if it can perform measurements to obtain $a_{A,\psi}(t',t)$ (with $t'\le t$) with an error smaller than $\epsilon$ in a time
 \be
 \label{Polytime}
 T = {\rm poly}(N,t,1/\epsilon).
 \ee
Note that this basically requires an efficient procedure to evolve according to the Hamiltonian and the possibility of performing interferometric measurements. 

{\bf Result:} If a quantum simulator can efficiently prepare $\psi$ and measure $A$, with $\|A\|_{\infty}\le 1$, then for any $\epsilon,\delta>0$, one can always find
 \be
 \label{Einterval}
 E\in [E_\psi - r\sigma_\psi, E_\psi + r\sigma_\psi]
 \ee
with $r=[3\log[2(1+2\sigma_\psi^2/\delta^2)]]^{1/2}$, so that one can obtain (\ref{Adeltapsi}) up to an additive error $\epsilon$
in a time
 \be
 \label{Polytime2}
 T = {\rm poly}(N,1/\delta,1/\epsilon),
 \ee
 including the cost of finding the value $E$.

Note that for $\delta={\rm poly}(1/N)$, the result can still be obtained in polynomial time. Furthermore, since $\sigma_\psi\le N/2$, $E$ will differ from $E_\psi$ by at most a constant times $\sigma_\psi\log^{1/2}(N)$. If $\sigma_\psi \propto O(\sqrt{N})$ as it occurs in states with finite correlation length and local Hamiltonians, this difference will only scale as
$\sqrt{N \log N}$.  
The reader may wonder why we need to introduce an interval, instead of simply fixing $E$ to some value, for instance $E=E_\psi$. The reason is that the spectrum of $H$ is discrete and it may well be that $\langle \psi| P_{\delta}(E) |\psi\rangle$ is exponentially small in $N$. As we will show in Appendix \ref{AppendixEfficient}, this issue can be avoided if we are allowed to vary $E$ in a small interval. We are not aware of any classical algorithm that can achieve this scaling.

The result can be proven by expressing the numerator and denominator of (\ref{Adeltapsi1},\ref{Adeltapsi2}) as a function of (\ref{ApsiA}) by means of (\ref{Cosfilt}), and then showing that both, as well as their quotient, can be computed with the required accuracy in a time (\ref{Polytime2}). Here, we show it explicitly only for (\ref{Adeltapsi1}), but it can be done in the same way for (\ref{Adeltapsi2}). We define the notation
 \bse
 \bea
 \label{Athermal}
  {p} &=& \langle \psi|[A P_\delta(E) + P_\delta(E) A]|\psi\rangle,\\
  {q} &=& 2\langle \psi| P_\delta(E) |\psi\rangle,
 \eea
 \ese
and denote by ${\Delta p,~\Delta q}$ the bounds on their error; that is, if the measured values are ${\tilde p}$, ${\tilde q}$, they fulfill ${|\tilde p-p|<\Delta p}$ (and analogously for ${q}$). Let us first argue that if we require ${\Delta p}$ to scale polynomially with $N^{-1}$, $\delta$ and $\epsilon$, we can reach it with the quantum simulator in a time (\ref{Polytime2}). The reason is that, using (\ref{Cosfilt}) and (\ref{Mt}) we will only need to determine the $2 x N/\delta$ values $a_{A,\psi}(t_m)$, and each of them will require to run the quantum simulator for a time $\le 2 x / \delta$.
Furthermore, we will have to repeat the procedure a number of times (\ref{Polytime2}) in order to reduce the error. The time to perform each of those tasks also scales in the same way, given that, by assumption, the quantum simulator can efficiently measure $A$. Thus, we conclude (\ref{Polytime2}). An analogous argument applies to ${\Delta q}$.

For a given $E$, the total error ${|\tilde p/\tilde q - p/q|}$, will be upper bounded by
 \be
 {
 \frac{\Delta p + p\Delta q/q}{q-\Delta q} \le \frac{\Delta x + 2\Delta q/q}{q-\Delta q},
 }
 \ee
as long as ${\Delta q< q}$, and where we have used that ${p\le 2}$. One can readily check that if ${\Delta p=\epsilon q/3}$ and ${\Delta q=\epsilon q^2/6}$, then the error will be bounded by $\epsilon$. Thus, the problem is reduced to proving that
 \be
 \label{q}
 {q}={\rm poly}(N,1/\delta),
 \ee
since in this case, ${\Delta p}$ and ${\Delta q}$ will scale polynomially with $N^{-1}$, $\delta$ and $\epsilon$ which, as we argued, can be accomplished with (\ref{Polytime2}). As for the cost of finding the value $E$, we show in Appendix \ref{AppendixEfficient} that there always exists an interval of energies of size $\Delta E\ge \delta^2/6N$
 within the interval (\ref{Einterval}) where $q\ge (1/4)\cdot[\delta^2/(\delta^2+2\sigma_\psi^2)]^{3/2}$. Since $\sigma_\psi\le N/2$, we have that in that interval ${q}$ fulfills (\ref{q}). Thus, the procedure consists in dividing the interval (\ref{Einterval}) in
$24Nr\sigma_\psi/\delta^2$ equal slices and picking an energy $E$ in each of them. At least one of them is then guaranteed to fulfill (\ref{q}) and thus we will be able to determine (\ref{Adeltapsi}) with an error smaller than $\epsilon$ in a time (\ref{Polytime2}).

\section{Practical computations}
\label{practical-computation}

In practice, one can use the quantum simulator much more efficiently than what has been presented in the previous section, and also employ it to access other physical properties. In this section we propose and analyze several algorithms to compute different quantities related to the microcanonical and canonical quantum statistical ensembles. We will also illustrate them with some examples for the model of Section \ref{Model}.

\subsection{Local density of states}

The simplest quantity is a broadened version of the local density of states,
 \be
 \label{LDOE}
 D_{\delta,\psi}(E) = \langle \psi | P_\delta(E) |\psi\rangle
 \ee
which (up to a factor) converges to that quantity in the limit $\delta\to 0$. Using (\ref{Cosfilt}), we can express
 \be
 \label{DpsiE0}
  D_{\delta,\psi}(E) \approx  \sum_{m=-R}^{R}
  c_m a_\psi(t_m)e^{i(E-E_\psi) t_m}
 \ee
with (\ref{Mt}).

As before, our algorithm uses the quantum simulator to determine $a_\psi(t_m)$. The method can be made more efficient by noticing that norm of $X_0=(H-E_\psi)/(\tilde r\sigma_\psi)$ will be bounded by one in the subspace where the state $\psi$ has most of its weight if we choose $\tilde r\sim 1$. Thus, we can use the expansion (\ref{Cosfilt}) with $X=X_0$ and $M=\tilde r^2 \sigma_\psi^2/\delta^2$ to obtain (\ref{LDOE}), but now with $R= x \tilde r \sigma_\psi/\delta$ and $t_m=2m/(\tilde r\sigma_\psi)$. For $\sigma_\psi \le \sqrt{N}$ the number of required measurements will significantly decrease with respect to (\ref{Mt}). In that case, we
denote $\tilde r\sigma_\psi=r \sqrt{N}$ so that
 \be
\label{Mt2}
 R= x r \sqrt{N}/\delta, \quad t_m=2m/(r\sqrt{N}).
 \ee

\begin{figure}[tbp]
\includegraphics[width=1.0\linewidth]{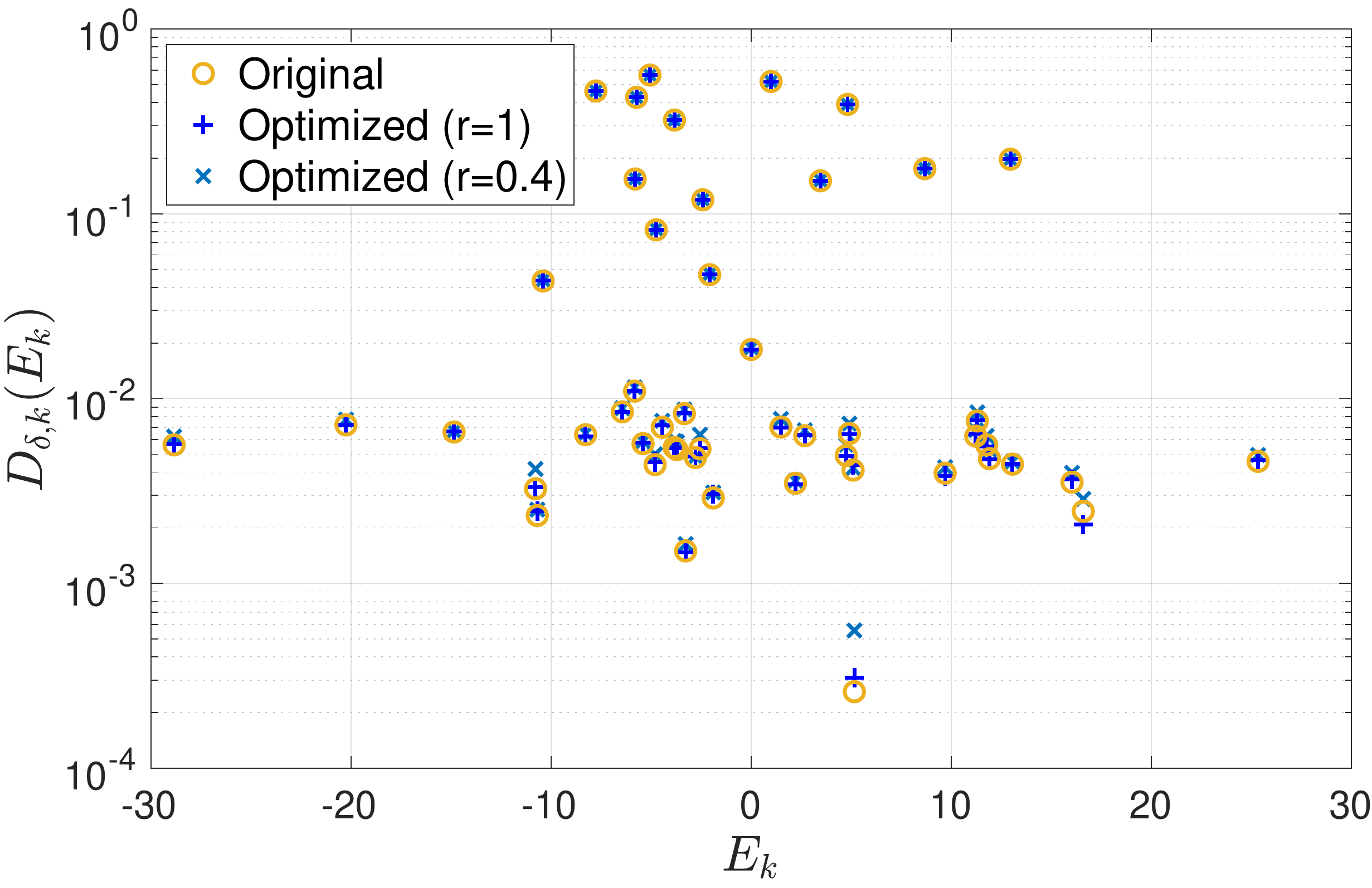}
\caption{Local density of states as a function of the mean energy for the Hamiltonian (\ref{Hamfermions}) for $N=100$ spins, $g=1$, $h=2$, $\delta=0.1$, and 50 randomly chosen states $|k\rangle$. The circles have been computed with (\ref{Mt}) whereas the cross and plus symbols correspond to (\ref{Mt2}) with $r=0.4$ and $r=1$, respectively. In all cases $x=3$.}
\label{fig1}
\end{figure}

In Fig.~\ref{fig1} we have plotted $D_{\delta,k}(E_k)$ in logarithmic scale for $N=100$ spins, $\delta=0.1$, and the Hamiltonian (\ref{Hamfermions}) with $g=1$, $h=2$, for 50 randomly generated states $|k\rangle$ (\ref{kkk}). We have taken $x=3$ and compared the results of the original \eqref{Mt} (circles) and the optimized  alternatives \eqref{Mt2} (crosses for $r=0.4$, plus symbols for $r=1$). For this value of $\delta$, the improved method yielding \eqref{Mt2} with $r=0.4$ corresponds to 120 measurements of $a_\psi(t)$ and a maximum value of $t=60$. For $\delta=1$ it requires 12 measurements with a maximum value of $t=6$, which is very reasonable for present experiments. We observe that for $r=1$ one already obtains an error of the order of $10^{-3}$, whereas for $r=0.4$ it is about $10^{-2}$, which is what one could expect with imperfect devices.

\subsection{Microcanonical observables}

\begin{figure}[tbp]
\includegraphics[width=1.0\linewidth]{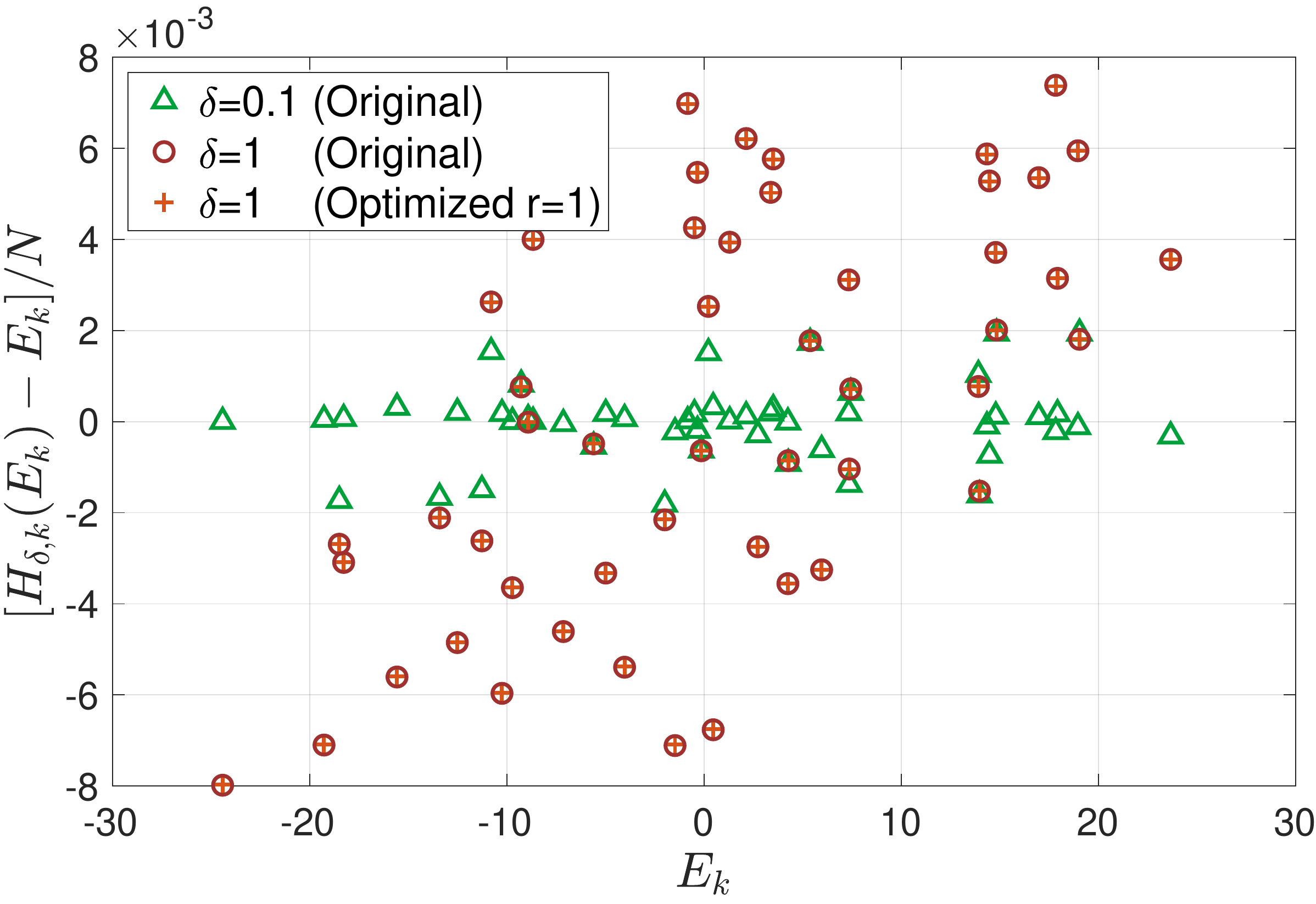}
\caption{Expectation value of the energy for the same model and parameters as in Fig.~\ref{fig1}. The data show the results of \eqref{Mt} for $\delta=0.1$ (triangle) and $1$ (circles),
and \eqref{Mt2} for $\delta=1$ with $r=1$ (plus symbols).}
\label{fig2}
\end{figure}

The very same simplified program can be applied to (\ref{Adeltapsi}), as we can also choose different values of $r$ to make the procedure much more efficient. In Fig.~\ref{fig2} we have plotted (\ref{Adeltapsi1}) for the model (\ref{Hamfermions}) with $g=1$, $h=2$, and we have chosen the energy as an observable, i.e. $A=H/N$. Note that the microcanonical expectation value of $A=H$ at $E$ is only trivial (i.e., equals to $E$ exactly) in the limit $\delta\to 0$. Here, our purpose is to investigate the convergence of $H_{\delta,\psi}(E)$ with respect to $\delta$. We have again chosen 50 random states $|k\rangle$, and subtracted their mean energy $E_k/N$ to optimize the visualization, since $H_{\delta,k}(E_k)\sim E_k$. We have taken $\delta=0.1$ for the triangles, while $\delta=1$ for plus symbols and circles. Only for the latter we have chosen the more efficient version (\ref{Mt2}), with $r=1$. We observe a clear difference (at the percent level) for different values of $\delta$, however the value $r=1$ is sufficient to obtain reliable results. 

An interesting question in this context is to what extent one can recover the microcanonical expectation value by decreasing $\delta$.
This makes sense if the system is sufficiently large and fulfills the eigenstate thermalization hypothesis (ETH) \cite{Srednicki1994,reviewETH}, which ensures the convergence of the procedure in the thermodynamic limit.
The question of how narrow the energy support of a pure state needs to be in order to recover microcanonical expectation values has been analyzed in Ref. \cite{Dymarsky2017,Banuls} for one-dimensional systems. In~\cite{Dymarsky2017} it is concluded that, 
for generic systems, $\delta$ needs to decrease with $N$, 
 while \cite{Banuls} gives evidence that for local observables it may suffice that $\delta\sim 1/\log(N)$. In Appendix \ref{app:ed_non_int} we analyze this question for a non-integrable model and up to $N=28$ spins, and give further evidence for the need to decrease $\delta$ with $N$. This can be qualitatively understood as follows. Let us denote by $|E\rangle$ the eigenvectors of $H$ with energy $E$. In case the ETH applies, the diagonal matrix elements of physical observables (which can also include, for instance,  correlation functions.) in the energy basis rapidly converge to the microcanonical expectation value, while off-diagonal elements vanish exponentially fast \cite{Srednicki1994}. However, when one has a superposition of an exponential number of eigenstates around some energy, the sum of the off-diagonal terms does not need to converge even though the diagonal terms do. This is also the reason why $\delta$ must decrease with $N$, since $P_\delta(E)|\psi\rangle$ contains superpositions of $|E\rangle$ and thus the expectation value depends on off-diagonal elements. 

\section{Quantum assisted Monte Carlo algorithms for microcanonical and canonical observables}
\label{Sec:QAMC}

The considerations at the end of the previous section suggest 
 an alternative strategy that can be used for microcanonical and canonical observables.   
 At a high level, the algorithms we propose in this section can be seen as a quantum version of the classical quantum Monte Carlo algorithms: they use classical Monte Carlo to sample different initial states, while the quantum device assists with the computation of sampling probabilities and the measurements of observables.

Different to the algorithm presented in Sec.~\ref{efficient-computation}, the quantum assisted sampling discussed here is not proven to be efficient. However, when compared to the first algorithm from an experimental point of view, these methods offer the potential advantage of requiring shorter coherence times, at the cost of increased number of measurements in the quantum device. Moreover,  the sampling probabilities are always positive and the sign problem in classical quantum Monte Carlo~\cite{WieseTroyer} is circumvented. 
We also present numerical experiments  for a $N=100$ Ising Hamiltonian, to demonstrate the viability of these methods on near-term devices. The results show that physical quantities can be obtained accurately even in the presence of certain level of noise.

\subsection{Microcanonical observables}

 The quantity
 \be
 \label{AMicro}
 A_\delta(E) = \frac{{\rm tr}[A P_\delta(E)]}{{\rm tr}[P_\delta(E)]}
 \ee
may converge to the microcanonical expectation value even for constant $\delta$ since, by definition, $P_\delta(E)$ is diagonal in the eigenbasis of $H$ and thus no off-diagonal element appears in the expectation value. This is indeed observed in Ref. \cite{Yilun}, where tensor network techniques are used in order to compute quantities closely related to (\ref{AMicro}) with energy resolution corresponding to larger values of $\delta$ than those required by (\ref{Adeltapsi}).
In fact, that a constant $\delta$ suffices also follows from the fact that, if the thermodynamic limit $N\to\infty$ exists for the observable $A$, then in that limit the expectation value of $A$ in almost all eigenstates of $H$ should coincide if the corresponding energies fulfill $|E_N-E_N'|/N\to 0$.  The intuitive reason is that in that limit, only intensive quantities matter. Therefore, what is relevant is $\delta/N$, so that as long as it vanishes, one should obtain the thermodynamic value.

In principle, one could obtain (\ref{AMicro}) in the very same way as in (\ref{Adeltapsi1}). This can be seen by noticing that
 \be
 \label{AdeltaE}
 A_\delta(E)=A_{\delta,\Phi}(E)
 \ee
where $\Phi$ is a maximally entangled state of each spin with an auxiliary one
 \be
 |\Phi\rangle =\frac{1}{2^{N/2}} \left[|0,0\rangle + |1,1\rangle\right]^{\otimes N}.
 \ee
That is, just one has to add an auxiliary qubit for each existing one and prepare an entangled state of each pair. Thus, from (\ref{AdeltaE},\ref{Adeltapsi1}) it follows that one could compute the numerator and denominator independently with the help of the quantum simulator, and then the quotient. However, we face here the problem that the denominator will typically decrease exponentially with $N$. For local Hamiltonians, the reason is that $\sigma_\Phi\propto \sqrt{N}$ and therefore, for any extensive value of the energy, $E=e N$, $\langle \Phi|[P_\delta(E)\otimes \Id]|\Phi\rangle \sim \exp{(-cN)}$ for some $c=O(1)$. This makes this procedure impracticable already for $N\agt  20$.

In the following, we propose an algorithm to circumvent, at least in part, this issue. Let us re-express (\ref{AMicro}) with the help of an (over)complete basis of states fulfilling
 \be
 \label{Id}
 \int d{\mu_\psi} |\psi\rangle\langle \psi| = \Id,
 \ee
where $d \mu_\psi$ is a measure in the basis set. For instance, we can take an orthonormal basis of product states $|p_1,p_2,\ldots,p_N\rangle$ or all product states (\ref{product}), in which case \be
 \label{mup}
 d\mu_p= \prod_{n=1}^N d\Omega_n= \frac{1}{(4\pi)^N}\prod_{n=1}^N  \sin(\theta_n) d\theta_n d\varphi_n.
 \ee
Inserting (\ref{Id}) in (\ref{AMicro}) and using definitions (\ref{Adeltapsi1}) and (\ref{AMicro}), we can rewrite
 \be
\label{AMicro2}
A_\delta(E) =\frac{ \int d\mu_\psi D_{\delta,\psi}(E) A_{\delta,\psi}(E)}{\int d\mu_\psi D_{\delta,\psi}(E) }.
 \ee
This quantity can be computed using Monte Carlo algorithms so long as one is able to compute $D_{\delta,\psi}(E)$ and $A_{\delta,\psi}(E)$. But these two can be computed using the quantum simulator. 
More concretely, one can implement importance sampling according to a distribution proportional to $D_{\delta,\psi}(E)$. For instance, using a basis of product states, as mentioned above, this can be achieved by a Metropolis-Hastings algorithm in which once a move is proposed, the quantum simulator is used to estimate the corresponding value of $D_{\delta,\psi}(E)$, and thus  determine the acceptance rate. The Monte Carlo algorithm 
 circumvents not only the burden of summing over all states, but also the need to measure $D_{\delta,\psi}(E)$ with an exponentially small accuracy,
 because $D_{\delta,\psi}(E)$  needs to be evaluated only for states $\psi$ for which it is not negligible.
  We also emphasize that if the observable $A$ is chosen so that $A|\psi\rangle=\lambda|\psi\rangle$, and $\lambda$ can be classically computed, then one only has to determine $D_{\delta,\psi}(E)$ with the quantum simulator.

We have tested this algorithm with the Hamiltonian (\ref{Hamfermions}) and the simplest Monte Carlo algorithm that changes one spin at a time for the sampling.
Notice that our goal is to demonstrate the viability of the approach, rather than the competitiveness of the algorithm, since that would require optimizing the sampling methods and other parameters. We have chosen the Hamiltonian so that we can compare the results with an exact calculation for $N\sim 100$ spins. We have implemented a Metropolis algorithm that takes a random state $|k\rangle$ (\ref{kkk}), according to a probability proportional to $D_{\delta,k}(E)$. We have then computed the magnetization $A=M$ with
 \be
 \label{Magne}
 M=\frac{1}{2N} \sum_{n=1}^N (\sigma_{n,z}+1)
 \ee
as a function of $E$. Since the model is exactly solvable, we have also computed $A_\delta(E)$ directly using the method of Appendix \ref{AppendixIsing}. This direct numerical calculation requires the computation of very large and small numbers, so that one can easily run into precision problems. In fact, for some plots we can only provide the exact result for some values of $E$, since otherwise our exact method did not give consistent values.

\begin{figure}[tbp]
\centering
\includegraphics[width=0.98\linewidth]{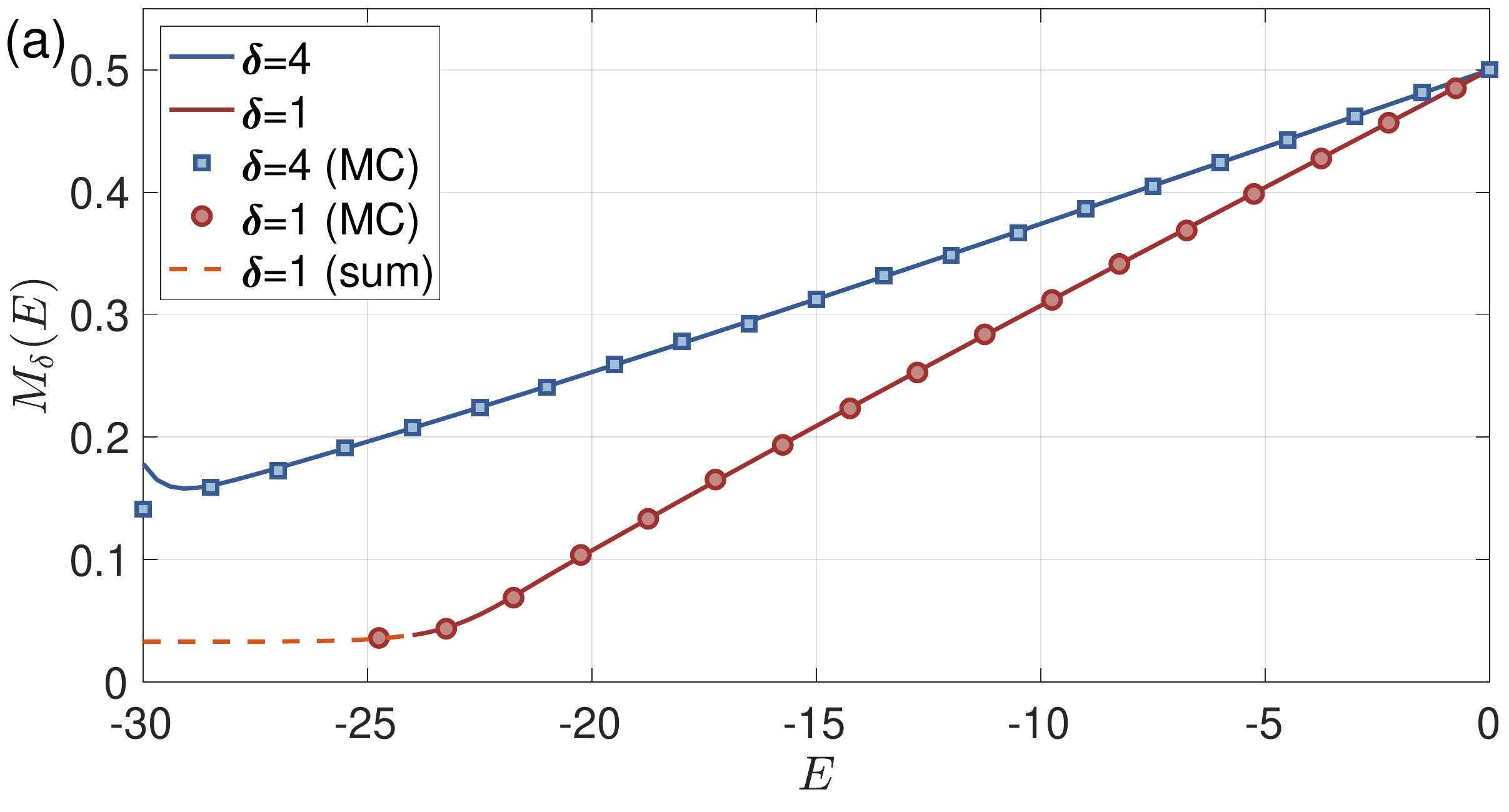}
\includegraphics[width=1\linewidth]{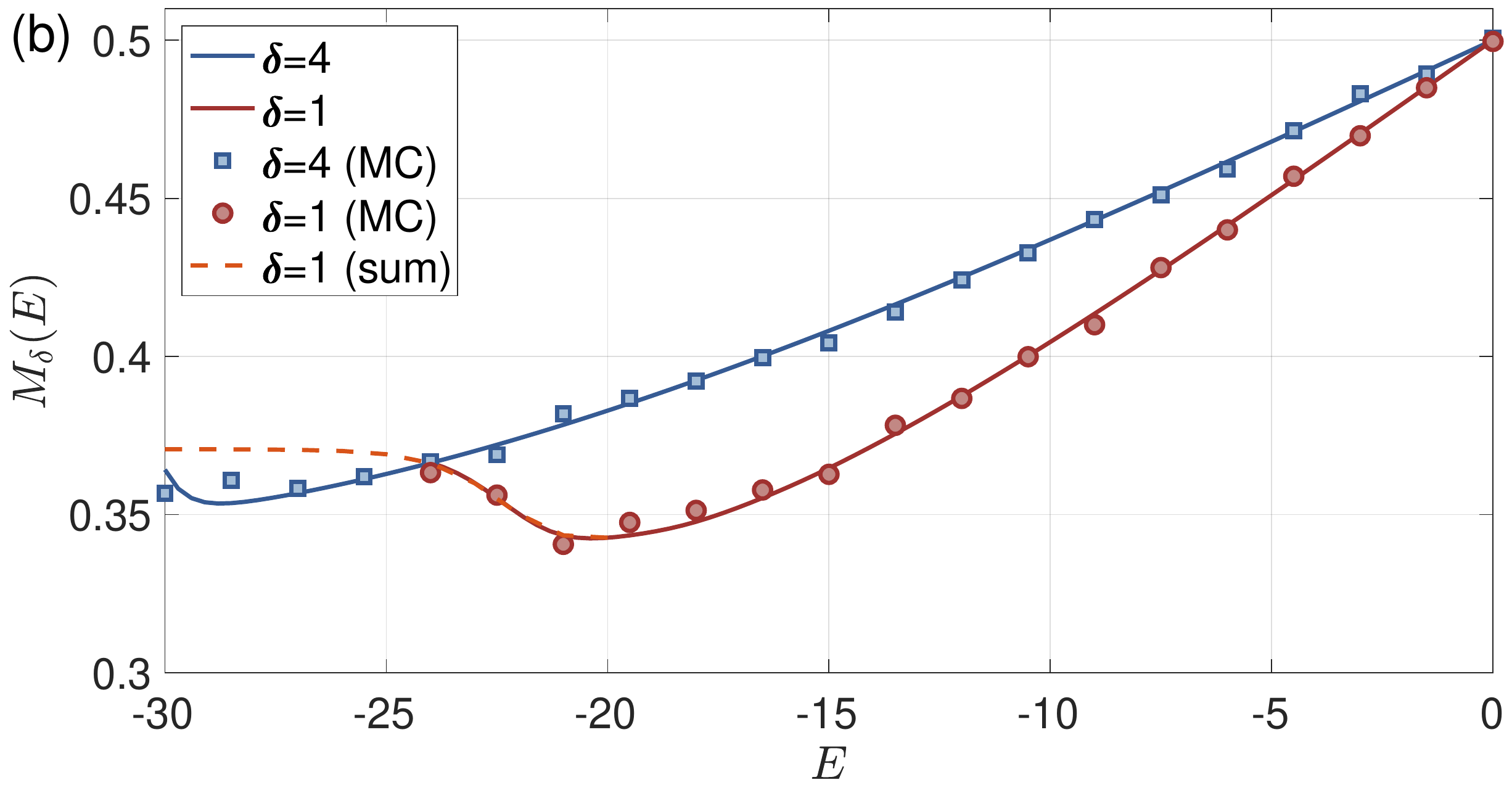}
\caption{Magnetization computed according to (\ref{AMicro}) for Hamiltonian (\ref{Hamfermions}) and $N=20$ spins computed exactly (solid lines) and by the Monte Carlo method described in the text with $10^5$ samples per point (symbols). The Monte Carlo simulations were done for the full expression \eqref{Mt} with $x=3$, for the different values of $\delta$ shown in the legend. Upper figure: $g=1$, $h=2$; Lower figure: $g=2$, $h=1$.}
\label{fig3}
\end{figure}

In Fig.~\ref{fig3}(a) we plot $M_\delta(E)$ for $N=20$ spins, $g=1$, $h=2$ and $\delta=1,\, 4$ (red and blue lines) obtained with the numerical computation. The symbols are obtained with the Monte Carlo method with $\delta=1$ (circles) and $\delta=4$ (squares). We have checked that the results for $\delta=1$ and $\delta=0.1$ are almost indistinguishable, and this is why we plot only the first ones. We have sampled $10^5$ times for each point. In Fig.~\ref{fig3}(b) we plot the same for $g=1$ and $h=2$, also showing very good results. Notice that the lower curve terminates at around $E_k\sim -25$; the reason is that the lowest energy that can be reached with the states (\ref{kkk}) is $E_{
p,{\rm min}}=-20,-14.39$ for Fig.~\ref{fig3}(a,b) respectively, and thus, at lower energies, $D_{\delta,\psi}(E)$ becomes very small for any state $\psi$. Thus, not only the Monte Carlo method but also the exact one encounters convergence problems for such low energies and this is why they are not plotted. We have used an exact summation to obtain the dotted line. This is only possible because we only have $N=20$ spins in this figure.

\begin{figure}[tbp]
\centering
\includegraphics[width=1.0\linewidth]{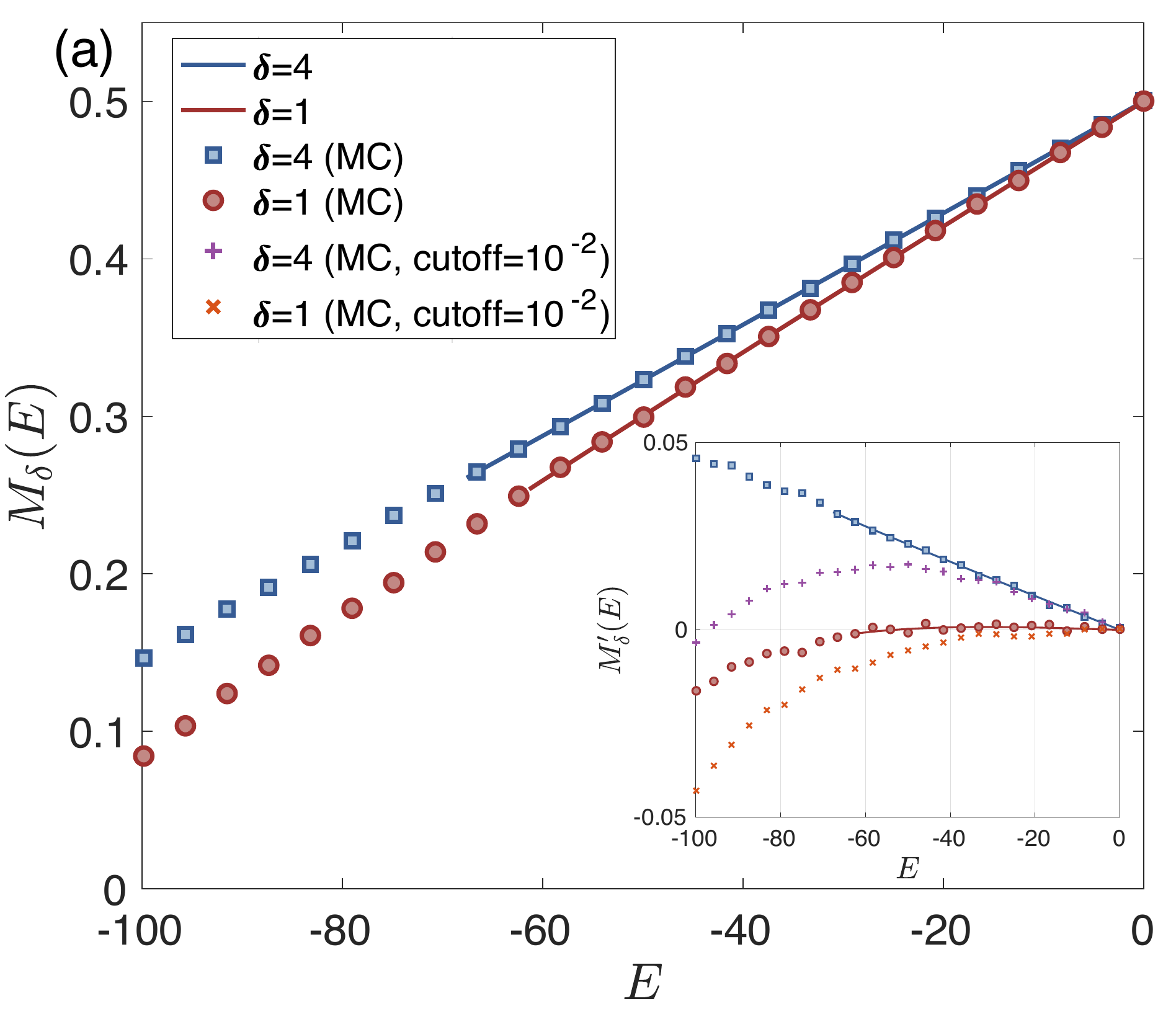}
\includegraphics[width=1.0\linewidth]{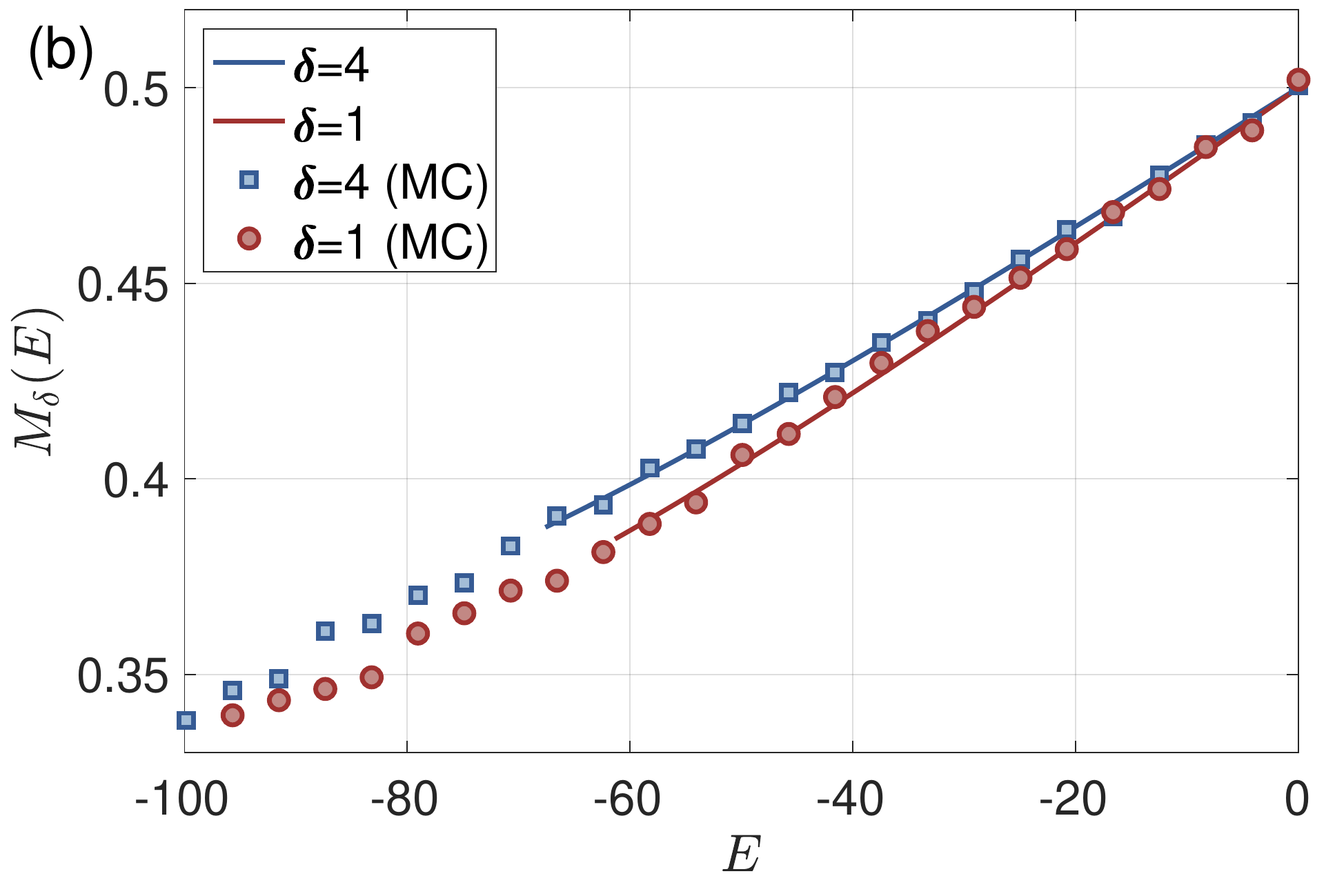}
\caption{Magnetization as a function of energy, as in Fig.~\ref{fig3} but with $N=100$. For the figure in the inset, we have subtracted 
$1/2+0.004E/N$ in order to make the curves more visible. There,
the crosses and plus symbols indicate the values obtained by setting a cutoff of $10^{-2}$ in the sampling procedure to resemble an experiment. The upper figure corresponds to $g=1$, $h=2$, whereas the lower to $g=2$, $h=1$.}
\label{fig4}
\end{figure}

In Fig.~\ref{fig4} we have considered larger systems, with $N=100$. In Fig.~\ref{fig4}(a) we depict the magnetization for $g=1$ and $h=2$. For a more convenient graphical representation of the data, we plot in the inset the modified quantity
$M'_\delta(E)=M_\delta(E)-1/2-0.004E/N$ as a function of $E$. The solid lines indicate the exact results for $\delta=1,\,4$ (red and blue), while the symbols show the Monte Carlo results (circles and crosses for $\delta=1$, squares and plus symbols for $\delta=4$).
As before, we have sampled $10^5$ times for each point. The cross and plus symbols in the inset indicate the results when a cutoff of $10^{-2}$ is set for $D_{\delta,k}(E)$.
That is, in the Metropolis method, as soon as we compute it and obtain a smaller value, we set it to zero. In practice, this intends to resemble an experiment where this quantity has been obtained to that precision. The solid lines are exact results, which we have only computed up to some values of $E$, since otherwise we encountered precision problems. As one can gather from the plots, the Monte Carlo results resemble well the corresponding values. For $E<-60$ the exact method encounters precision problems, while this happens for the Monte Carlo method only for $E<-100= E_{p,{\rm min}}$. When decreasing $\delta$ even further, the results are practically indistinguishable from those of $\delta=1$. However, the program gets unstable for energies $E\alt E_{p,{\rm min}}$ since, as expected, the values of $D_{\delta,k}(E)$ become extremely small due to the Gaussian dependence. 

\subsection{Canonical observable:}

Now we show how, using similar ideas, one can also compute canonical observables, i.e.,
 \be
 A(\beta) = \frac{{\tr}(e^{-\beta H}A)}{{\rm tr}(e^{-\beta H})}
 \ee
where $\beta$ is the inverse temperature. We use the fact that for sufficiently small $\delta$, we can approximate
 \be
 e^{-\beta H} \approx \int_{E_0}^{E_1} dE e^{-\beta E} P_\delta(E)
 \ee
where $E_{rhs)0,1}=E_{0,1}'\mp y \delta$ with $E_0'$ ($E_1'$) the lowest (highest) eigenvalue of $H$, and $y\gg 1$. This motivates the definition
 \be
 A_\delta(\beta) = \frac{\int_{E_0}^{E_1} dE \;e^{-\beta E} {\rm tr}[P_\delta(E)A]}
{\int_{E_0}^{E_1} dE \;e^{-\beta E}{\rm tr}[P_\delta(E)]}
 \ee
which converges to the canonical value for $\delta\to 0$.
Using (\ref{AMicro2}), we have
 \be
 \label{MacroMC}
 A_\delta(\beta) = \frac{\int_{E_0}^{E_1} dE \;e^{-\beta E} \int d\mu_\psi \; D_{\delta,\psi}(E) A_{\delta,\psi}(E) }
{\int_{E_0}^{E_1} dE \;e^{-\beta E} \int d\mu_\psi \; D_{\delta,\psi}(E)}.
 \ee

The quantum algorithm proceeds in the same way as before, by using the quantum simulator to recover $a_{\psi}(t_m)$ and $a_{A,\psi}(t_m)$, and then a classical computation to do the rest. In particular, Monte Carlo samples both the states $\psi$ and the energies $E$ with a probability proportional to $e^{-\beta E} D_{\delta,\psi}(E)$. For observables that are diagonal in the chosen basis, it is also possible to sample only over states, with the integrated probability $\int dE e^{-\beta E} D_{\delta,\psi}(E)$, which can be reconstructed using the quantum simulator in the same way as for the microcanonical observables, since the energy dependence can be integrated analytically. 

\begin{figure}[tbp]
\centering
\includegraphics[width=1.0\linewidth]{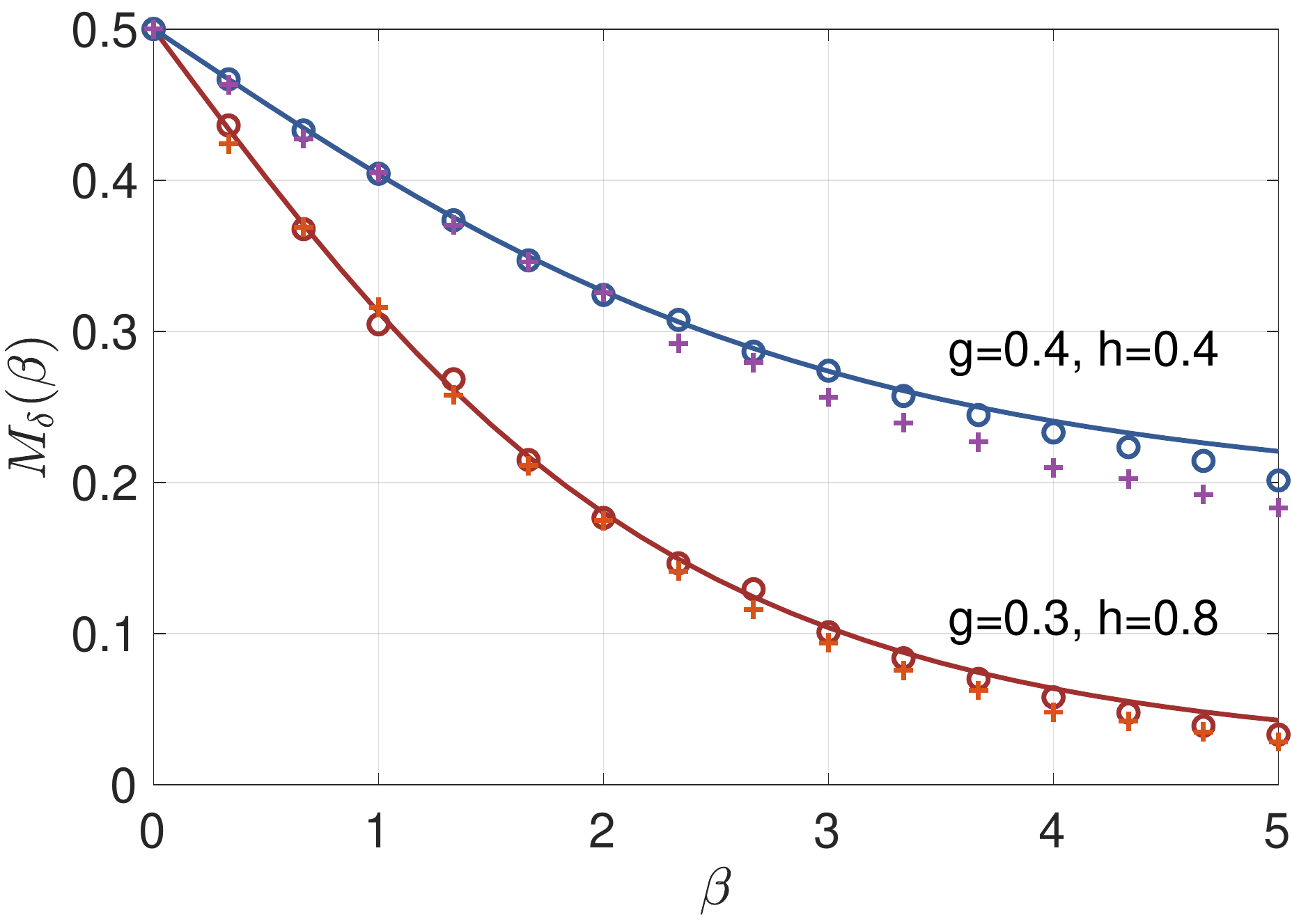}
\caption{Magnetization as a function of the inverse temperature for the Hamiltonian (\ref{Hamfermions}) for $N=100$ spins, $g=0.3$, and  $h=0.8$ (lower curve) and $g=h=0.4$ (upper curve). The solid line represents the exact value $M(\beta)$, whereas the symbols are obtained with the Monte Carlo method, with $10^5$ samples per point and $\delta=1$ and $x=3$. We have discretized the values of the energy from $-N$ to $N$ in intervals of $0.5$. The symbols '+' are obtained by setting a cutoff equals to $10^{-2}$ in $D_{\delta,k}(E)$.}
\label{fig5}
\end{figure}

We have performed Monte Carlo computations and shown the results in Fig.~\ref{fig5}. We have plotted the magnetization (\ref{Magne}) as a function of the inverse temperature $\beta$ for the Hamiltonian (\ref{Hamfermions}) with $N=100$ spins and $g=0.3$, $h=0.8$ (lower curve) and $g=0.4$ and $h=0.4$ (upper curve). The solid line is the exact result (\ref{Abeta1}) computed with the formulas given in Appendix \ref{AppendixIsing}. Note that here there is no problem with the precision, as most of the products in the numerator and denominator cancel and one ends up with a simple sum (\ref{Aexact2}), and that the result is independent of $\delta$ [since the definition (\ref{Abeta1}) is, too]. The symbols are obtained with the Monte Carlo simulation for $\delta=1$ and $x=3$. We have discretized the integral in energy appearing in (\ref{MacroMC}) by taking $E$ from $-N$ to $N$ in steps of $0.5$, and sampling only those values, although the Metropolis algorithm only took values around certain energies, $E\pm 5$, for each value of $\beta$. For the Monte Carlo methods we took $E_{0,1}=\pm 3N/2$, but those values were never reached in the Metropolis sampling. From the figure we conclude that the result converges very well already for $\delta=1$, and also that the performance of the Monte Carlo method is very good. One can observe that for larger values of $\beta$, there is a little bias towards smaller values of the magnetization. In the inset of Fig.~\ref{fig4}, we also display the results carried out with the Monte Carlo computation with a cutoff $10^{-2}$, still rendering competitive results. We note that the fact that the upper curve corresponds to the critical point $g=h$ does not have relevant consequences at the temperatures considered here. For lower temperatures, the program does not give reliable results as it has to scan energies for which $D_{\delta,\psi}(E)$ becomes very small.

\section{Summary and Outlook}

We have proposed and analyzed two types of quantum algorithms to characterize quantum many-body states in finite energies intervals and temperatures. They are based on the cosine-filter, which when applied to a state, reduces its variance to a predetermined value around a given energy. However, instead of preparing the filtered state, our algorithms compute different quantities that allow one to reconstruct expectation values of observables or other magnitudes. The algorithms we proposed are quite flexible - they can either be implemented on digital quantum computers (e.g., based on superconducting qubits, trapped ions, or Rydberg atoms) and on analog quantum simulators (e.g., based on cold atoms, trapped ions, photons), as long as they can perform certain kinds of interferometric measurements.
We have also presented in Appendix \ref{AppendixMeas} more practical approaches which, in some cases, possess clear advantages with respect to such measurements.

We have shown how our first algorithm can be used to to efficiently compute expectation values over filtered states, as long as we can prepare the initial state. The algorithm requires a time that scales polynomially with the system size, $N$, the inverse variance, and the inverse error. We have also shown how it can be simplified in practice, leading to a practicable method for present or planned quantum simulators. The simulator has to be run for times $\le 6/\delta$ (where we took $x\sim 3$ and $r\sim 1$) and therefore for $\delta =O(1)$ this should be feasible for existing devices. The price one has to pay is that one has to perform many more measurements. This number can be estimated by taking into account 
that we need to compute sums of the form \eqref{DpsiE0}, where we add $2 R$ terms so that to have a total error of the order of $\epsilon$, this will require of the order of $R\epsilon^{-2}\approx 3\sqrt{N}/\delta \epsilon^2$ measurements.
 For $N=100$, $\delta=1$ and $\epsilon=10^{-2}$ this yields of the order of $3\cdot 10^5$ measurements. Actually, by taking into account that $c_m$ becomes small, using smaller $x\sim 1$, exploiting symmetries and other optimizations, we expect that it is possible to reduce significantly this number, to the order of $10^4$.

The second algorithm we introduce proposes a way to combine the first algorithm with Monte Carlo simulations in order to obtain the expectation value of microcanonical and canonical observables. The resulting methods also rely on the ability to prepare certain states (e.g. product states) efficiently, to run the quantum simulator for times of the order of $6/\delta$ and to perform interferometric measurements. Our numerical simulations of these algorithms for a simple model indicate that with several tens of thousands of samplings one can obtain reliable results, at least in some energy regimes. Those algorithms can also be significantly improved by using standard Monte Carlo strategies to speed up convergence.
These quantum assisted sampling algorithms open new possibilities to study thermal properties of quantum many-body systems with near-term quantum devices.

There are other modifications that may help to improve the algorithms. First, one could take a different expansion of the filter. For instance, one can use instead of the cosine-filter, Chebyshev expansions for a quantum computer \cite{Ge,Banuls,aless2020spectral,rall2020quantum}, a low-pass filter, or choose $c_m$ differently for an analog simulator to better adapt to the specific errors in which it may incur. Second, in the sampling, there is information that can be collected to investigate other physical questions. For instance, the spins or energy samples that are used in the Monte Carlo methods can be used to estimate other quantities directly (like higher moments), without the need to make other computations. Another possibility is to use the adiabatic or variational algorithms to prepare states with small variance to start with, which would make the algorithms more efficient, or allow us to access energies which are out of the scope of product states.

We have formulated most of our results for a many-body spin Hamiltonian with local interactions. However, they can be equally applied to fermionic systems, systems with longer-range interactions, or disordered setups, as long as the quantum device can emulate the corresponding Hamiltonians. Also, the ideas developed here can be easily adapted to measure other quantities of interest, like Green functions or structure factors.

\acknowledgments

We  acknowledge  support  from  the  ERC  Advanced Grant QUENOCOBA under the EU Horizon 2020 program  (grant  agreement  742102), the Deutsche Forschungsgemeinschaft (DFG, German Research Foundation) under "Germany Excellence Strategy" EXC-2111 – 390814868, and the project number 414325145 in the framework of the Austrian Science Fund (FWF): SFB F71.

\newpage

\appendix

\section{Ways to retrieve \texorpdfstring{$a_{A,\psi}(t)$}{aApsi(t)}}
\label{AppendixMeas}

In this appendix, we describe different methods to retrieve $a_\psi(t)$ and $a_{A,\psi}(t)$ from analog quantum simulators. Those typically possess severe limitations in the available operations as compared with quantum computers, and thus they may not be able to carry out the interferometric measurements that are needed to obtain those quantities. The methods adapt to different situations and have different requirements.

We start out by briefly reviewing the standard method based on conditional dynamics where, during the evolution, all the qubits have to be coupled to an extra one, called the control qubit. Then, we analyze a closely related method where the interaction with the control qubit only needs to occur at the beginning and at the end of the evolution, although at the expense of using other internal states. The third method does not require a control qubit but the possibility of creating certain cat-like states, as well as an extra internal state; the latter can be avoided for certain kind of Hamiltonians for which one can prepare one eigenstate efficiently, like, for instance, Hamiltonians of Heisenberg type. The fourth method builds on the previous one and does not need cat states. This is thus much simpler to implement in practice and may be more robust against decoherence. However, it requires more measurements. The last method applies to Hamiltonians with some special symmetries, like XY or Hubbard models, and can be very practical.

\subsection{Conditional dynamics}

The conceptually simplest way to obtain $a_{A,\psi}(t)$ is to carry out conditional dynamics \cite{Ekert2002} depending on the state of one of the qubits, the control qubit, which we call $c$. This corresponds to the operation
 \bse
 \label{conditional}
 \bea
 U|0\rangle_c \otimes|\psi\rangle &=& |0\rangle_c \otimes e^{-iHt}|\psi\rangle,\\
 U|1\rangle_c \otimes|\psi\rangle &=& |1\rangle_c \otimes|\psi\rangle,
 \eea
 \ese
If we denote by $H_c$ the Hadamard transformation on the
control qubit, then one can first produce the state
 \be
 H_c^\dagger UH_c |0\rangle_c\otimes |\psi\rangle.
 \ee
By then measuring the control qubit in the computational basis, and the observable $A$ on the rest, one can retrieve
(\ref{ApsiA}). This method requires the ability to couple all the qubits to the control one during the evolution in order to apply (\ref{conditional}), which may be difficult in practice. However, it can be very naturally applied in trapped ion simulators \cite{GardinerCiracZoller,ZollerNatPhys2019}, since all ions are coupled to the same phonon bus, and in Rydberg atoms in optical lattices, as one atom in a Rydberg state can influence the dynamics of the rest \cite{Lukin2001}. In fact, very efficient techniques have been proposed to perform this kind of dynamics and measurements using that implementation \cite{Pupillo2010, Jau2016, Glaetzle2017}.

\subsection{Additional internal states}

\begin{figure}[tbp]
\includegraphics[width=0.8\linewidth]{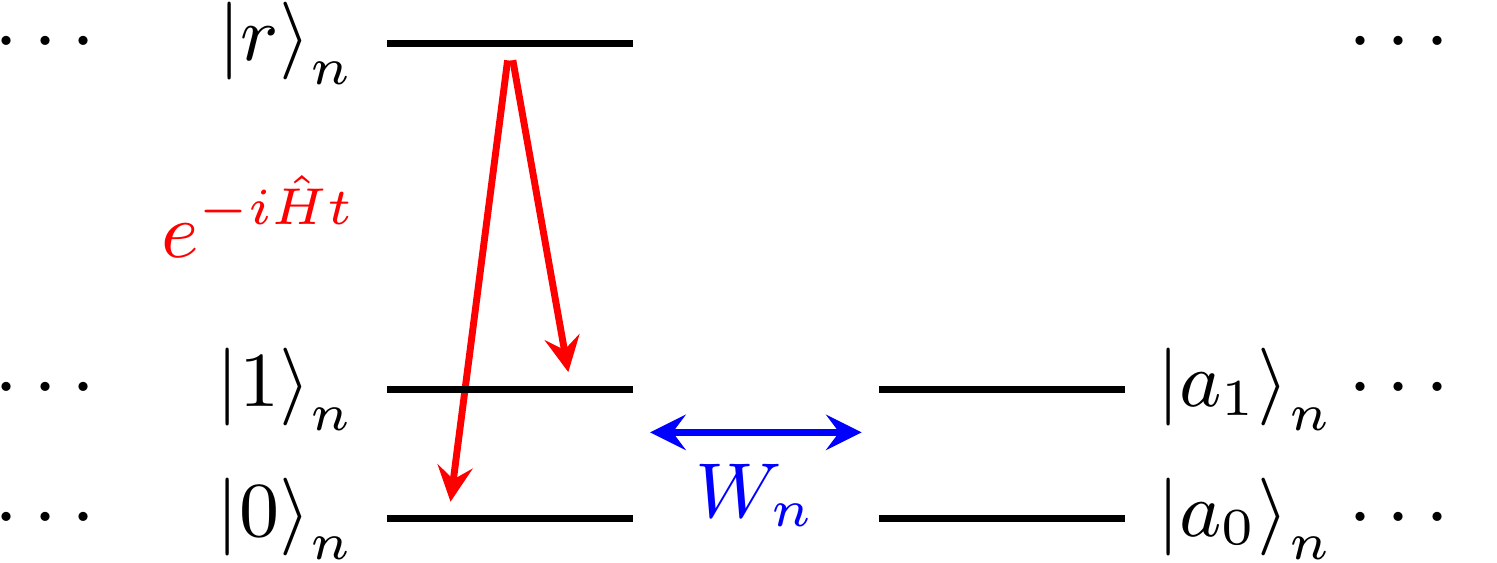}
\caption{Internal level scheme for the interferometric measurement: the simulation acts on levels $|0\rangle$ and $|1\rangle$, while the other two are used to generate the cat-like state.}
\label{figRydberg}
\end{figure}

The next method does not require to couple all the systems to the control qubit during the interaction, but makes use of extra levels in each of the systems. Instead of qubits, one uses four-level systems where, apart from the qubit states $|0\rangle$ and $|1\rangle$, there are other two
$|a_{0,1}\rangle$ (see Fig.~\ref{figRydberg}) that are idle with respect to the action of the Hamiltonian; that is, $\tilde U=e^{-iHt}$ only acts non-trivially in the subspace spanned by $|0\rangle$ and $|1\rangle$. Let us assume that one has the possibility of adding extra 2-body operations
 \bse
 \bea
 W_n |0\rangle_c\otimes |i\rangle_n &=& |0\rangle_c\otimes |i\rangle_n,\\
 W_n |a_0\rangle_c\otimes |i\rangle_n &=& |a_0\rangle_c\otimes |a_i\rangle_n,
 \eea
 \ese
as well as the Hadamard $H_c$ between levels $|0\rangle_c$ and
$|a_0\rangle_c$. Then, one can implement
 \be
   H_c^\dagger W^\dagger \tilde U WH_c |0\rangle_c\otimes |\psi\rangle
 \ee
where $W=\otimes_{n=1}^N W_n$. As before, by measuring in the basis $|0\rangle_c,|a_0\rangle_c$ the control qubit, and the observable $A$ on the rest, then one can also obtain (\ref{ApsiA}).

\subsection{Cat-like states}

We show now that neither an ancilla system nor additional internal states are required as long as one can prepare some cat-like states. This has been achieved with different setups already \cite{Laflamme1998NMRGHZ,Neumann2009GHZNV,Leibfried2005GHZ6,
Monz2011GHZ14,Dicarlo2010GHZ3,Song2017GHZ10,Wang2018GHZ18,
Friis2018GHZ20,LukinScienceSchCats,IBMGHZ18}.

Let us first assume that $H$ has an eigenstate $\varphi$, i.e
 \be
 H|\varphi\rangle=\lambda|\varphi\rangle
 \ee
such that, for a given state $\psi$, one can efficiently apply an operator $V(\theta)$, fulfilling
 \be
 \label{right}
 V(\theta) |0,\ldots,0\rangle = \frac{1}{\sqrt{2}} (|\varphi\rangle + e^{i\theta} |\psi\rangle).
 \ee
In general, $\psi$ and $\varphi$ may be very different, in the sense that they can be distinguished by measuring few qubits \cite{Korsbakken2007}, and thus the state on the right of (\ref{right}) will be a cat-like state. In some cases, $\varphi$ is just a product state. This occurs, for instance, for the Heisenberg Hamiltonian (in any dimension)
 \be
 H = \sum_{n,m} J_{n,m} \vec\sigma_{n} \cdot \vec\sigma_{m} + \sum_n h_n\sigma_{n,z}.
 \ee
Indeed, for arbitrary $J_{n,m}$ and $h_n$, $|\varphi\rangle=|0,\ldots,0\rangle$ is an eigenstate. Here $\vec \sigma=(\sigma_x,\sigma_y,\sigma_z)$ is a vector of Pauli operators and $\sigma_{n,z}$ is the one that acts on the $n$-th qubit.

Under the above assumption, one can prepare
 \be
 \label{A8}
 |\Phi(\theta,\theta')\rangle=V(\theta')^\dagger e^{-iHt} V(\theta) |0,\ldots,0\rangle
 \ee
and measure in the computational basis to obtain $a_\psi(t)$. In order to show that, without loss of generality we set $\lambda=0$ as we can always subtract it from the Hamiltonian. The probability of obtaining the state
$|0,\ldots,0\rangle$ is
 \be
 \left|\langle  0,\ldots,0|\Phi(\theta,\theta')\rangle \right|^2 = \frac{1}{4} \left|e^{-i\alpha}+ a_\psi(t)e^{i\alpha} + 2r\cos(\beta)  \right|^2
 \ee
where $\langle \psi|\varphi\rangle=r e^{-i\gamma}$, $r\ge 0$, $\alpha=(\theta-\theta')/2$ and $\beta=(\theta+\theta')/2+\gamma$. Note that, apart from $a_\psi(t)$, $r$ and $\gamma$ are, in principle, unknown. If they can be computed classically (e.g., if we deal with product states), then one can directly obtain $a_\psi(t)$ by measuring this quantity for $\beta=0$ and $\alpha=0,\pi/2$ and $\pi/4$. If not, one can proceed as follows. First, set $\alpha=0$; by changing $\theta+\theta'$ one can find when $\beta=0$ or $\pi$, as the expression is maximal or minimal. Then, one can also identify when $\beta=\pi/2$. Using these three values, it is possible to recover the real part of $a_\psi(t)$. By choosing $\alpha=\pi/2$ and proceeding in the same way, one can then get the imaginary part. The procedure to measure $a_{A,\psi}(t)$ is the same, but: (i) one has to apply the operator $A$ before the measurement; (ii) one has to compute classically (or measure with the simulator) $\langle \varphi|A|\varphi\rangle$. The first task can be obtained as follows, as long as one can use the simulator to evolve according to the dynamics generated by $A$. Then, one just applies the operator $\exp({-i A \delta t})\approx \Id -i A\delta t$ for a short time $\delta t\ll \|A\|_\infty$ and measures as before. If $A$ is not Hermitian, one can always write it as a sum $A=A_r + i A_i$, where $A_r=A_r^\dagger$ and $A_i=A_i^\dagger$ and measure these two independently. In practice, there may be simpler procedures that do not involve small times, for instance, if $A^2=\Id$.

So far we have requested that $H$ has an eigenstate that can be easily prepared. This condition can always be satisfied so long as one has an extra level available on each system, $|a_0\rangle$, where the Hamiltonian does not act. In that case, taking $|\varphi\rangle=|a_0,\ldots,a_0\rangle$ one has $H|\varphi\rangle=0$ and thus it is an eigenstate.

\subsection{Product states}

The procedure presented above can be further simplified if one wants to measure $a_{A,\psi}(t)$ for states of the form
 \be
 |\psi\rangle=\prod_{n=1}^N \sigma_{n,\alpha_n}|\varphi\rangle,
 \ee
where $\sigma_{\alpha_n}$ is a Pauli operator. For instance, if $\varphi$ is a product state this allows one to obtain (\ref{ApsiA}) for any product state. Also, it can be easily extended to states that are connected with products of other simple operators.

The idea is to sequentially measure $a_{A,\varphi_m}(t)$ (for $m=1$ to $m=N$), where
 \be
 |\varphi_m\rangle =\prod_{n=1}^m\sigma_{n,\alpha_n}|\varphi\rangle
  = \sigma_{m,\alpha_m}|\varphi_{m-1}\rangle,
 \ee
$\varphi_0=\varphi$ and $\varphi_N=\psi$. Once $a_{\varphi_{m-1}}(t)$ is obtained, one can start out with $\varphi_{m-1}$ and obtain $a_{A,\varphi_{m}}(t)$ since the procedure explained right after (\ref{A8}) can be applied if $a_\varphi(t)$ is known (it does not have to be equal to one).

With single qubit operations only acting on the $m$-th qubit, one can prepare
 \be
 V_m(\theta) |\varphi_{m-1}\rangle = \frac{1}{\sqrt{2}}( |\varphi_{m-1}\rangle + e^{i\theta} |\varphi_{m}\rangle)
 \ee
and follow the same procedure as before. This method does not require the preparation of cat-like states, as the two states building the superposition  at any given step differ just on one qubit. 
However, it requires many more measurements since one has to obtain $a_{A,\varphi_n}(t)$ for $n\le m$. Note that one has to obtain, at most, $N$ of those quantities. Note also that this procedure may have some practical restrictions as the errors accumulate.

\subsection{Symmetric Hamiltonians}

A special class of Hamiltonians for which one can use a simplified procedure contains those for which there exists a unitary operator, $R$, and a Hermitian operator, $T$, such that
 \be
 \label{HRRH}
 R^\dagger H R=-H + T
 \ee
with $[T,H]=0$ 
and 
for which $\psi$ is an eigenstate, i.e.
 \be
 \label{eq:propPsi}
 R|\psi\rangle=\lambda|\psi\rangle,\quad T|\psi\rangle=\mu|\psi\rangle.
 \ee

Given such $R$, $T$ and $\psi$, then
 \[
 a_\psi(t) =\langle \psi|R^\dagger e^{-iHt} R|\psi\rangle  =
 \langle \psi| e^{i(H-T)t} |\psi\rangle =
\overline{a_\psi(t)}e^{-i\mu t}.
 \]
Thus, $a_\psi(t)e^{i\mu t/2}$ is real. Moreover, if 
\be
R^\dagger A R = \pm A, 
\label{eq:rar_pm}
\ee
and additionally either $\psi$ is an eigenstate of $A$ (with eigenvalue 1 for simplicity), or $[A,H]=0$,
then $a_{A,\psi}(t)e^{i\mu t/2}$ is also real or purely imaginary, depending on the sign in \eqref{eq:rar_pm}.
To determine $a_{A,\psi}(t)$, then, one only needs the absolute value $|a_{A,\psi}(t)|$ and a sign. Since
\be
 |a_{A,\psi}(t)|^2 = |\langle \psi|A e^{-iHt}|\psi\rangle|^2,
 \ee
the former can be found just by letting the system evolve and then measuring in a basis that contains $A|\psi\rangle$.
The sign change can be inferred if we track when the absolute value becomes zero as follows. Because we have a finite system,  $a_{A,\psi}(t)$ will be an analytic function of $t$ so that we can Taylor expand it when it is near zero. Suppose $a_{A,\psi}(t)$ approaches zero at a time $t_0$, then one can expand it to the leading order: $a_{A,\psi}(t)\approx \alpha (t-t_0)^n+O((t-t_0)^{n+1})$. If $n$ is odd, then $a_{A,\psi}(t)$ will change sign and if it is even it will not. Since we are measuring the square of $a_{A,\psi}(t)$, $|\langle \psi|A e^{-iHt}|\psi\rangle|^2$, we can figure out what is $2n$, and thus $n$. In practice, $n=1$ or 2, so that this can be obtained more easily.

Hence, here we only require individual measurements, available in several labs worldwide that possess analog quantum computers \cite{BlochManyBodyLocalin1D,Parsons2016,Chiu2019,MonroeTimeCrystal,BlattEE,DeLeseleuc2019}

In the rest of this section, we show physically relevant examples in which one  can find the operators $R$ and $T$ with specially simple structure, and a whole basis of common product eigenstates $\psi$ can be used to run the algorithms from Sect.~\ref{practical-computation}.
There are many textbook Hamiltonians and states fulfilling properties \eqref{HRRH}, \eqref{eq:propPsi}. For instance, if one has a bipartite lattice with sublattices A and B, and $N$ sites in total, and a Hamiltonian of the form
 \bea
  \label{HXY}
 H &=& \sum_{n\in A,m\in B} \left[ J_{n,m,x} \sigma_{n,x} \sigma_{m,x}  + J_{n,m,y} \sigma_{n,y} \sigma_{m,y}\right] \nonumber\\ &&+ \sum_{n=1}^N h_n \sigma_{n,z}
 \eea
then
 \be
 R=\otimes_{n\in A} \sigma_{n,x} \otimes_{m\in B} \sigma_{m,y}
 \ee
fulfills (\ref{HRRH}) with $T=0$. Therefore, if one chooses $\psi$ as an eigenstate of $R$ (e.g., a product state), the requirements are satisfied. The operator $R$ is, up to single qubit rotations, a parity operator.

The Hubbard model in a bipartite lattice
 \bea
 H &=& \sum_{n\in A,m\in B}\sum_{\sigma=\uparrow,\downarrow} J_{n,m,\sigma} \left[ a^\dagger_{n,\sigma}a_{m,\sigma} + a^\dagger_{m,\sigma}a_{n,\sigma}\right]\nonumber\\
&& + \sum_{n=1}^N U a^\dagger_{n,\uparrow}a_{n,\uparrow}
a^\dagger_{n,\downarrow}a_{n,\downarrow}
 \eea
with $a^\dagger$ and $a$, fermionic creation and annihilation operators, also fulfills (\ref{HXY}). In such a case,
\bse
 \bea
 R &=& e^{i \pi \sum_{n\in A} a^\dagger_{n,\downarrow}a_{n,\downarrow}} 
\prod_{n} (a_{n,\uparrow} +  a_{n,\uparrow}^\dagger),\\
 T &=& \sum_{n=1}^N U a^\dagger_{n,\downarrow} a_{n,\downarrow}.
 \eea
 \ese
We can find a whole orthogonal basis of common eigenstates of $R$ and $T$ with simple structure in the following way.
In particular, if the number of sites $N$ is even, we can divide the lattice in $N/2$ disjoint pairs of sites, $S_j=(n_j,m_j)$. For each pair $S_j$ we choose a state $\ket{\phi_{0,j}}$ which is the vacuum of spin up, and an arbitrary Fock state of spin down modes, and define
\bea
\ket{\varphi_j^{\pm}}&=&(a^\dagger_{n_j,\uparrow}\pm i a^\dagger_{m_j,\uparrow})\ket{\phi_{0,j}},  \\
\ket{\phi_j^{\pm}}&=&(1\pm i a^\dagger_{n_j,\uparrow} a^\dagger_{m_j,\uparrow})\ket{\phi_{0,j}}.
\eea
Then, an orthogonal basis of common eigenvectors $\ket{\psi}$ can be formed as all possible products of one of these factors for each pair of sites. 

\section{Ising model}
\label{AppendixIsing}

In this appendix we give some formulas that we have used in our numerical illustrations regarding the Ising model. Let us consider $N$ fermionic modes and a Hamiltonian
 \be
 \label{HIsing}
  H = \frac{g}{2} \sum_{n=1}^N (a_n+a_{n}^\dagger) (a_{n+1}- a_{n+1}^\dagger) + h \sum_n (a_n^\dagger a_n-1/2),
 \ee
where $a_n$ are annihilation operators of the vacuum $|{\rm vac}\rangle$ and we have taken periodic boundary conditions for the fermions, $a_{N+1}=a_1$. Through the Jordan-Wigner transformation, this corresponds to the Ising-like Hamiltonian
 \be
  H = \frac{g}{2} \sum_{n=1}^N \sigma_{n,x}\sigma_{n+1,x} + \frac{h}{2} \sum_{n=1}^N \sigma_{n,z}
 \ee
where $\sigma_{x,z}$ are Pauli operators, with appropriate boundary conditions.

As usual, we first perform a Fourier transform with (we assume even $N$)
 \begin{eqnarray*}
 b_k &=& \frac{1}{\sqrt{N}} \sum_{n=1}^N e^{i2\pi n k /N} a_n, \\
 a_n &=& \frac{1}{\sqrt{N}} \sum_{k=-N/2+1}^{N/2} e^{-i2\pi n k /N} b_k,
 \end{eqnarray*}
The new Hamiltonian is
 \be
 H = \sum_{k=0}^{N/2} \tilde H_k
 \ee
where
 \begin{eqnarray*}
 \tilde H_k &=& x_k (b_k^\dagger b_k + b_{-k}^\dagger b_{-k}-1) + i y_k
 (b_k b_{-k} - b_{-k}^\dagger b_{k}^\dagger) ,\\
 \tilde H_0 &=& x_0 (b_0^\dagger b_0-1/2),\\
 \tilde H_{N/2} &=& x_{N/2} (b_{N/2}^\dagger b_{N/2}-1/2),
 \end{eqnarray*}
and
 \begin{eqnarray*}
 x_k&=& h+g \cos(2\pi k/N), \\
 y_k&=& g \sin(2\pi k/N).
 \end{eqnarray*}
Note that ${\rm tr}({\tilde H_k})=0$.

We now proceed as follows: for each value of $k=1,\ldots,N/2-1$ we write $H_k=\tilde H_k$ as a $4\times 4$ matrix in the basis
 \begin{eqnarray*}
 |1\rangle &=& |{\rm vac}\rangle,\\
 |2\rangle &=& b_{k}^\dagger b_{-k}^\dagger|{\rm vac}\rangle,\\
 |3\rangle &=& b_{k}^\dagger |{\rm vac}\rangle,\\
 |4\rangle &=& b_{-k}^\dagger |{\rm vac}\rangle,
 \end{eqnarray*}
and also define $H_{0}=\tilde H_0+\tilde H_{N/2}$ and write it as
a $4\times 4$ matrix in the basis
 \begin{eqnarray*}
 |1\rangle &=& |{\rm vac}\rangle,\\
 |2\rangle &=& b_{N/2}^\dagger b_{0}^\dagger|{\rm vac}\rangle\\
 |3\rangle &=& b_{N/2}^\dagger|{\rm vac}\rangle,\\
 |4\rangle &=&  b_{0}^\dagger  |{\rm vac}\rangle.
  \end{eqnarray*}
We obtain
 \be
 H = \sum_{k=0}^{N/2-1} H_k
 \ee
where
 \begin{eqnarray*}
H_k &=& [x_k \sigma_z + y_k \sigma_y] \oplus 0,\\
 H_{0} &=& [x_{0,+} \sigma_z]\oplus [x_{0,-} \sigma_z],
 \end{eqnarray*}
with
 \be
 x_{0,\pm} = (x_0\pm x_{N/2})/2,
 \ee
 and 
$ \sigma_z=\big(\begin{smallmatrix}
  -1 & 0\\
  0 & 1
\end{smallmatrix}\big)$.
The direct sum structure corresponds to the subspaces $\{|1\rangle,|2\rangle\}$ and $\{|3\rangle,|4\rangle\}$. Thus, the problem is reduced to $N/2$ non-interacting 4-level systems.

We can easily compute the eigenvalues of $H_k$. For $k\ne 0$, they are given by
$E^{1,2}_{k,\pm}= \pm z_k$ and $E^{3,4}_{k,\pm}=0$ (doubly degenerate), where
 \be
 z_k = \sqrt{x_k^2+y_k^2},
 \ee
whereas for $k=0$ by $E^{1,2}_{0,\pm}=\pm x_{0,+}$ and
$E^{3,4}_{0,\pm}=\pm x_{0,-}$.

The expectation value of the Hamiltonian in a product state $|p\rangle=|p_1,\ldots,p_{N/2}\rangle$ with $p_k=1,\ldots,4$ is
 \be
 E_p=\langle p| H |p\rangle = \sum_{k=0}^{N/2-1} \langle p_k|H_k|p_k\rangle = \sum_{k=0}^{N/2-1} E_{p_k}
 \ee
where
 \begin{eqnarray*}
 E_{k,1}&=& \langle 1|H_k|1\rangle = -x_k, \\
 E_{k,2}&=&\langle 2|H_k|2\rangle = x_k,\\
 E_{k,3}&=&\langle 3|H_k|3\rangle = 0, \\
 E_{k,4}&=&\langle 4|H_k|4\rangle = 0,
 \end{eqnarray*}
and the rest zero, for $k=1,\ldots,N/2-1$ and
 \begin{eqnarray*}
 E_{0,1}&=&\langle 1|{H_{0}}|1\rangle =- x_{0,+}, \\
 E_{0,2}&=&\langle 2|{H_{0}}|2\rangle = x_{0,+}, \\
 E_{0,3}&=&\langle 3|{H_{0}}|3\rangle = - x_{0,-},\\
 E_{0,4}&=&\langle 4|{H_{0}}|4\rangle = x_{0,-}.
 \end{eqnarray*}

In order to evaluate the expressions required for the simulation, we need to compute
 \be
\label{pkeiH}
 \langle p|e^{i  Ht} |p\rangle = \prod_{k=1}^{N/2}
 \langle p_k|e^{i H_k t}|p_k\rangle.
 \ee
We find
 \bse
 \label{expiHkt1}
 \begin{eqnarray}
 \langle 1|e^{i H_kt}|1\rangle &=& \cos(z_kt)
- i \sin(z_kt) x_k/z_k,\nonumber \\
 \langle 2|e^{iH_kt}|2\rangle &=& \cos(z_kt)
+ i  \sin(z_kt) x_k/z_k,\nonumber\\
 \langle 3|e^{iH_kt}|3\rangle &=& \langle 4|e^{iH_kt}|4\rangle=1,
 \end{eqnarray}
for $k\ne 0$, whereas
 \be
 \label{expiH0t1}
 \langle n|e^{i H_{0}t}|n\rangle =e^{i E_{0,n} t}
 \ee
\ese
for $n=1,\ldots,4$.

Although we are interested here in finite energies and temperatures, we now briefly discuss the zero temperature behavior (i.e., the ground state). In the basis introduced here, it can be written as
 \be
 \prod_{k} (\alpha_k |1\rangle_k + \beta_k |2\rangle_k)
 \ee
with eigenvalue $E_0=-\sum_k z_k$. The coefficients $\alpha_k$ and $\beta_k$ depend on $g$ and $h$. In particular, for $h\gg g$ we have $\alpha_k\sim 1$ and $\beta_k\sim 0$, whereas for $h\ll g$ they change with $k$, and thus one gets superpositions of many configurations. At $g=h$, $z_{N/2-1}\to 0$ as $N\to\infty$, so that the gap closes and there is a quantum phase transition.

\subsection{Microcanonical average}

We compute now some of the quantities that are used in the numerical illustrations. For the sake of simplicity, let us assume that
 \be
\label{Ak}
 A= \sum_{k=0}^{N/2-1} A_k,
 \ee
where $A_k$ only depends on $b_{\pm k}$ for $k\ne 0$ and on $b_{0,N/2}$ for $k=0$. { Mapping back to qubits, such operator $A$ is that can be decomposed into a sum of local operators, of which the magnetization we used in Eq. (\ref{Magne}) is an example.}

Let us start with
 \[
 \langle p|P_{\delta}(E)|p\rangle = \sum_{m=-R}^R c_m e^{i2mE/N}
 a_{p}(t_m)
 \]
where $c_m$ are defined with respect to $\delta$ and $R$ is given in (\ref{Mt}) or (\ref{Mt2}). We have
 \[
 a_{p}(t_m)=\langle p|e^{-i2mH/N}|p\rangle=\prod_{k} \langle p_k| e^{-i2mH_k/N}|p_k\rangle
 \]
which can be readily computed using (\ref{expiHkt1}).

We can also compute (\ref{AMicro}) directly,
 \be
 A(E) = \frac{\sum_{m=-R}^R c_m e^{i 2mE/N} n_m}
 {\sum_{m=-R}^R c_m e^{i 2mE/N} d_m}
 \ee
where
 \begin{eqnarray*}
 n_m &=& \frac{1}{2^N} {\rm tr}(e^{-i 2m H/N} A)\\
&=& \frac{1}{2^N} \sum_
{k=0}^{N/2-1} \left[ \prod_{q=0,\, q\ne k}^{N/2-1}
{\rm tr}\left(e^{-i2mH_q/N}\right)\right] \left[ {\rm tr}( e^{-i2mH_k/N} A_k) \right],\\
 d_m &=& \frac{1}{2^N}{\rm tr}(e^{-i2mH/N})= \frac{1}{2^N} \prod_{k=0}^{N/2-1}
 {\rm tr}( e^{-i2mH_k/N}).
 \end{eqnarray*}
 
Defining
  \begin{eqnarray*}
 r_{m,k} &=& \frac{1}{4}\tr(e^{-i2mH_k/N}),\\
 s_{m,k}^{(A)} &=& \frac{1}{4}\tr(A_k e^{-i2mH_k/N}),
 \end{eqnarray*}
 we can write
  \begin{eqnarray*}
 n_m &=& \sum_{k=0}^{N/2-1} \left[ \prod_{q=0,\, q\ne k}^{N/2-1} r_{m,q} \right] s_{m,k},\\
 d_m &=& \prod_{k=0}^{N/2-1}
 r_{m,k}.
 \end{eqnarray*}
Using \eqref{expiHkt1}, \eqref{expiH0t1} we have
 \[
 r_{m,k} =\cos^2\left(\frac{m z_k}{N}\right),
 \]
 for $k=1,\ldots,N/2-1$ and
 \begin{eqnarray*}
 r_{m,0} &=& \frac{1}{2}[\cos(2m x_{0,+}/N)+\cos(2m x_{0,-}/N)],
 \end{eqnarray*}
 while the values of $s_{m,k}$ depend on the observable.

\subsection{Canonical average}

In a similar way we may compute the canonical average
   \be
\label{Abeta1}
 A(\beta) = {\tr}(e^{-\beta H}A)/{\rm tr}(e^{-\beta H}).
 \ee
We have
 \begin{eqnarray*}
 {\rm tr}(e^{-\beta H}) &=& 2^N\prod_k \tilde r_k,\\
 {\rm tr}(e^{-\beta H}A ) &=& 2^N \sum_k \left[\prod_{q\ne k} \tilde r_q \right] \tilde s_k= {\rm tr}(e^{-\beta H}) \sum_k \frac{\tilde s_k}{\tilde r_k},
 \end{eqnarray*}
and thus
 \be
 \label{Aexact2}
 A(\beta)=  \sum_k \frac{\tilde s_k}{\tilde r_k},
 \ee
where
 \begin{eqnarray*}
 \tilde r_k &=& \frac{1}{4} \tr (e^{-\beta H_k}), \\
  \tilde s_k^{(A)} &=& \frac{1}{4} \tr (A_k e^{-\beta H_k}).
 \end{eqnarray*}
Using \eqref{expiHkt1} and \eqref{expiH0t1},
 \begin{eqnarray*}
 \tilde r_0 &=& \frac{1}{2}[\cosh(\beta x_{0,+})+ \cosh(\beta x_{0,-})],\\
 \tilde r_k &=& \cosh^2(\beta z_k/2).
 \end{eqnarray*}

\section{Efficient computation}
\label{AppendixEfficient}

Here we prove the statement that was used to show that one can compute (\ref{Adeltapsi1}) efficiently. Namely, we show that for any state $\psi$, 
and for any $\delta\le N/\sqrt{2}$, there exists an interval of width at least $\delta^2/6N$ contained in an energy window $|E-E_{\psi}|\le r \sigma_\psi$ (\ref{Einterval})
such that, for any $E$ in that interval,
 \be
 \label{nE2}
 n(E):=\langle \psi| P_\delta(E) |\psi\rangle \ge \frac{1}{4} \left( \frac{\delta^2}{\delta^2+2\sigma_\psi^2}\right )^{3/2}, 
 \ee
and
 \be
 \label{rrr}
 r=[3\log[2(1+2\sigma_\psi^2/\delta^2)]]^{1/2}.
 \ee
We emphasize that for our purposes we do not need a tight bound, so that we will be very rough when bounding different quantities with the goal of obtaining simple expressions. 
Moreover, although $n(E)$ involves the cosine filter \eqref{Cosinefilt}, it is simpler to bound the result of the Gaussian approximation \eqref{Gaussfiltapprox}.
Specifically, we prove that there is an energy interval of radius at least $\tilde{\delta}^2/3N$, contained in the window (\ref{Einterval}) and 
with its center inside the spectral limits of $H$, such that for any energy inside this interval,
 \be
 \label{nE}
 \tilde n(E):=\langle \psi| e^{-(H-E)^2/2\tilde \delta^2}|\psi\rangle \ge \frac{1}{4(1+\sigma_{\psi}^2/\tilde{\delta}^2)^{3/2}} 
 \ee
where $\tilde \delta=\delta/\sqrt{2}$. Since $\cos(x)>e^{-x^2}$ for $|x|<1.3$, if $E$ fulfills
 \be
 \label{HEN}
 \|(H-E)/N \|_\infty < 1.3,
 \ee
we will have that $n(E)\ge \tilde n(E)$. 
Because  $\| H\|_{\infty}=N/2$, and the center of the interval is within the limits of the spectrum of $H$, we can ensure that 
$\|H-E\|\le N+\delta^2/6N\le13 N/12$, and thus
(\ref{HEN}) is satisfied for $E$,
so that bounding the Gaussian approximation is enough for the desired result.
 
Let us denote by
 \be
 \label{Tr}
 T(r)= \int_{E_\psi-r\sigma_\psi}^{E_\psi+r\sigma_\psi} dE \; \tilde n(E) \frac{e^{-(E-E_\psi)^2/2\sigma_\psi^2}}{\sqrt{2\pi}\sigma_\psi}.
 \ee
First, we will show that for (\ref{rrr}),
$T(r)$ is upper and lower bounded by some quantity, which will indicate that there exists some $E$ in the interval such that $\tilde n(E)$ is sufficiently large. Then, by upper bounding the derivative of $\tilde n(E)$, we will conclude that there is a neighbourhood of that $E$ where $\tilde n(E)$ fulfills the required condition.

We write $T(r)=T_0 - T_1(r)$, where
 \be
 \label{T0}
  T_0 = \int_{-\infty}^{\infty} dE \;\tilde n(E) \frac{e^{-(E-E_\psi)^2/2\sigma_\psi^2}}{\sqrt{2\pi}\sigma_\psi}
 \ee
and $T_1(r)=T(r)-T_0$. Given that $\tilde n(E)\le 1$, we have 
 \bea
 |T_1(r)| & \le & \sqrt{\frac{2}{\pi}} \int_{r}^{\infty} dE e^{-E^2/2} ={\rm erfc} (r/\sqrt{2})\\
 & \le & \sqrt{\frac{2}{\pi}}\frac{e^{-r^2/2}}{r} \le e^{-r^2/2},
 \eea
 where the last step uses that for  \eqref{rrr}, $r\le \sqrt{2/\pi}$. 
We can perform the integration in (\ref{T0}) explicitly using (\ref{nE}) with (\ref{rrr}), to get
 \be
 T_0 = \frac{\langle \psi| e^{-(H-E_\psi)^2/[2(\sigma_\psi^2+ \tilde \delta^2)]} |\psi\rangle}{\sqrt{1+\sigma_\psi^2/\tilde\delta^2}}.
 \ee
Using $e^{-x^2}\ge 1-2 x^2$ we obtain
 \be
 T_0 \ge (1+\sigma_\psi^2/\tilde \delta^2)^{-3/2}.
 \ee
 Thus, putting things together we obtain the lower bound
 \be
 T(r) \ge   (1+\sigma_{\psi}^2/\tilde \delta^2)^{-3/2} - e^{-r^2/2}\ge
\frac{1}{2} (1+\sigma_{\psi}^2/\tilde \delta^2)^{-3/2},
 \ee
where we have chosen (\ref{rrr})
and used that $r^2/2> \log[2(1+\sigma_\psi^2/\tilde \delta^2)^{3/2}]$. 

In order to bound $T(r)$ from above, let us denote by $E_0$ the value where $\tilde n(E)$ attains its maximum within the interval (\ref{Einterval}). 
Note that $E_0$ must be within $[E_{\min},E_{\max}]$ (since, if it was outside, choosing the closest extreme eigenvalue of $H$ 
would yield a larger value for $\tilde{n}$). We then have 
\be
T(r)\le \tilde n(E_0) \frac{1}{\sqrt{2 \pi} \sigma} \int_{E_{\psi}-r\sigma_{\psi}}^{E_{\psi}+r\sigma_{\psi}} dE e^{-(E-E_{\psi})^2/2 \sigma_{\psi}^2}\le \tilde n(E_0), 
\ee
so that
 \be
 \tilde n(E_0)\ge  \frac{1}{2} (1+\sigma_{\psi}^2/\tilde \delta^2)^{-3/2}.
 \ee

It is now possible to show that in a neighbourhood of $E_0$ of radius $\delta^2/3N$, $\tilde{n}$ takes sufficiently large values,
$\tilde n(E)\ge \tilde n(E_0)/2$.
For that, we just have to bound the derivative of $\tilde{n}$ within the interval (\ref{Einterval}). 
Let us call $\tilde n'_{\rm max}$ the maximal value of the derivative of $\tilde n(E)$ with respect to $E$ in that interval, which occurs at some $E=E_1$.
Although $E_1$ could lie outside the limits of the spectrum of $H$, if that is the case, 
and using that $x\ e^{-x^2/2}$ has a maximum at $x=1$, it is easy to see that the maximum cannot occur more than $\tilde{\delta}$ away from the edges, 
so that $\|H-E_1\|_{\infty} \le N+\tilde{\delta}\le 3N/2$.
Then we can bound,
 \be
|\tilde n'_{\rm max}| \le  \frac{3N \tilde n(E_1)}{2 \tilde\delta^2} \le \frac{3N \tilde n(E_0)}{2 \tilde \delta^2}.
 \ee
Finally, for any value of energy such that
 \be
 |E-E_0|\le \frac{\tilde \delta^2}{3N}
 \ee
we will have
 \be
 \tilde n(E_0)-\tilde n(E) \le |n'_{\max}| \cdot |E_0-E| \le \frac{\tilde n(E_0)}{2},
 \ee
and thus
 \bea
 \tilde n(E) &=& \tilde n(E_0) - |\tilde n(E_0)-\tilde n(E)| \nonumber \\ &\ge & \frac{\tilde n(E_0)}{2} 
 \ge \frac{1}{4}(1+\sigma_{\psi}^2/\tilde \delta^2)^{-3/2}.
 \eea

Let us finish this appendix with a remark on the difficulty of finding the state $\psi$ given a prescribed energy. As we mentioned at the end of Sec. \ref{initial-states}, this can be analyzed, for instance, through mean field theory for product states, or with matrix product states in one-dimensional problems. Regarding rigorous bounds, for product states it was shown by Lieb in \cite{lieb1973} that for any local Hamiltonian with spectrum contained in $[-N,N]$, one can efficiently find a product state with an energy $-DN$, where $D > 1/9$. While this is a theoretical bound, we expect that for most relevant Hamiltonians $D$ will be much larger (in fact, one can find much tighter bounds for specific models, like those used in this paper). Furthermore, as emphasized in the main text, one does not necessarily have to use product states, which gives access to even larger values of $D$.

\section{Convergence to the microcanonical and canonical values for a non-integrable model}
\label{app:ed_non_int}

\begin{figure}
\centering
\includegraphics[width=0.4\textwidth]{{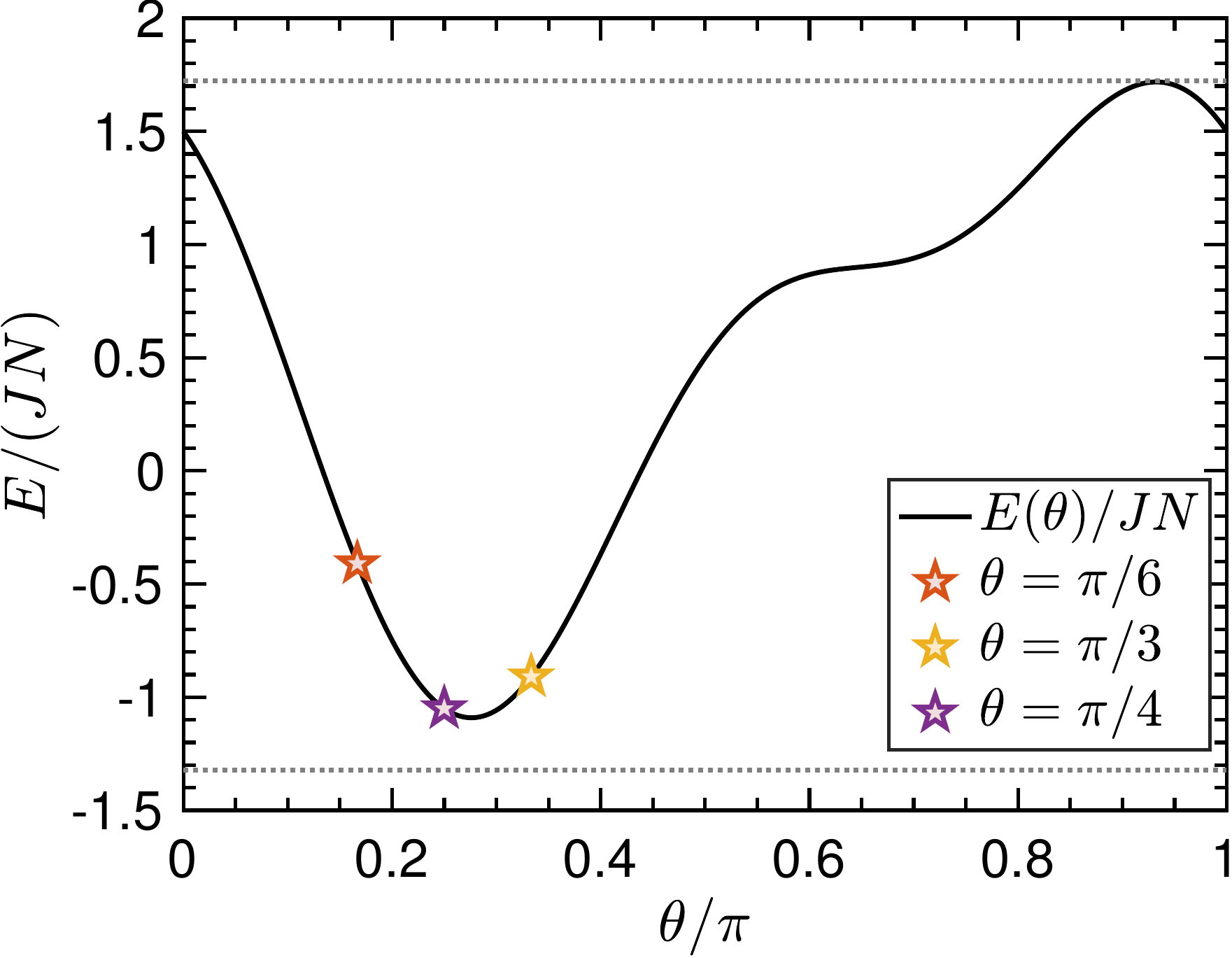}}
\caption{Energy density in the non-integrable Ising chain \eqref{eq:HamIsing} in the thermodynamic limit. The solid line indicates the energy density of translationally invariant real
product states, while the dashed horizontal lines indicate the energy densities of the ground and maximally excited states (estimated numerically with MPS).
The colored symbols indicate the states chosen in our numerical simulations.
\label{fig:ELNI}}
\end{figure}

\begin{figure*}[tbp]
\centering
  \begin{tabular}{ccc}
\includegraphics[width=.3\linewidth]{{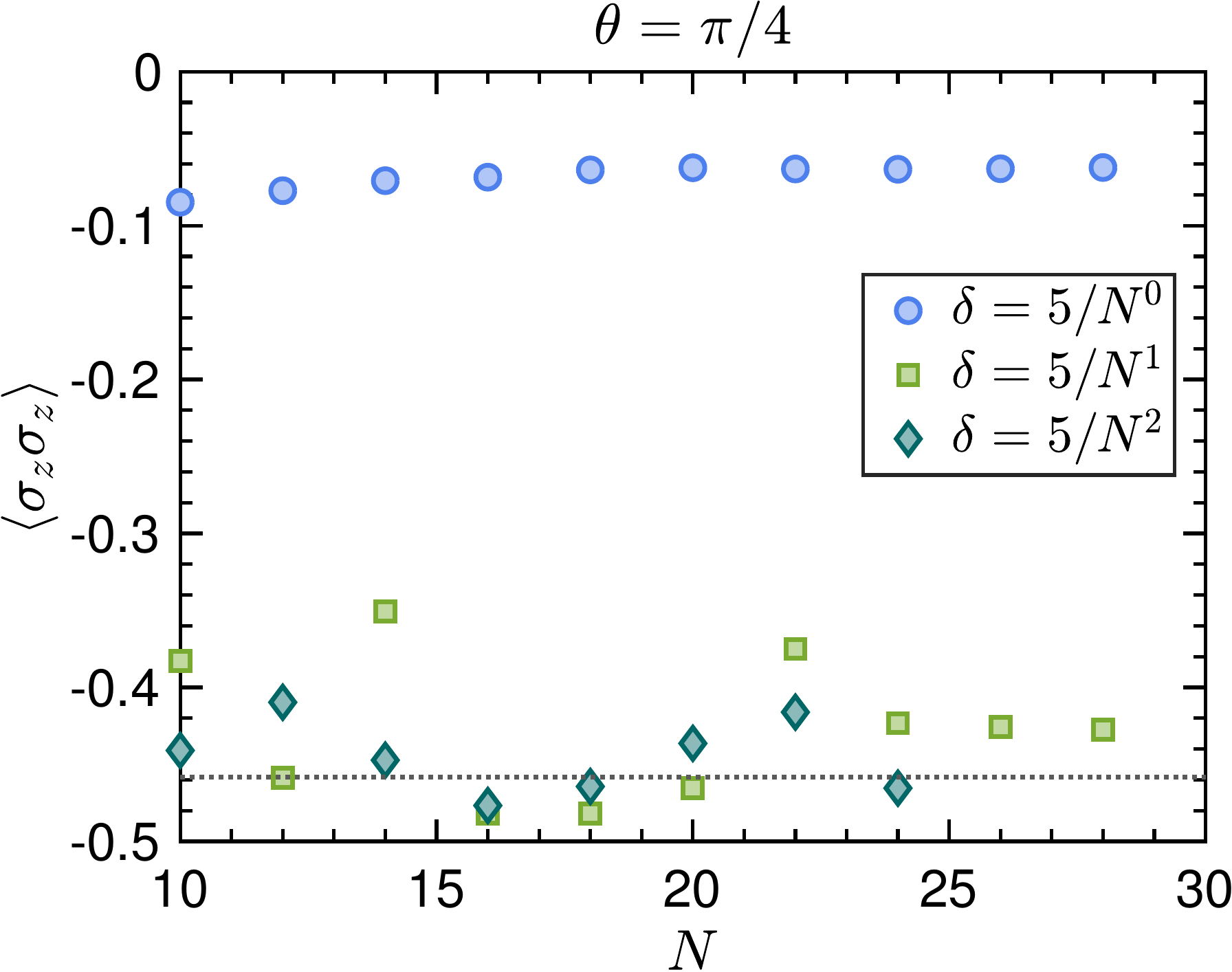}} &
\includegraphics[width=.3\linewidth]{{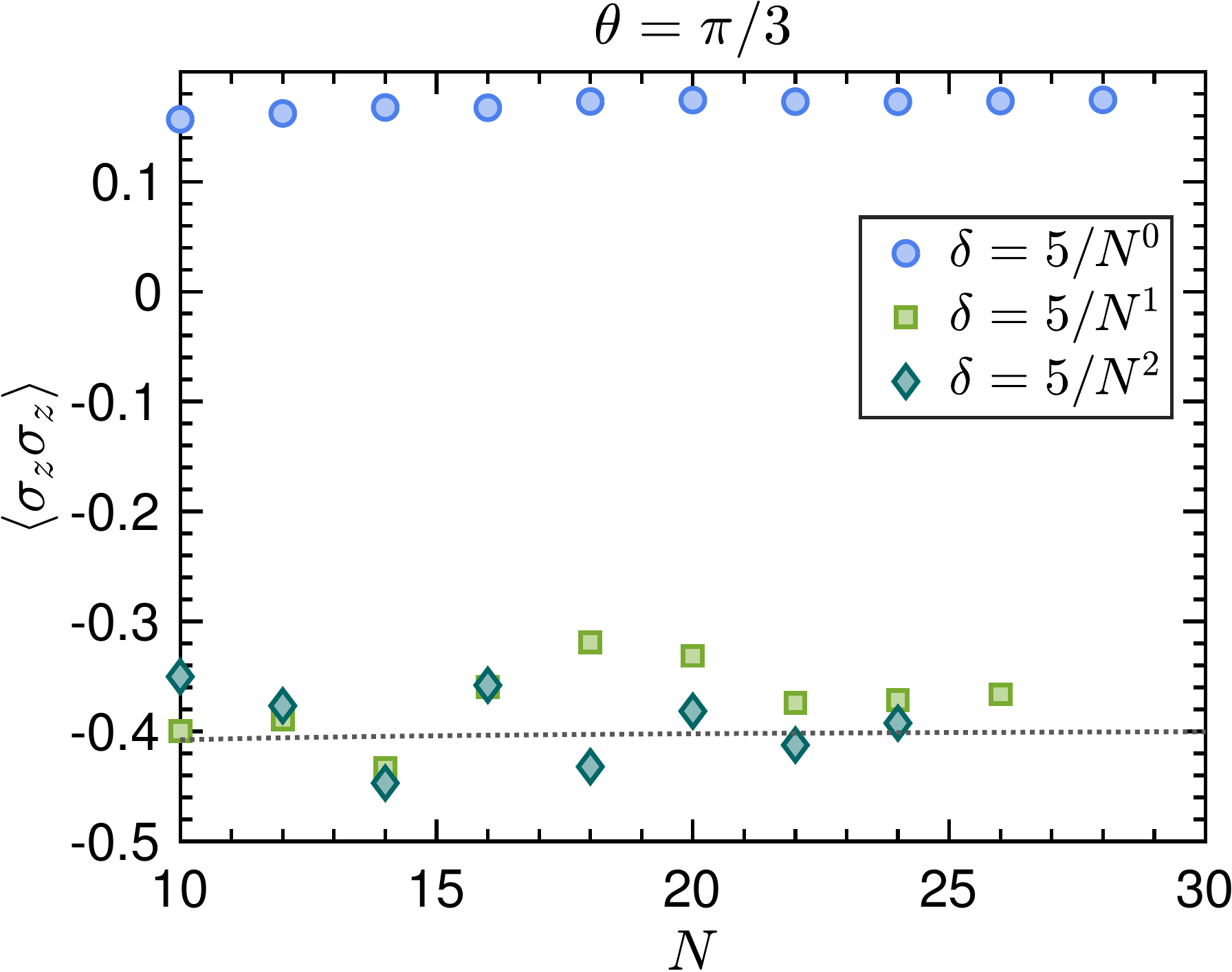}} &
\includegraphics[width=.3\linewidth]{{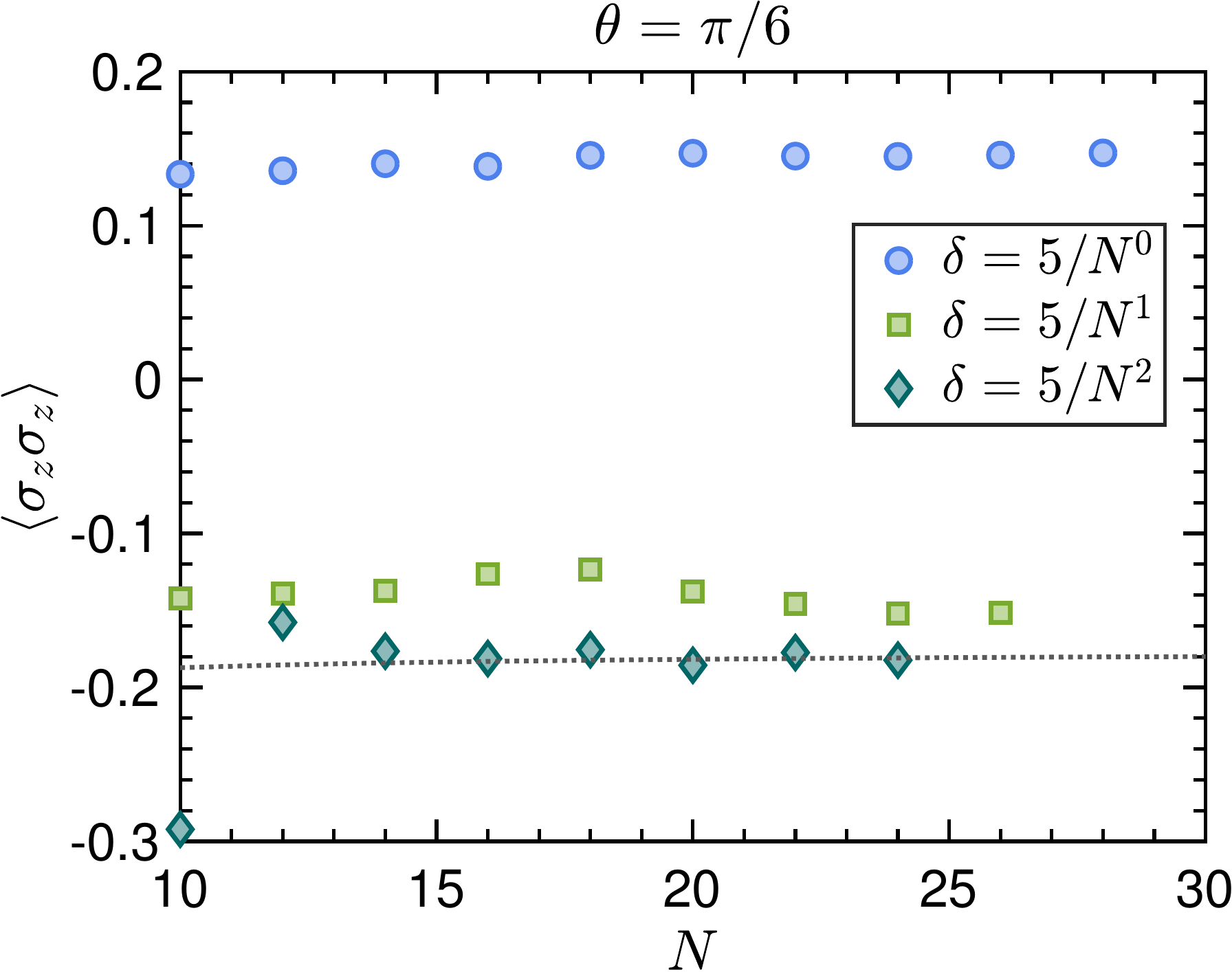}} \\
\includegraphics[width=.3\linewidth]{{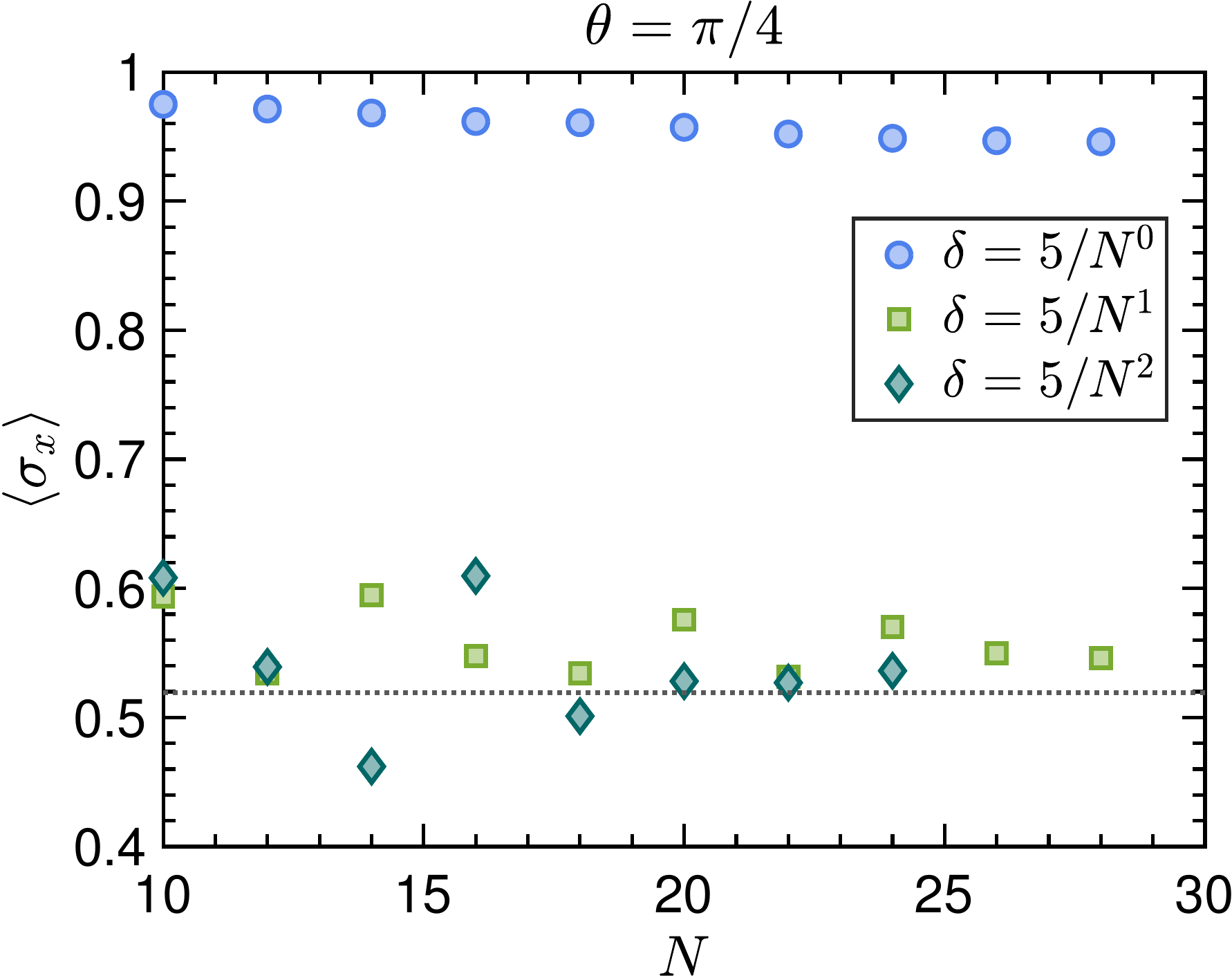}} &
\includegraphics[width=.3\linewidth]{{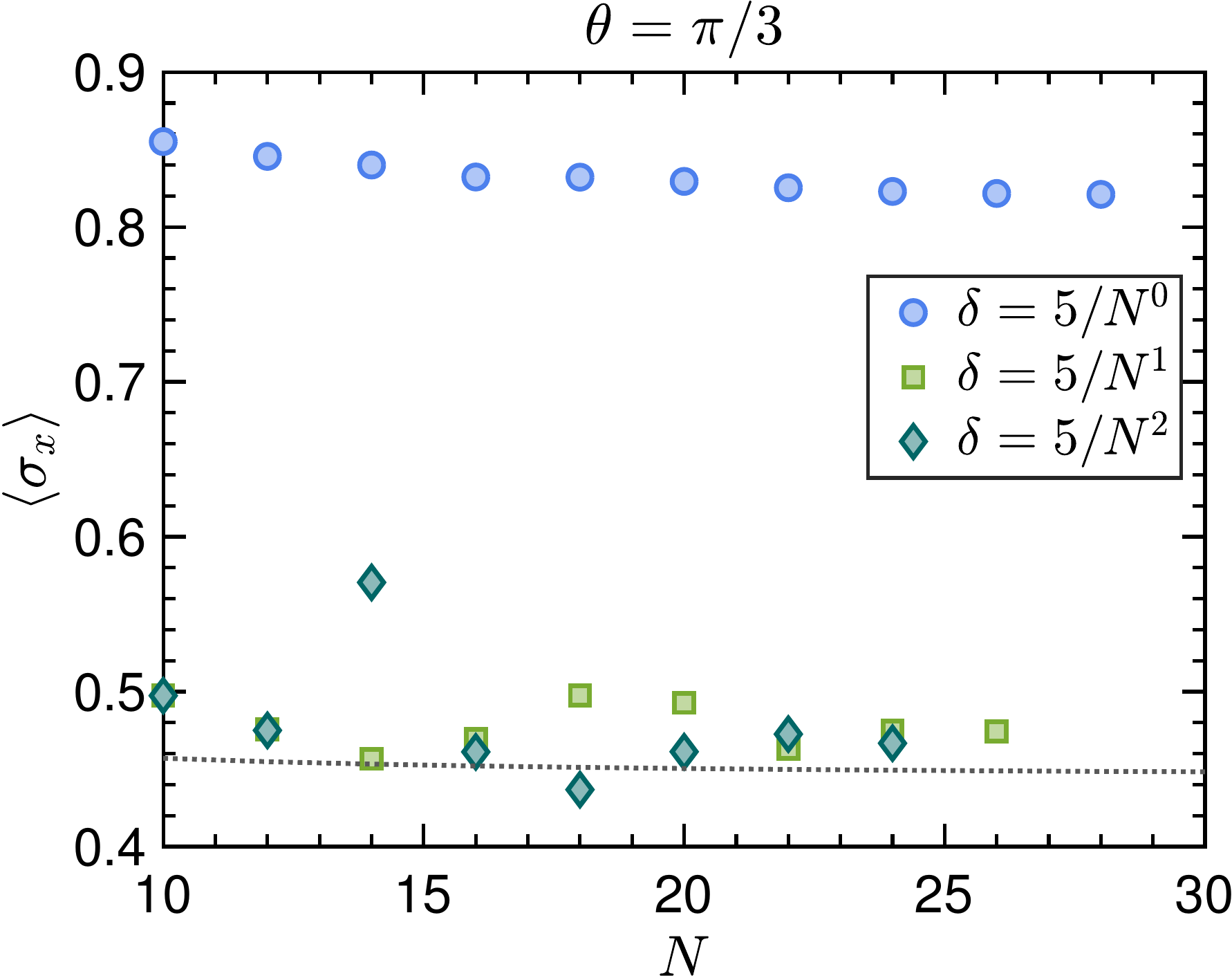}} &
\includegraphics[width=.3\linewidth]{{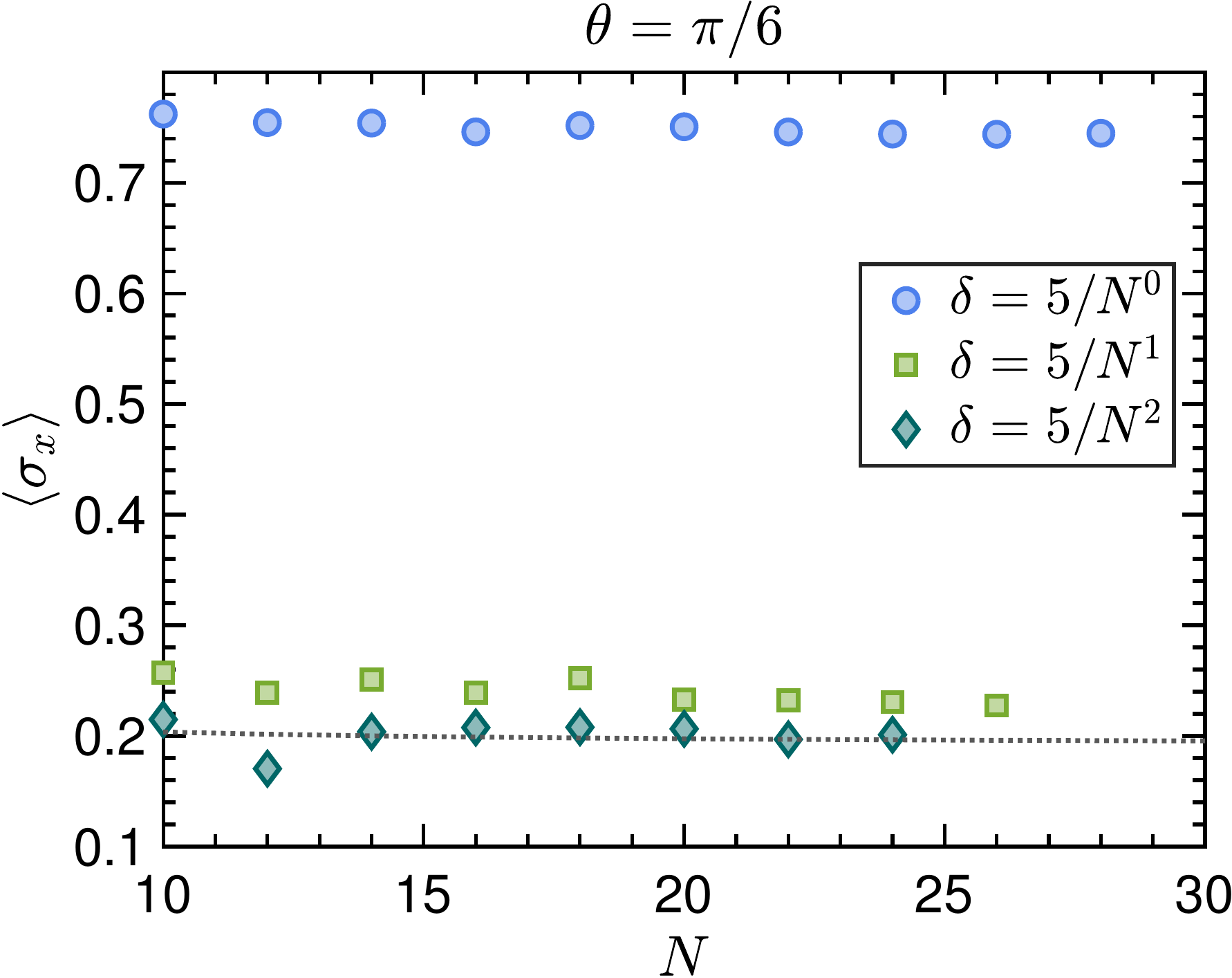}}
\end{tabular}
\caption{Convergence of Eq. \eqref{Adeltapsi2} to the microcanonical values for several local observables, in the non-integrable Ising model \eqref{eq:HamIsing}.
Each column corresponds to a real translationally invariant product state (in order of increasing energy density), determined by $\theta=\pi/4$ (left), $\pi/3$ (center) and $\pi/6$ (right).
The upper row illustrates the results for the 2-site observable $\sigma_z\otimes \sigma_z$, and the lower row for $\sigma_x$, measured, in both cases, in the center of the chain.
The dashed lines indicate the microcanonical values in the thermodynamic limit corresponding to the same energy density.
Our results show that $\delta \propto 1/N^2$ converges fast to the microcanonical expectation value, and for $\delta \propto 1/N$ the values are reasonably close.
}
\label{fig:convergenceDelta}
\end{figure*}

In this appendix we  numerically investigate the convergence to the microcanonical and canonical values of the quantities defined in the main text using exact diagonalization.
In particular, we show how a polynomially decreasing $\delta\sim \mathrm{poly}(1/N)$ seems to be enough for the quantity $A'_{\delta,\psi}(E)$ \eqref{Adeltapsi2}
to converge to the true microcanonical expectation value.

We consider the Ising model in a tilted field, described by the Hamiltonian
\begin{equation}
H_{\text{Ising}}=J\left[\sum_{n=1}^{N-1}\sigma_{n,z}\sigma_{n+1,z}+h\sum_{n=1}^{N} \sigma_{n,z}+g\sum_{n=1}^{N} \sigma_{n,x}\right],
\label{eq:HamIsing}
\end{equation}
which is in general non-integrable, except in the limits \(g=0\) (classical) and \(h=0\) (transverse field Ising model).
In the following, we choose a strongly non-integrable point \(h=0.5,g=-1.05\)~\cite{kim2013ballistic},
  and we have taken $J=1$. Notice that this corresponds to a different normalization for $H/N$ than the one used in the main text.
However, it is enough to ensure that \eqref{Cosinefilt} filters out energies much farther than the width $\delta$ for all the states analyzed here (see appendix~\ref{app:cos_gaussian}), so that it allows us to study how the microcanonical values are approached as the width decreases.

We consider real translationally invariant product states, which can be parametrized as
$\ket{\Psi(\theta)}=\ket{p(\theta)}^{\otimes N}$,
where $\ket{p(\theta)}=\cos{\theta}\ket{0}+\sin{\theta}\ket{1}$.
In the thermodynamic limit, these states have energy density $E/N=\cos^2(2\theta)+h \cos(2\theta)+g \sin(2\theta)$, ranging over most of the energy
band (see Fig.~\ref{fig:ELNI}).

We choose three values of $\theta$ corresponding to states in the lower part of the energy band, namely
$\theta_1=\pi/4$, for which $E_1/(JN)=g=-1.05$,
$\theta_2=\pi/3$, for which $E_2/(JN)=-0.909$,
and
$\theta_3=\pi/6$, with $E_3/(JN)=-0.409$.
For each of these states, and for several local observables,
we compute exactly the expression
 $A'_{\delta,\psi}(E=E_{\psi})$ from Eq. \eqref{Adeltapsi2}
 for different values of $\delta$.
 In order to check that it is enough to decrease $\delta$ polynomially with the system size,
 we run the calculations
for system sizes $10\leq N \leq 28$,
and $\delta\propto N^{s}$, for $s=0,-1,-2$.
The results, for observables $A=\sigma_{[N/2],z} \otimes \sigma_{[N/2+1],z}$
and $A=\sigma_{[N/2],x}$, are shown in Fig.~\ref{fig:convergenceDelta}.
As reference, we estimate the microcanonical expectation values in the thermodynamic limit
using uniform MPS \cite{Verstraete2008a} (more concretely, we approximate the canonical ensemble at the
same energy density in the thermodynamic limit as a matrix product operator, in which the observables
can be easily computed, and use the fact that in this limit, both ensembles are equivalent).

Our results indicate that, although a constant value of $\delta$ is not enough for $A'_{\delta,\psi}$ to approximate the microcanonical
value, when $\delta$ decreases as $1/N^2$, the expectation values indeed converge. For $\delta\propto 1/N$ we observe that the values are reasonably close, and they would be compatible with a slower convergence.


\section{Approximating the Cosine filter as Gaussian}
\label{app:cos_gaussian}

We want to bound the absolute value of the difference
\be
f_M(x)=e^{-Mx^2/2}-\cos^M x.
\ee
Since both terms are even, we can consider only $x\geq 0$.
If $|x|<\pi/2$, $e^{-x^2/2}\geq \cos x$,
and both terms are positive, so also
$ 0 \leq \cos^M x \leq  e^{-Mx^2/2}$, and
\be
| f_M(x)| =f_M(x) \leq e^{-Mx^2/2}.
\ee
The bound actually holds for slightly larger $x$, as long as $e^{-x^2/2}\geq |\cos x|$,
which is true up to $x_1\approx 0.566 \pi$.

Actually, the Gaussian form approximates the cosine also beyond this value, since for $x_1\le x \le \pi-\mu$, it holds
\be
|f_M(x)|=\cos^M x-e^{-Mx^2/2}\le \cos^M x \le \cos^M \mu,
\ee
where we have assumed that $M$ is even, as in the text. 
Thus the difference decreases exponentially with $M$ up to $x=\pi-\mu$.

At very small $x$ it is more useful to 
use the Taylor expansion.
There we can notice that
at very small $|x|$ the difference vanishes as $f_M(x)\approx \frac{M}{12} x^4 +O(x^6)$, and, in fact,
 $ f_M(x)\le \frac{M}{12} x^4$.

\bibliography{QSFT}

\begin{thebibliography}{87}%
\makeatletter
\providecommand \@ifxundefined [1]{%
 \@ifx{#1\undefined}
}%
\providecommand \@ifnum [1]{%
 \ifnum #1\expandafter \@firstoftwo
 \else \expandafter \@secondoftwo
 \fi
}%
\providecommand \@ifx [1]{%
 \ifx #1\expandafter \@firstoftwo
 \else \expandafter \@secondoftwo
 \fi
}%
\providecommand \natexlab [1]{#1}%
\providecommand \enquote  [1]{``#1''}%
\providecommand \bibnamefont  [1]{#1}%
\providecommand \bibfnamefont [1]{#1}%
\providecommand \citenamefont [1]{#1}%
\providecommand \href@noop [0]{\@secondoftwo}%
\providecommand \href [0]{\begingroup \@sanitize@url \@href}%
\providecommand \@href[1]{\@@startlink{#1}\@@href}%
\providecommand \@@href[1]{\endgroup#1\@@endlink}%
\providecommand \@sanitize@url [0]{\catcode `\\12\catcode `\$12\catcode
  `\&12\catcode `\#12\catcode `\^12\catcode `\_12\catcode `\%12\relax}%
\providecommand \@@startlink[1]{}%
\providecommand \@@endlink[0]{}%
\providecommand \url  [0]{\begingroup\@sanitize@url \@url }%
\providecommand \@url [1]{\endgroup\@href {#1}{\urlprefix }}%
\providecommand \urlprefix  [0]{URL }%
\providecommand \Eprint [0]{\href }%
\providecommand \doibase [0]{https://doi.org/}%
\providecommand \selectlanguage [0]{\@gobble}%
\providecommand \bibinfo  [0]{\@secondoftwo}%
\providecommand \bibfield  [0]{\@secondoftwo}%
\providecommand \translation [1]{[#1]}%
\providecommand \BibitemOpen [0]{}%
\providecommand \bibitemStop [0]{}%
\providecommand \bibitemNoStop [0]{.\EOS\space}%
\providecommand \EOS [0]{\spacefactor3000\relax}%
\providecommand \BibitemShut  [1]{\csname bibitem#1\endcsname}%
\let\auto@bib@innerbib\@empty
\bibitem [{\citenamefont {Feynman}(1982)}]{feynman1982simulating}%
  \BibitemOpen
  \bibfield  {author} {\bibinfo {author} {\bibfnamefont {R.~P.}\ \bibnamefont
  {Feynman}},\ }\bibfield  {title} {\bibinfo {title} {{Simulating physics with
  computers}},\ }\href {https://doi.org/10.1007/BF02650179} {\bibfield
  {journal} {\bibinfo  {journal} {Int. J. Theor. Phys.}\ }\textbf {\bibinfo
  {volume} {21}},\ \bibinfo {pages} {467} (\bibinfo {year} {1982})}\BibitemShut
  {NoStop}%
\bibitem [{\citenamefont {Lloyd}(1996)}]{Lloyd96}%
  \BibitemOpen
  \bibfield  {author} {\bibinfo {author} {\bibfnamefont {S.}~\bibnamefont
  {Lloyd}},\ }\bibfield  {title} {\bibinfo {title} {{Universal quantum
  simulators}},\ }\href {https://doi.org/10.1126/science.273.5278.1073}
  {\bibfield  {journal} {\bibinfo  {journal} {Science}\ }\textbf {\bibinfo
  {volume} {273}},\ \bibinfo {pages} {1073} (\bibinfo {year}
  {1996})}\BibitemShut {NoStop}%
\bibitem [{\citenamefont {Cirac}\ and\ \citenamefont
  {Zoller}(2004)}]{CiracZollerPhysToday}%
  \BibitemOpen
  \bibfield  {author} {\bibinfo {author} {\bibfnamefont {J.~I.}\ \bibnamefont
  {Cirac}}\ and\ \bibinfo {author} {\bibfnamefont {P.}~\bibnamefont {Zoller}},\
  }\bibfield  {title} {\bibinfo {title} {{New Frontiers in Quantum Information
  With Atoms and Ions}},\ }\href {https://doi.org/10.1063/1.1712500} {\bibfield
   {journal} {\bibinfo  {journal} {Phys. Today}\ }\textbf {\bibinfo {volume}
  {57}},\ \bibinfo {pages} {38} (\bibinfo {year} {2004})}\BibitemShut {NoStop}%
\bibitem [{\citenamefont {Bauer}\ \emph {et~al.}(2020)\citenamefont {Bauer},
  \citenamefont {Bravyi}, \citenamefont {Motta},\ and\ \citenamefont {{Kin-Lic
  Chan}}}]{Bauer2001.03685}%
  \BibitemOpen
  \bibfield  {author} {\bibinfo {author} {\bibfnamefont {B.}~\bibnamefont
  {Bauer}}, \bibinfo {author} {\bibfnamefont {S.}~\bibnamefont {Bravyi}},
  \bibinfo {author} {\bibfnamefont {M.}~\bibnamefont {Motta}},\ and\ \bibinfo
  {author} {\bibfnamefont {G.}~\bibnamefont {{Kin-Lic Chan}}},\ }\bibfield
  {title} {\bibinfo {title} {{Quantum Algorithms for Quantum Chemistry and
  Quantum Materials Science}},\ }\href
  {https://doi.org/10.1021/acs.chemrev.9b00829} {\bibfield  {journal} {\bibinfo
   {journal} {Chem. Rev.}\ }\textbf {\bibinfo {volume} {120}},\ \bibinfo
  {pages} {12685} (\bibinfo {year} {2020})}\BibitemShut {NoStop}%
\bibitem [{\citenamefont {Georgescu}\ \emph {et~al.}(2014)\citenamefont
  {Georgescu}, \citenamefont {Ashhab},\ and\ \citenamefont
  {Nori}}]{GeorgescuRMP2014}%
  \BibitemOpen
  \bibfield  {author} {\bibinfo {author} {\bibfnamefont {I.~M.}\ \bibnamefont
  {Georgescu}}, \bibinfo {author} {\bibfnamefont {S.}~\bibnamefont {Ashhab}},\
  and\ \bibinfo {author} {\bibfnamefont {F.}~\bibnamefont {Nori}},\ }\bibfield
  {title} {\bibinfo {title} {{Quantum simulation}},\ }\href
  {https://doi.org/10.1103/RevModPhys.86.153} {\bibfield  {journal} {\bibinfo
  {journal} {Rev. Mod. Phys.}\ }\textbf {\bibinfo {volume} {86}},\ \bibinfo
  {pages} {153} (\bibinfo {year} {2014})}\BibitemShut {NoStop}%
\bibitem [{\citenamefont {Bloch}\ \emph {et~al.}(2008)\citenamefont {Bloch},
  \citenamefont {Dalibard},\ and\ \citenamefont {Zwerger}}]{RMP_BlochDalibard}%
  \BibitemOpen
  \bibfield  {author} {\bibinfo {author} {\bibfnamefont {I.}~\bibnamefont
  {Bloch}}, \bibinfo {author} {\bibfnamefont {J.}~\bibnamefont {Dalibard}},\
  and\ \bibinfo {author} {\bibfnamefont {W.}~\bibnamefont {Zwerger}},\
  }\bibfield  {title} {\bibinfo {title} {{Many-body physics with ultracold
  gases}},\ }\href {https://doi.org/10.1103/RevModPhys.80.885} {\bibfield
  {journal} {\bibinfo  {journal} {Rev. Mod. Phys.}\ }\textbf {\bibinfo {volume}
  {80}},\ \bibinfo {pages} {885} (\bibinfo {year} {2008})}\BibitemShut
  {NoStop}%
\bibitem [{\citenamefont {Gross}\ and\ \citenamefont
  {Bloch}(2017)}]{Gross2017a}%
  \BibitemOpen
  \bibfield  {author} {\bibinfo {author} {\bibfnamefont {C.}~\bibnamefont
  {Gross}}\ and\ \bibinfo {author} {\bibfnamefont {I.}~\bibnamefont {Bloch}},\
  }\bibfield  {title} {\bibinfo {title} {{Quantum simulations with ultracold
  atoms in optical lattices}},\ }\href
  {https://doi.org/10.1126/science.aal3837} {\bibfield  {journal} {\bibinfo
  {journal} {Science}\ }\textbf {\bibinfo {volume} {357}},\ \bibinfo {pages}
  {995} (\bibinfo {year} {2017})}\BibitemShut {NoStop}%
\bibitem [{\citenamefont {Blatt}\ and\ \citenamefont
  {Roos}(2012)}]{BlattRossNatPhys2012}%
  \BibitemOpen
  \bibfield  {author} {\bibinfo {author} {\bibfnamefont {R.}~\bibnamefont
  {Blatt}}\ and\ \bibinfo {author} {\bibfnamefont {C.~F.}\ \bibnamefont
  {Roos}},\ }\bibfield  {title} {\bibinfo {title} {{Quantum simulations with
  trapped ions}},\ }\href {https://doi.org/10.1038/nphys2252} {\bibfield
  {journal} {\bibinfo  {journal} {Nat. Phys.}\ }\textbf {\bibinfo {volume}
  {8}},\ \bibinfo {pages} {277} (\bibinfo {year} {2012})}\BibitemShut {NoStop}%
\bibitem [{\citenamefont {Saffman}(2016)}]{Saffman2016b}%
  \BibitemOpen
  \bibfield  {author} {\bibinfo {author} {\bibfnamefont {M.}~\bibnamefont
  {Saffman}},\ }\bibfield  {title} {\bibinfo {title} {{Quantum computing with
  atomic qubits and Rydberg interactions: progress and challenges}},\ }\href
  {https://doi.org/10.1088/0953-4075/49/20/202001} {\bibfield  {journal}
  {\bibinfo  {journal} {J. Phys. B At. Mol. Opt. Phys.}\ }\textbf {\bibinfo
  {volume} {49}},\ \bibinfo {pages} {202001} (\bibinfo {year}
  {2016})}\BibitemShut {NoStop}%
\bibitem [{\citenamefont {Barthelemy}\ and\ \citenamefont
  {Vandersypen}(2013)}]{VandersypenAnnalenderPhysik}%
  \BibitemOpen
  \bibfield  {author} {\bibinfo {author} {\bibfnamefont {P.}~\bibnamefont
  {Barthelemy}}\ and\ \bibinfo {author} {\bibfnamefont {L.~M.~K.}\ \bibnamefont
  {Vandersypen}},\ }\bibfield  {title} {\bibinfo {title} {{Quantum Dot Systems:
  a versatile platform for quantum simulations}},\ }\href
  {https://doi.org/10.1002/andp.201300124} {\bibfield  {journal} {\bibinfo
  {journal} {Ann. Phys.}\ }\textbf {\bibinfo {volume} {525}},\ \bibinfo {pages}
  {808} (\bibinfo {year} {2013})}\BibitemShut {NoStop}%
\bibitem [{\citenamefont {Lamata}\ \emph {et~al.}(2018)\citenamefont {Lamata},
  \citenamefont {Parra-Rodriguez}, \citenamefont {Sanz},\ and\ \citenamefont
  {Solano}}]{LamataAdvancesinPhys2018}%
  \BibitemOpen
  \bibfield  {author} {\bibinfo {author} {\bibfnamefont {L.}~\bibnamefont
  {Lamata}}, \bibinfo {author} {\bibfnamefont {A.}~\bibnamefont
  {Parra-Rodriguez}}, \bibinfo {author} {\bibfnamefont {M.}~\bibnamefont
  {Sanz}},\ and\ \bibinfo {author} {\bibfnamefont {E.}~\bibnamefont {Solano}},\
  }\bibfield  {title} {\bibinfo {title} {{Digital-analog quantum simulations
  with superconducting circuits}},\ }\href
  {https://doi.org/10.1080/23746149.2018.1457981} {\bibfield  {journal}
  {\bibinfo  {journal} {Adv. Phys. X}\ }\textbf {\bibinfo {volume} {3}},\
  \bibinfo {pages} {1457981} (\bibinfo {year} {2018})}\BibitemShut {NoStop}%
\bibitem [{\citenamefont {Hartmann}(2016)}]{HartmannJofOptics2018}%
  \BibitemOpen
  \bibfield  {author} {\bibinfo {author} {\bibfnamefont {M.~J.}\ \bibnamefont
  {Hartmann}},\ }\bibfield  {title} {\bibinfo {title} {{Quantum simulation with
  interacting photons}},\ }\href
  {https://doi.org/10.1088/2040-8978/18/10/104005} {\bibfield  {journal}
  {\bibinfo  {journal} {J. Opt.}\ }\textbf {\bibinfo {volume} {18}},\ \bibinfo
  {pages} {104005} (\bibinfo {year} {2016})}\BibitemShut {NoStop}%
\bibitem [{\citenamefont {Schreiber}\ \emph {et~al.}(2015)\citenamefont
  {Schreiber}, \citenamefont {Hodgman}, \citenamefont {Bordia}, \citenamefont
  {Luschen}, \citenamefont {Fischer}, \citenamefont {Vosk}, \citenamefont
  {Altman}, \citenamefont {Schneider},\ and\ \citenamefont
  {Bloch}}]{BlochManyBodyLocalin1D}%
  \BibitemOpen
  \bibfield  {author} {\bibinfo {author} {\bibfnamefont {M.}~\bibnamefont
  {Schreiber}}, \bibinfo {author} {\bibfnamefont {S.~S.}\ \bibnamefont
  {Hodgman}}, \bibinfo {author} {\bibfnamefont {P.}~\bibnamefont {Bordia}},
  \bibinfo {author} {\bibfnamefont {H.~P.}\ \bibnamefont {Luschen}}, \bibinfo
  {author} {\bibfnamefont {M.~H.}\ \bibnamefont {Fischer}}, \bibinfo {author}
  {\bibfnamefont {R.}~\bibnamefont {Vosk}}, \bibinfo {author} {\bibfnamefont
  {E.}~\bibnamefont {Altman}}, \bibinfo {author} {\bibfnamefont
  {U.}~\bibnamefont {Schneider}},\ and\ \bibinfo {author} {\bibfnamefont
  {I.}~\bibnamefont {Bloch}},\ }\bibfield  {title} {\bibinfo {title}
  {{Observation of many-body localization of interacting fermions in a
  quasirandom optical lattice}},\ }\href
  {https://doi.org/10.1126/science.aaa7432} {\bibfield  {journal} {\bibinfo
  {journal} {Science}\ }\textbf {\bibinfo {volume} {349}},\ \bibinfo {pages}
  {842} (\bibinfo {year} {2015})}\BibitemShut {NoStop}%
\bibitem [{\citenamefont {Choi}\ \emph {et~al.}(2016)\citenamefont {Choi},
  \citenamefont {Hild}, \citenamefont {Zeiher}, \citenamefont {Schauss},
  \citenamefont {Rubio-Abadal}, \citenamefont {Yefsah}, \citenamefont
  {Khemani}, \citenamefont {Huse}, \citenamefont {Bloch},\ and\ \citenamefont
  {Gross}}]{BlochMBLin2D}%
  \BibitemOpen
  \bibfield  {author} {\bibinfo {author} {\bibfnamefont {J.-y.}\ \bibnamefont
  {Choi}}, \bibinfo {author} {\bibfnamefont {S.}~\bibnamefont {Hild}}, \bibinfo
  {author} {\bibfnamefont {J.}~\bibnamefont {Zeiher}}, \bibinfo {author}
  {\bibfnamefont {P.}~\bibnamefont {Schauss}}, \bibinfo {author} {\bibfnamefont
  {A.}~\bibnamefont {Rubio-Abadal}}, \bibinfo {author} {\bibfnamefont
  {T.}~\bibnamefont {Yefsah}}, \bibinfo {author} {\bibfnamefont
  {V.}~\bibnamefont {Khemani}}, \bibinfo {author} {\bibfnamefont {D.~A.}\
  \bibnamefont {Huse}}, \bibinfo {author} {\bibfnamefont {I.}~\bibnamefont
  {Bloch}},\ and\ \bibinfo {author} {\bibfnamefont {C.}~\bibnamefont {Gross}},\
  }\bibfield  {title} {\bibinfo {title} {{Exploring the many-body localization
  transition in two dimensions}},\ }\href
  {https://doi.org/10.1126/science.aaf8834} {\bibfield  {journal} {\bibinfo
  {journal} {Science}\ }\textbf {\bibinfo {volume} {352}},\ \bibinfo {pages}
  {1547} (\bibinfo {year} {2016})}\BibitemShut {NoStop}%
\bibitem [{\citenamefont {Parsons}\ \emph {et~al.}(2016)\citenamefont
  {Parsons}, \citenamefont {Mazurenko}, \citenamefont {Chiu}, \citenamefont
  {Ji}, \citenamefont {Greif},\ and\ \citenamefont {Greiner}}]{Parsons2016}%
  \BibitemOpen
  \bibfield  {author} {\bibinfo {author} {\bibfnamefont {M.~F.}\ \bibnamefont
  {Parsons}}, \bibinfo {author} {\bibfnamefont {A.}~\bibnamefont {Mazurenko}},
  \bibinfo {author} {\bibfnamefont {C.~S.}\ \bibnamefont {Chiu}}, \bibinfo
  {author} {\bibfnamefont {G.}~\bibnamefont {Ji}}, \bibinfo {author}
  {\bibfnamefont {D.}~\bibnamefont {Greif}},\ and\ \bibinfo {author}
  {\bibfnamefont {M.}~\bibnamefont {Greiner}},\ }\bibfield  {title} {\bibinfo
  {title} {{Site-resolved measurement of the spin-correlation function in the
  Fermi-Hubbard model}},\ }\href {https://doi.org/10.1126/science.aag1430}
  {\bibfield  {journal} {\bibinfo  {journal} {Science}\ }\textbf {\bibinfo
  {volume} {353}},\ \bibinfo {pages} {1253} (\bibinfo {year}
  {2016})}\BibitemShut {NoStop}%
\bibitem [{\citenamefont {Chiu}\ \emph {et~al.}(2019)\citenamefont {Chiu},
  \citenamefont {Ji}, \citenamefont {Bohrdt}, \citenamefont {Xu}, \citenamefont
  {Knap}, \citenamefont {Demler}, \citenamefont {Grusdt}, \citenamefont
  {Greiner},\ and\ \citenamefont {Greif}}]{Chiu2019}%
  \BibitemOpen
  \bibfield  {author} {\bibinfo {author} {\bibfnamefont {C.~S.}\ \bibnamefont
  {Chiu}}, \bibinfo {author} {\bibfnamefont {G.}~\bibnamefont {Ji}}, \bibinfo
  {author} {\bibfnamefont {A.}~\bibnamefont {Bohrdt}}, \bibinfo {author}
  {\bibfnamefont {M.}~\bibnamefont {Xu}}, \bibinfo {author} {\bibfnamefont
  {M.}~\bibnamefont {Knap}}, \bibinfo {author} {\bibfnamefont {E.}~\bibnamefont
  {Demler}}, \bibinfo {author} {\bibfnamefont {F.}~\bibnamefont {Grusdt}},
  \bibinfo {author} {\bibfnamefont {M.}~\bibnamefont {Greiner}},\ and\ \bibinfo
  {author} {\bibfnamefont {D.}~\bibnamefont {Greif}},\ }\bibfield  {title}
  {\bibinfo {title} {{String patterns in the doped Hubbard model}},\ }\href
  {https://doi.org/10.1126/science.aav3587} {\bibfield  {journal} {\bibinfo
  {journal} {Science}\ }\textbf {\bibinfo {volume} {365}},\ \bibinfo {pages}
  {251} (\bibinfo {year} {2019})}\BibitemShut {NoStop}%
\bibitem [{\citenamefont {Labuhn}\ \emph {et~al.}(2016)\citenamefont {Labuhn},
  \citenamefont {Barredo}, \citenamefont {Ravets}, \citenamefont {{De
  L{\'{e}}s{\'{e}}leuc}}, \citenamefont {Macr{\`{i}}}, \citenamefont {Lahaye},\
  and\ \citenamefont {Browaeys}}]{Labuhn2016}%
  \BibitemOpen
  \bibfield  {author} {\bibinfo {author} {\bibfnamefont {H.}~\bibnamefont
  {Labuhn}}, \bibinfo {author} {\bibfnamefont {D.}~\bibnamefont {Barredo}},
  \bibinfo {author} {\bibfnamefont {S.}~\bibnamefont {Ravets}}, \bibinfo
  {author} {\bibfnamefont {S.}~\bibnamefont {{De L{\'{e}}s{\'{e}}leuc}}},
  \bibinfo {author} {\bibfnamefont {T.}~\bibnamefont {Macr{\`{i}}}}, \bibinfo
  {author} {\bibfnamefont {T.}~\bibnamefont {Lahaye}},\ and\ \bibinfo {author}
  {\bibfnamefont {A.}~\bibnamefont {Browaeys}},\ }\bibfield  {title} {\bibinfo
  {title} {{Tunable two-dimensional arrays of single Rydberg atoms for
  realizing quantum Ising models}},\ }\href
  {https://doi.org/10.1038/nature18274} {\bibfield  {journal} {\bibinfo
  {journal} {Nature}\ }\textbf {\bibinfo {volume} {534}},\ \bibinfo {pages}
  {667} (\bibinfo {year} {2016})}\BibitemShut {NoStop}%
\bibitem [{\citenamefont {Bernien}\ \emph {et~al.}(2017)\citenamefont
  {Bernien}, \citenamefont {Schwartz}, \citenamefont {Keesling}, \citenamefont
  {Levine}, \citenamefont {Omran}, \citenamefont {Pichler}, \citenamefont
  {Choi}, \citenamefont {Zibrov}, \citenamefont {Endres}, \citenamefont
  {Greiner}, \citenamefont {Vuleti{\'{c}}},\ and\ \citenamefont
  {Lukin}}]{Bernien2017a}%
  \BibitemOpen
  \bibfield  {author} {\bibinfo {author} {\bibfnamefont {H.}~\bibnamefont
  {Bernien}}, \bibinfo {author} {\bibfnamefont {S.}~\bibnamefont {Schwartz}},
  \bibinfo {author} {\bibfnamefont {A.}~\bibnamefont {Keesling}}, \bibinfo
  {author} {\bibfnamefont {H.}~\bibnamefont {Levine}}, \bibinfo {author}
  {\bibfnamefont {A.}~\bibnamefont {Omran}}, \bibinfo {author} {\bibfnamefont
  {H.}~\bibnamefont {Pichler}}, \bibinfo {author} {\bibfnamefont
  {S.}~\bibnamefont {Choi}}, \bibinfo {author} {\bibfnamefont {A.~S.}\
  \bibnamefont {Zibrov}}, \bibinfo {author} {\bibfnamefont {M.}~\bibnamefont
  {Endres}}, \bibinfo {author} {\bibfnamefont {M.}~\bibnamefont {Greiner}},
  \bibinfo {author} {\bibfnamefont {V.}~\bibnamefont {Vuleti{\'{c}}}},\ and\
  \bibinfo {author} {\bibfnamefont {M.~D.}\ \bibnamefont {Lukin}},\ }\bibfield
  {title} {\bibinfo {title} {{Probing many-body dynamics on a 51-atom quantum
  simulator}},\ }\href {https://doi.org/10.1038/nature24622} {\bibfield
  {journal} {\bibinfo  {journal} {Nature}\ }\textbf {\bibinfo {volume} {551}},\
  \bibinfo {pages} {579} (\bibinfo {year} {2017})}\BibitemShut {NoStop}%
\bibitem [{\citenamefont {Zhang}\ \emph
  {et~al.}(2017{\natexlab{a}})\citenamefont {Zhang}, \citenamefont {Pagano},
  \citenamefont {Hess}, \citenamefont {Kyprianidis}, \citenamefont {Becker},
  \citenamefont {Kaplan}, \citenamefont {Gorshkov}, \citenamefont {Gong},\ and\
  \citenamefont {Monroe}}]{Zhang2017}%
  \BibitemOpen
  \bibfield  {author} {\bibinfo {author} {\bibfnamefont {J.}~\bibnamefont
  {Zhang}}, \bibinfo {author} {\bibfnamefont {G.}~\bibnamefont {Pagano}},
  \bibinfo {author} {\bibfnamefont {P.~W.}\ \bibnamefont {Hess}}, \bibinfo
  {author} {\bibfnamefont {A.}~\bibnamefont {Kyprianidis}}, \bibinfo {author}
  {\bibfnamefont {P.}~\bibnamefont {Becker}}, \bibinfo {author} {\bibfnamefont
  {H.}~\bibnamefont {Kaplan}}, \bibinfo {author} {\bibfnamefont {A.~V.}\
  \bibnamefont {Gorshkov}}, \bibinfo {author} {\bibfnamefont {Z.-X.}\
  \bibnamefont {Gong}},\ and\ \bibinfo {author} {\bibfnamefont
  {C.}~\bibnamefont {Monroe}},\ }\bibfield  {title} {\bibinfo {title}
  {{Observation of a many-body dynamical phase transition with a 53-qubit
  quantum simulator}},\ }\href {https://doi.org/10.1038/nature24654} {\bibfield
   {journal} {\bibinfo  {journal} {Nature}\ }\textbf {\bibinfo {volume}
  {551}},\ \bibinfo {pages} {601} (\bibinfo {year}
  {2017}{\natexlab{a}})}\BibitemShut {NoStop}%
\bibitem [{\citenamefont {Zhang}\ \emph
  {et~al.}(2017{\natexlab{b}})\citenamefont {Zhang}, \citenamefont {Hess},
  \citenamefont {Kyprianidis}, \citenamefont {Becker}, \citenamefont {Lee},
  \citenamefont {Smith}, \citenamefont {Pagano}, \citenamefont {Potirniche},
  \citenamefont {Potter}, \citenamefont {Vishwanath} \emph
  {et~al.}}]{MonroeTimeCrystal}%
  \BibitemOpen
  \bibfield  {author} {\bibinfo {author} {\bibfnamefont {J.}~\bibnamefont
  {Zhang}}, \bibinfo {author} {\bibfnamefont {P.}~\bibnamefont {Hess}},
  \bibinfo {author} {\bibfnamefont {A.}~\bibnamefont {Kyprianidis}}, \bibinfo
  {author} {\bibfnamefont {P.}~\bibnamefont {Becker}}, \bibinfo {author}
  {\bibfnamefont {A.}~\bibnamefont {Lee}}, \bibinfo {author} {\bibfnamefont
  {J.}~\bibnamefont {Smith}}, \bibinfo {author} {\bibfnamefont
  {G.}~\bibnamefont {Pagano}}, \bibinfo {author} {\bibfnamefont {I.-D.}\
  \bibnamefont {Potirniche}}, \bibinfo {author} {\bibfnamefont {A.~C.}\
  \bibnamefont {Potter}}, \bibinfo {author} {\bibfnamefont {A.}~\bibnamefont
  {Vishwanath}}, \emph {et~al.},\ }\bibfield  {title} {\bibinfo {title}
  {{Observation of a discrete time crystal}},\ }\href
  {https://doi.org/10.1038/nature21413} {\bibfield  {journal} {\bibinfo
  {journal} {Nature}\ }\textbf {\bibinfo {volume} {543}},\ \bibinfo {pages}
  {217} (\bibinfo {year} {2017}{\natexlab{b}})}\BibitemShut {NoStop}%
\bibitem [{\citenamefont {Arute}\ \emph {et~al.}(2019)\citenamefont {Arute},
  \citenamefont {Arya}, \citenamefont {Babbush}, \citenamefont {Bacon},
  \citenamefont {Bardin}, \citenamefont {Barends}, \citenamefont {Biswas},
  \citenamefont {Boixo}, \citenamefont {Brandao}, \citenamefont {Buell} \emph
  {et~al.}}]{Arute2019}%
  \BibitemOpen
  \bibfield  {author} {\bibinfo {author} {\bibfnamefont {F.}~\bibnamefont
  {Arute}}, \bibinfo {author} {\bibfnamefont {K.}~\bibnamefont {Arya}},
  \bibinfo {author} {\bibfnamefont {R.}~\bibnamefont {Babbush}}, \bibinfo
  {author} {\bibfnamefont {D.}~\bibnamefont {Bacon}}, \bibinfo {author}
  {\bibfnamefont {J.~C.}\ \bibnamefont {Bardin}}, \bibinfo {author}
  {\bibfnamefont {R.}~\bibnamefont {Barends}}, \bibinfo {author} {\bibfnamefont
  {R.}~\bibnamefont {Biswas}}, \bibinfo {author} {\bibfnamefont
  {S.}~\bibnamefont {Boixo}}, \bibinfo {author} {\bibfnamefont {F.~G.}\
  \bibnamefont {Brandao}}, \bibinfo {author} {\bibfnamefont {D.~A.}\
  \bibnamefont {Buell}}, \emph {et~al.},\ }\bibfield  {title} {\bibinfo {title}
  {{Quantum supremacy using a programmable superconducting processor}},\ }\href
  {https://doi.org/10.1038/s41586-019-1666-5} {\bibfield  {journal} {\bibinfo
  {journal} {Nature}\ }\textbf {\bibinfo {volume} {574}},\ \bibinfo {pages}
  {505} (\bibinfo {year} {2019})}\BibitemShut {NoStop}%
\bibitem [{\citenamefont {Kokail}\ \emph {et~al.}(2019)\citenamefont {Kokail},
  \citenamefont {Maier}, \citenamefont {van Bijnen}, \citenamefont {Brydges},
  \citenamefont {Joshi}, \citenamefont {Jurcevic}, \citenamefont {Muschik},
  \citenamefont {Silvi}, \citenamefont {Blatt}, \citenamefont {Roos},\ and\
  \citenamefont {Zoller}}]{ZollerRoosBlattNature2019}%
  \BibitemOpen
  \bibfield  {author} {\bibinfo {author} {\bibfnamefont {C.}~\bibnamefont
  {Kokail}}, \bibinfo {author} {\bibfnamefont {C.}~\bibnamefont {Maier}},
  \bibinfo {author} {\bibfnamefont {R.}~\bibnamefont {van Bijnen}}, \bibinfo
  {author} {\bibfnamefont {T.}~\bibnamefont {Brydges}}, \bibinfo {author}
  {\bibfnamefont {M.~K.}\ \bibnamefont {Joshi}}, \bibinfo {author}
  {\bibfnamefont {P.}~\bibnamefont {Jurcevic}}, \bibinfo {author}
  {\bibfnamefont {C.~A.}\ \bibnamefont {Muschik}}, \bibinfo {author}
  {\bibfnamefont {P.}~\bibnamefont {Silvi}}, \bibinfo {author} {\bibfnamefont
  {R.}~\bibnamefont {Blatt}}, \bibinfo {author} {\bibfnamefont {C.~F.}\
  \bibnamefont {Roos}},\ and\ \bibinfo {author} {\bibfnamefont
  {P.}~\bibnamefont {Zoller}},\ }\bibfield  {title} {\bibinfo {title}
  {{Self-verifying variational quantum simulation of lattice models}},\ }\href
  {https://doi.org/10.1038/s41586-019-1177-4} {\bibfield  {journal} {\bibinfo
  {journal} {Nature}\ }\textbf {\bibinfo {volume} {569}},\ \bibinfo {pages}
  {355} (\bibinfo {year} {2019})}\BibitemShut {NoStop}%
\bibitem [{\citenamefont {Brydges}\ \emph {et~al.}(2019)\citenamefont
  {Brydges}, \citenamefont {Elben}, \citenamefont {Jurcevic}, \citenamefont
  {Vermersch}, \citenamefont {Maier}, \citenamefont {Lanyon}, \citenamefont
  {Zoller}, \citenamefont {Blatt},\ and\ \citenamefont {Roos}}]{BlattEE}%
  \BibitemOpen
  \bibfield  {author} {\bibinfo {author} {\bibfnamefont {T.}~\bibnamefont
  {Brydges}}, \bibinfo {author} {\bibfnamefont {A.}~\bibnamefont {Elben}},
  \bibinfo {author} {\bibfnamefont {P.}~\bibnamefont {Jurcevic}}, \bibinfo
  {author} {\bibfnamefont {B.}~\bibnamefont {Vermersch}}, \bibinfo {author}
  {\bibfnamefont {C.}~\bibnamefont {Maier}}, \bibinfo {author} {\bibfnamefont
  {B.~P.}\ \bibnamefont {Lanyon}}, \bibinfo {author} {\bibfnamefont
  {P.}~\bibnamefont {Zoller}}, \bibinfo {author} {\bibfnamefont
  {R.}~\bibnamefont {Blatt}},\ and\ \bibinfo {author} {\bibfnamefont {C.~F.}\
  \bibnamefont {Roos}},\ }\bibfield  {title} {\bibinfo {title} {Probing
  r{\'e}nyi entanglement entropy via randomized measurements},\ }\href
  {https://doi.org/10.1126/science.aau4963} {\bibfield  {journal} {\bibinfo
  {journal} {Science}\ }\textbf {\bibinfo {volume} {364}},\ \bibinfo {pages}
  {260} (\bibinfo {year} {2019})}\BibitemShut {NoStop}%
\bibitem [{\citenamefont {G{\"{a}}rttner}\ \emph {et~al.}(2017)\citenamefont
  {G{\"{a}}rttner}, \citenamefont {Bohnet}, \citenamefont {Safavi-Naini},
  \citenamefont {Wall}, \citenamefont {Bollinger},\ and\ \citenamefont
  {Rey}}]{Garttner2017}%
  \BibitemOpen
  \bibfield  {author} {\bibinfo {author} {\bibfnamefont {M.}~\bibnamefont
  {G{\"{a}}rttner}}, \bibinfo {author} {\bibfnamefont {J.~G.}\ \bibnamefont
  {Bohnet}}, \bibinfo {author} {\bibfnamefont {A.}~\bibnamefont
  {Safavi-Naini}}, \bibinfo {author} {\bibfnamefont {M.~L.}\ \bibnamefont
  {Wall}}, \bibinfo {author} {\bibfnamefont {J.~J.}\ \bibnamefont
  {Bollinger}},\ and\ \bibinfo {author} {\bibfnamefont {A.~M.}\ \bibnamefont
  {Rey}},\ }\bibfield  {title} {\bibinfo {title} {{Measuring out-of-time-order
  correlations and multiple quantum spectra in a trapped-ion quantum magnet}},\
  }\href {https://doi.org/10.1038/nphys4119} {\bibfield  {journal} {\bibinfo
  {journal} {Nat. Phys.}\ }\textbf {\bibinfo {volume} {13}},\ \bibinfo {pages}
  {781} (\bibinfo {year} {2017})}\BibitemShut {NoStop}%
\bibitem [{\citenamefont {de~L{\'{e}}s{\'{e}}leuc}\ \emph
  {et~al.}(2019)\citenamefont {de~L{\'{e}}s{\'{e}}leuc}, \citenamefont
  {Lienhard}, \citenamefont {Scholl}, \citenamefont {Barredo}, \citenamefont
  {Weber}, \citenamefont {Lang}, \citenamefont {B{\"{u}}chler}, \citenamefont
  {Lahaye},\ and\ \citenamefont {Browaeys}}]{DeLeseleuc2019}%
  \BibitemOpen
  \bibfield  {author} {\bibinfo {author} {\bibfnamefont {S.}~\bibnamefont
  {de~L{\'{e}}s{\'{e}}leuc}}, \bibinfo {author} {\bibfnamefont
  {V.}~\bibnamefont {Lienhard}}, \bibinfo {author} {\bibfnamefont
  {P.}~\bibnamefont {Scholl}}, \bibinfo {author} {\bibfnamefont
  {D.}~\bibnamefont {Barredo}}, \bibinfo {author} {\bibfnamefont
  {S.}~\bibnamefont {Weber}}, \bibinfo {author} {\bibfnamefont
  {N.}~\bibnamefont {Lang}}, \bibinfo {author} {\bibfnamefont {H.~P.}\
  \bibnamefont {B{\"{u}}chler}}, \bibinfo {author} {\bibfnamefont
  {T.}~\bibnamefont {Lahaye}},\ and\ \bibinfo {author} {\bibfnamefont
  {A.}~\bibnamefont {Browaeys}},\ }\bibfield  {title} {\bibinfo {title}
  {{Observation of a symmetry-protected topological phase of interacting bosons
  with Rydberg atoms}},\ }\href {https://doi.org/10.1126/science.aav9105}
  {\bibfield  {journal} {\bibinfo  {journal} {Science}\ }\textbf {\bibinfo
  {volume} {365}},\ \bibinfo {pages} {775} (\bibinfo {year}
  {2019})}\BibitemShut {NoStop}%
\bibitem [{\citenamefont {Wei}\ \emph {et~al.}(2020)\citenamefont {Wei},
  \citenamefont {Lauer}, \citenamefont {Srinivasan}, \citenamefont
  {Sundaresan}, \citenamefont {McClure}, \citenamefont {Toyli}, \citenamefont
  {McKay}, \citenamefont {Gambetta},\ and\ \citenamefont {Sheldon}}]{IBMGHZ18}%
  \BibitemOpen
  \bibfield  {author} {\bibinfo {author} {\bibfnamefont {K.~X.}\ \bibnamefont
  {Wei}}, \bibinfo {author} {\bibfnamefont {I.}~\bibnamefont {Lauer}}, \bibinfo
  {author} {\bibfnamefont {S.}~\bibnamefont {Srinivasan}}, \bibinfo {author}
  {\bibfnamefont {N.}~\bibnamefont {Sundaresan}}, \bibinfo {author}
  {\bibfnamefont {D.~T.}\ \bibnamefont {McClure}}, \bibinfo {author}
  {\bibfnamefont {D.}~\bibnamefont {Toyli}}, \bibinfo {author} {\bibfnamefont
  {D.~C.}\ \bibnamefont {McKay}}, \bibinfo {author} {\bibfnamefont {J.~M.}\
  \bibnamefont {Gambetta}},\ and\ \bibinfo {author} {\bibfnamefont
  {S.}~\bibnamefont {Sheldon}},\ }\bibfield  {title} {\bibinfo {title}
  {{Verifying multipartite entangled Greenberger-Horne-Zeilinger states via
  multiple quantum coherences}},\ }\href
  {https://doi.org/10.1103/PhysRevA.101.032343} {\bibfield  {journal} {\bibinfo
   {journal} {Phys. Rev. A}\ }\textbf {\bibinfo {volume} {101}},\ \bibinfo
  {pages} {032343} (\bibinfo {year} {2020})}\BibitemShut {NoStop}%
\bibitem [{\citenamefont {Haah}\ \emph {et~al.}(2018)\citenamefont {Haah},
  \citenamefont {Hastings}, \citenamefont {Kothari},\ and\ \citenamefont
  {Low}}]{HaahIEEE2018}%
  \BibitemOpen
  \bibfield  {author} {\bibinfo {author} {\bibfnamefont {J.}~\bibnamefont
  {Haah}}, \bibinfo {author} {\bibfnamefont {M.}~\bibnamefont {Hastings}},
  \bibinfo {author} {\bibfnamefont {R.}~\bibnamefont {Kothari}},\ and\ \bibinfo
  {author} {\bibfnamefont {G.~H.}\ \bibnamefont {Low}},\ }\bibfield  {title}
  {\bibinfo {title} {{Quantum Algorithm for Simulating Real Time Evolution of
  Lattice Hamiltonians}},\ }in\ \href {https://doi.org/10.1109/FOCS.2018.00041}
  {\emph {\bibinfo {booktitle} {2018 IEEE 59th Annu. Symp. Found. Comput.
  Sci.}}},\ Vol.\ \bibinfo {volume} {2018-Octob}\ (\bibinfo  {publisher}
  {IEEE},\ \bibinfo {year} {2018})\ pp.\ \bibinfo {pages}
  {350--360}\BibitemShut {NoStop}%
\bibitem [{\citenamefont {Kempe}\ \emph {et~al.}(2006)\citenamefont {Kempe},
  \citenamefont {Kitaev},\ and\ \citenamefont {Regev}}]{KitaevQMAHard}%
  \BibitemOpen
  \bibfield  {author} {\bibinfo {author} {\bibfnamefont {J.}~\bibnamefont
  {Kempe}}, \bibinfo {author} {\bibfnamefont {A.}~\bibnamefont {Kitaev}},\ and\
  \bibinfo {author} {\bibfnamefont {O.}~\bibnamefont {Regev}},\ }\bibfield
  {title} {\bibinfo {title} {The complexity of the local hamiltonian problem},\
  }\href {https://doi.org/10.1007/978-3-540-30538-5_31} {\bibfield  {journal}
  {\bibinfo  {journal} {SIAM Journal on Computing}\ }\textbf {\bibinfo {volume}
  {35}},\ \bibinfo {pages} {1070} (\bibinfo {year} {2006})}\BibitemShut
  {NoStop}%
\bibitem [{\citenamefont {Kitaev}(1997)}]{KitaevRussMathSurveys1997}%
  \BibitemOpen
  \bibfield  {author} {\bibinfo {author} {\bibfnamefont {A.~Y.}\ \bibnamefont
  {Kitaev}},\ }\bibfield  {title} {\bibinfo {title} {{Quantum computations:
  algorithms and error correction}},\ }\href
  {https://doi.org/10.1070/RM1997v052n06ABEH002155} {\bibfield  {journal}
  {\bibinfo  {journal} {Russ. Math. Surv.}\ }\textbf {\bibinfo {volume} {52}},\
  \bibinfo {pages} {1191} (\bibinfo {year} {1997})}\BibitemShut {NoStop}%
\bibitem [{\citenamefont {Abrams}\ and\ \citenamefont
  {Lloyd}(1997)}]{AbramsLloydPRL97}%
  \BibitemOpen
  \bibfield  {author} {\bibinfo {author} {\bibfnamefont {D.~S.}\ \bibnamefont
  {Abrams}}\ and\ \bibinfo {author} {\bibfnamefont {S.}~\bibnamefont {Lloyd}},\
  }\bibfield  {title} {\bibinfo {title} {{Simulation of Many-Body Fermi Systems
  on a Universal Quantum Computer}},\ }\href
  {https://doi.org/10.1103/PhysRevLett.79.2586} {\bibfield  {journal} {\bibinfo
   {journal} {Phys. Rev. Lett.}\ }\textbf {\bibinfo {volume} {79}},\ \bibinfo
  {pages} {2586} (\bibinfo {year} {1997})}\BibitemShut {NoStop}%
\bibitem [{\citenamefont {Abrams}\ and\ \citenamefont
  {Lloyd}(1999)}]{AbramsLloydPRL99}%
  \BibitemOpen
  \bibfield  {author} {\bibinfo {author} {\bibfnamefont {D.~S.}\ \bibnamefont
  {Abrams}}\ and\ \bibinfo {author} {\bibfnamefont {S.}~\bibnamefont {Lloyd}},\
  }\bibfield  {title} {\bibinfo {title} {{Quantum Algorithm Providing
  Exponential Speed Increase for Finding Eigenvalues and Eigenvectors}},\
  }\href {https://doi.org/10.1103/PhysRevLett.83.5162} {\bibfield  {journal}
  {\bibinfo  {journal} {Phys. Rev. Lett.}\ }\textbf {\bibinfo {volume} {83}},\
  \bibinfo {pages} {5162} (\bibinfo {year} {1999})}\BibitemShut {NoStop}%
\bibitem [{\citenamefont {Aspuru-Guzik}\ \emph {et~al.}(2006)\citenamefont
  {Aspuru-Guzik}, \citenamefont {Dutoi}, \citenamefont {Love},\ and\
  \citenamefont {Head-Gordon}}]{AspuruDutoiLoveScience2005}%
  \BibitemOpen
  \bibfield  {author} {\bibinfo {author} {\bibfnamefont {A.}~\bibnamefont
  {Aspuru-Guzik}}, \bibinfo {author} {\bibfnamefont {A.~D.}\ \bibnamefont
  {Dutoi}}, \bibinfo {author} {\bibfnamefont {P.~J.}\ \bibnamefont {Love}},\
  and\ \bibinfo {author} {\bibfnamefont {M.}~\bibnamefont {Head-Gordon}},\
  }\bibfield  {title} {\bibinfo {title} {{Simulated Quantum Computation of
  Molecular Energies}},\ }\href {https://doi.org/10.1126/science.1113479}
  {\bibfield  {journal} {\bibinfo  {journal} {Science}\ }\textbf {\bibinfo
  {volume} {309}},\ \bibinfo {pages} {1704} (\bibinfo {year}
  {2006})}\BibitemShut {NoStop}%
\bibitem [{\citenamefont {Poulin}\ and\ \citenamefont
  {Wocjan}(2009)}]{PoulinWocjanPRL09}%
  \BibitemOpen
  \bibfield  {author} {\bibinfo {author} {\bibfnamefont {D.}~\bibnamefont
  {Poulin}}\ and\ \bibinfo {author} {\bibfnamefont {P.}~\bibnamefont
  {Wocjan}},\ }\bibfield  {title} {\bibinfo {title} {{Sampling from the Thermal
  Quantum Gibbs State and Evaluating Partition Functions with a Quantum
  Computer}},\ }\href {https://doi.org/10.1103/PhysRevLett.103.220502}
  {\bibfield  {journal} {\bibinfo  {journal} {Phys. Rev. Lett.}\ }\textbf
  {\bibinfo {volume} {103}},\ \bibinfo {pages} {220502} (\bibinfo {year}
  {2009})}\BibitemShut {NoStop}%
\bibitem [{\citenamefont {Ge}\ \emph {et~al.}(2019)\citenamefont {Ge},
  \citenamefont {Tura},\ and\ \citenamefont {Cirac}}]{Ge}%
  \BibitemOpen
  \bibfield  {author} {\bibinfo {author} {\bibfnamefont {Y.}~\bibnamefont
  {Ge}}, \bibinfo {author} {\bibfnamefont {J.}~\bibnamefont {Tura}},\ and\
  \bibinfo {author} {\bibfnamefont {J.~I.}\ \bibnamefont {Cirac}},\ }\bibfield
  {title} {\bibinfo {title} {{Faster ground state preparation and
  high-precision ground energy estimation with fewer qubits}},\ }\href
  {https://doi.org/10.1063/1.5027484} {\bibfield  {journal} {\bibinfo
  {journal} {J. Math. Phys.}\ }\textbf {\bibinfo {volume} {60}},\ \bibinfo
  {pages} {022202} (\bibinfo {year} {2019})}\BibitemShut {NoStop}%
\bibitem [{\citenamefont {Lin}\ and\ \citenamefont {Tong}(2020)}]{Lin2020}%
  \BibitemOpen
  \bibfield  {author} {\bibinfo {author} {\bibfnamefont {L.}~\bibnamefont
  {Lin}}\ and\ \bibinfo {author} {\bibfnamefont {Y.}~\bibnamefont {Tong}},\
  }\bibfield  {title} {\bibinfo {title} {Near-optimal ground state
  preparation},\ }\href {https://doi.org/10.22331/q-2020-12-14-372} {\bibfield
  {journal} {\bibinfo  {journal} {{Quantum}}\ }\textbf {\bibinfo {volume}
  {4}},\ \bibinfo {pages} {372} (\bibinfo {year} {2020})}\BibitemShut {NoStop}%
\bibitem [{\citenamefont {Ba{\~{n}}uls}\ \emph {et~al.}(2020)\citenamefont
  {Ba{\~{n}}uls}, \citenamefont {Huse},\ and\ \citenamefont {Cirac}}]{Banuls}%
  \BibitemOpen
  \bibfield  {author} {\bibinfo {author} {\bibfnamefont {M.~C.}\ \bibnamefont
  {Ba{\~{n}}uls}}, \bibinfo {author} {\bibfnamefont {D.~A.}\ \bibnamefont
  {Huse}},\ and\ \bibinfo {author} {\bibfnamefont {J.~I.}\ \bibnamefont
  {Cirac}},\ }\bibfield  {title} {\bibinfo {title} {{Entanglement and its
  relation to energy variance for local one-dimensional Hamiltonians}},\ }\href
  {https://doi.org/10.1103/PhysRevB.101.144305} {\bibfield  {journal} {\bibinfo
   {journal} {Phys. Rev. B}\ }\textbf {\bibinfo {volume} {101}},\ \bibinfo
  {pages} {144305} (\bibinfo {year} {2020})}\BibitemShut {NoStop}%
\bibitem [{\citenamefont {Dalmonte}\ \emph {et~al.}(2018)\citenamefont
  {Dalmonte}, \citenamefont {Vermersch},\ and\ \citenamefont
  {Zoller}}]{ZollerNatPhys2019}%
  \BibitemOpen
  \bibfield  {author} {\bibinfo {author} {\bibfnamefont {M.}~\bibnamefont
  {Dalmonte}}, \bibinfo {author} {\bibfnamefont {B.}~\bibnamefont
  {Vermersch}},\ and\ \bibinfo {author} {\bibfnamefont {P.}~\bibnamefont
  {Zoller}},\ }\bibfield  {title} {\bibinfo {title} {{Quantum simulation and
  spectroscopy of entanglement Hamiltonians}},\ }\href
  {https://doi.org/10.1038/s41567-018-0151-7} {\bibfield  {journal} {\bibinfo
  {journal} {Nat. Phys.}\ }\textbf {\bibinfo {volume} {14}},\ \bibinfo {pages}
  {827} (\bibinfo {year} {2018})}\BibitemShut {NoStop}%
\bibitem [{\citenamefont {Farhi}\ \emph {et~al.}(2000)\citenamefont {Farhi},
  \citenamefont {Goldstone}, \citenamefont {Gutmann},\ and\ \citenamefont
  {Sipser}}]{farhi2000quantum}%
  \BibitemOpen
  \bibfield  {author} {\bibinfo {author} {\bibfnamefont {E.}~\bibnamefont
  {Farhi}}, \bibinfo {author} {\bibfnamefont {J.}~\bibnamefont {Goldstone}},
  \bibinfo {author} {\bibfnamefont {S.}~\bibnamefont {Gutmann}},\ and\ \bibinfo
  {author} {\bibfnamefont {M.}~\bibnamefont {Sipser}},\ }\href
  {http://arxiv.org/abs/quant-ph/0001106} {\bibinfo {title} {{Quantum
  Computation by Adiabatic Evolution}}} (\bibinfo {year} {2000}),\ \Eprint
  {https://arxiv.org/abs/0001106} {arXiv:0001106 [quant-ph]} \BibitemShut
  {NoStop}%
\bibitem [{\citenamefont {Aharonov}\ \emph {et~al.}(2008)\citenamefont
  {Aharonov}, \citenamefont {van Dam}, \citenamefont {Kempe}, \citenamefont
  {Landau}, \citenamefont {Lloyd},\ and\ \citenamefont
  {Regev}}]{aharonov2008adiabatic}%
  \BibitemOpen
  \bibfield  {author} {\bibinfo {author} {\bibfnamefont {D.}~\bibnamefont
  {Aharonov}}, \bibinfo {author} {\bibfnamefont {W.}~\bibnamefont {van Dam}},
  \bibinfo {author} {\bibfnamefont {J.}~\bibnamefont {Kempe}}, \bibinfo
  {author} {\bibfnamefont {Z.}~\bibnamefont {Landau}}, \bibinfo {author}
  {\bibfnamefont {S.}~\bibnamefont {Lloyd}},\ and\ \bibinfo {author}
  {\bibfnamefont {O.}~\bibnamefont {Regev}},\ }\bibfield  {title} {\bibinfo
  {title} {{Adiabatic Quantum Computation Is Equivalent to Standard Quantum
  Computation}},\ }\href {https://doi.org/10.1137/080734479} {\bibfield
  {journal} {\bibinfo  {journal} {SIAM Rev.}\ }\textbf {\bibinfo {volume}
  {50}},\ \bibinfo {pages} {755} (\bibinfo {year} {2008})}\BibitemShut
  {NoStop}%
\bibitem [{\citenamefont {Aharonov}\ and\ \citenamefont
  {Ta-Shma}(2003)}]{Aharonov2003}%
  \BibitemOpen
  \bibfield  {author} {\bibinfo {author} {\bibfnamefont {D.}~\bibnamefont
  {Aharonov}}\ and\ \bibinfo {author} {\bibfnamefont {A.}~\bibnamefont
  {Ta-Shma}},\ }\bibfield  {title} {\bibinfo {title} {{Adiabatic quantum state
  generation and statistical zero knowledge}},\ }in\ \href
  {https://doi.org/10.1145/780543.780546} {\emph {\bibinfo {booktitle} {Proc.
  thirty-fifth ACM Symp. Theory Comput. - STOC '03}}}\ (\bibinfo  {publisher}
  {ACM Press},\ \bibinfo {address} {New York, New York, USA},\ \bibinfo {year}
  {2003})\ p.~\bibinfo {pages} {20}\BibitemShut {NoStop}%
\bibitem [{\citenamefont {Peruzzo}\ \emph {et~al.}(2014)\citenamefont
  {Peruzzo}, \citenamefont {McClean}, \citenamefont {Shadbolt}, \citenamefont
  {Yung}, \citenamefont {Zhou}, \citenamefont {Love}, \citenamefont
  {Aspuru-Guzik}, \citenamefont {O'brien},\ and\ \citenamefont
  {O'Brien}}]{peruzzo2014variational}%
  \BibitemOpen
  \bibfield  {author} {\bibinfo {author} {\bibfnamefont {A.}~\bibnamefont
  {Peruzzo}}, \bibinfo {author} {\bibfnamefont {J.}~\bibnamefont {McClean}},
  \bibinfo {author} {\bibfnamefont {P.}~\bibnamefont {Shadbolt}}, \bibinfo
  {author} {\bibfnamefont {M.-h.}\ \bibnamefont {Yung}}, \bibinfo {author}
  {\bibfnamefont {X.-Q.}\ \bibnamefont {Zhou}}, \bibinfo {author}
  {\bibfnamefont {P.~J.}\ \bibnamefont {Love}}, \bibinfo {author}
  {\bibfnamefont {A.}~\bibnamefont {Aspuru-Guzik}}, \bibinfo {author}
  {\bibfnamefont {J.~L.}\ \bibnamefont {O'brien}},\ and\ \bibinfo {author}
  {\bibfnamefont {J.~L.}\ \bibnamefont {O'Brien}},\ }\bibfield  {title}
  {\bibinfo {title} {{A variational eigenvalue solver on a photonic quantum
  processor}},\ }\href {https://doi.org/10.1038/ncomms5213} {\bibfield
  {journal} {\bibinfo  {journal} {Nat. Commun.}\ }\textbf {\bibinfo {volume}
  {5}},\ \bibinfo {pages} {4213} (\bibinfo {year} {2014})}\BibitemShut
  {NoStop}%
\bibitem [{\citenamefont {Farhi}\ \emph {et~al.}(2014)\citenamefont {Farhi},
  \citenamefont {Goldstone},\ and\ \citenamefont {Gutmann}}]{farhi2014quantum}%
  \BibitemOpen
  \bibfield  {author} {\bibinfo {author} {\bibfnamefont {E.}~\bibnamefont
  {Farhi}}, \bibinfo {author} {\bibfnamefont {J.}~\bibnamefont {Goldstone}},\
  and\ \bibinfo {author} {\bibfnamefont {S.}~\bibnamefont {Gutmann}},\ }\href
  {http://arxiv.org/abs/1411.4028} {\bibinfo {title} {{A quantum approximate
  optimization algorithm}}} (\bibinfo {year} {2014}),\ \Eprint
  {https://arxiv.org/abs/1411.4028} {arXiv:1411.4028 [quant-ph]} \BibitemShut
  {NoStop}%
\bibitem [{\citenamefont {Chowdhury}\ and\ \citenamefont
  {Somma}(2017)}]{Chowdhury2017}%
  \BibitemOpen
  \bibfield  {author} {\bibinfo {author} {\bibfnamefont {A.~N.}\ \bibnamefont
  {Chowdhury}}\ and\ \bibinfo {author} {\bibfnamefont {R.~D.}\ \bibnamefont
  {Somma}},\ }\bibfield  {title} {\bibinfo {title} {{Quantum algorithms for
  gibbs sampling and hitting-time estimation}},\ }\href
  {https://doi.org/10.26421/QIC17.1-2} {\bibfield  {journal} {\bibinfo
  {journal} {Quantum Inf. Comput.}\ }\textbf {\bibinfo {volume} {17}},\
  \bibinfo {pages} {41} (\bibinfo {year} {2017})}\BibitemShut {NoStop}%
\bibitem [{\citenamefont {Temme}\ \emph {et~al.}(2011)\citenamefont {Temme},
  \citenamefont {Osborne}, \citenamefont {Vollbrecht}, \citenamefont {Poulin},\
  and\ \citenamefont {Verstraete}}]{Temme2011b}%
  \BibitemOpen
  \bibfield  {author} {\bibinfo {author} {\bibfnamefont {K.}~\bibnamefont
  {Temme}}, \bibinfo {author} {\bibfnamefont {T.~J.}\ \bibnamefont {Osborne}},
  \bibinfo {author} {\bibfnamefont {K.~G.}\ \bibnamefont {Vollbrecht}},
  \bibinfo {author} {\bibfnamefont {D.}~\bibnamefont {Poulin}},\ and\ \bibinfo
  {author} {\bibfnamefont {F.}~\bibnamefont {Verstraete}},\ }\bibfield  {title}
  {\bibinfo {title} {{Quantum Metropolis sampling}},\ }\href
  {https://doi.org/10.1038/nature09770} {\bibfield  {journal} {\bibinfo
  {journal} {Nature}\ }\textbf {\bibinfo {volume} {471}},\ \bibinfo {pages}
  {87} (\bibinfo {year} {2011})}\BibitemShut {NoStop}%
\bibitem [{\citenamefont {Motta}\ \emph {et~al.}(2020)\citenamefont {Motta},
  \citenamefont {Sun}, \citenamefont {Tan}, \citenamefont {O'Rourke},
  \citenamefont {Ye}, \citenamefont {Minnich}, \citenamefont {Brand{\~{a}}o},\
  and\ \citenamefont {Chan}}]{Motta2019}%
  \BibitemOpen
  \bibfield  {author} {\bibinfo {author} {\bibfnamefont {M.}~\bibnamefont
  {Motta}}, \bibinfo {author} {\bibfnamefont {C.}~\bibnamefont {Sun}}, \bibinfo
  {author} {\bibfnamefont {A.~T.~K.}\ \bibnamefont {Tan}}, \bibinfo {author}
  {\bibfnamefont {M.~J.}\ \bibnamefont {O'Rourke}}, \bibinfo {author}
  {\bibfnamefont {E.}~\bibnamefont {Ye}}, \bibinfo {author} {\bibfnamefont
  {A.~J.}\ \bibnamefont {Minnich}}, \bibinfo {author} {\bibfnamefont {F.~G.
  S.~L.}\ \bibnamefont {Brand{\~{a}}o}},\ and\ \bibinfo {author} {\bibfnamefont
  {G.~K.-L.}\ \bibnamefont {Chan}},\ }\bibfield  {title} {\bibinfo {title}
  {{Determining eigenstates and thermal states on a quantum computer using
  quantum imaginary time evolution}},\ }\href
  {https://doi.org/10.1038/s41567-019-0704-4} {\bibfield  {journal} {\bibinfo
  {journal} {Nat. Phys.}\ }\textbf {\bibinfo {volume} {16}},\ \bibinfo {pages}
  {205} (\bibinfo {year} {2020})}\BibitemShut {NoStop}%
\bibitem [{\citenamefont {Cohn}\ \emph {et~al.}(2020)\citenamefont {Cohn},
  \citenamefont {Yang}, \citenamefont {Najafi}, \citenamefont {Jones},\ and\
  \citenamefont {Freericks}}]{Cohn1812.03607}%
  \BibitemOpen
  \bibfield  {author} {\bibinfo {author} {\bibfnamefont {J.}~\bibnamefont
  {Cohn}}, \bibinfo {author} {\bibfnamefont {F.}~\bibnamefont {Yang}}, \bibinfo
  {author} {\bibfnamefont {K.}~\bibnamefont {Najafi}}, \bibinfo {author}
  {\bibfnamefont {B.}~\bibnamefont {Jones}},\ and\ \bibinfo {author}
  {\bibfnamefont {J.~K.}\ \bibnamefont {Freericks}},\ }\bibfield  {title}
  {\bibinfo {title} {Minimal effective gibbs ansatz: A simple protocol for
  extracting an accurate thermal representation for quantum simulation},\
  }\href {https://doi.org/10.1103/PhysRevA.102.022622} {\bibfield  {journal}
  {\bibinfo  {journal} {Phys. Rev. A}\ }\textbf {\bibinfo {volume} {102}},\
  \bibinfo {pages} {022622} (\bibinfo {year} {2020})}\BibitemShut {NoStop}%
\bibitem [{\citenamefont {Somma}\ \emph {et~al.}(2002)\citenamefont {Somma},
  \citenamefont {Ortiz}, \citenamefont {Gubernatis}, \citenamefont {Knill},\
  and\ \citenamefont {Laflamme}}]{Somma2002}%
  \BibitemOpen
  \bibfield  {author} {\bibinfo {author} {\bibfnamefont {R.~D.}\ \bibnamefont
  {Somma}}, \bibinfo {author} {\bibfnamefont {G.}~\bibnamefont {Ortiz}},
  \bibinfo {author} {\bibfnamefont {J.~E.}\ \bibnamefont {Gubernatis}},
  \bibinfo {author} {\bibfnamefont {E.}~\bibnamefont {Knill}},\ and\ \bibinfo
  {author} {\bibfnamefont {R.}~\bibnamefont {Laflamme}},\ }\bibfield  {title}
  {\bibinfo {title} {{Simulating physical phenomena by quantum networks}},\
  }\href {https://doi.org/10.1103/PhysRevA.65.042323} {\bibfield  {journal}
  {\bibinfo  {journal} {Phys. Rev. A}\ }\textbf {\bibinfo {volume} {65}},\
  \bibinfo {pages} {042323} (\bibinfo {year} {2002})}\BibitemShut {NoStop}%
\bibitem [{\citenamefont {Somma}(2019)}]{Somma2019}%
  \BibitemOpen
  \bibfield  {author} {\bibinfo {author} {\bibfnamefont {R.~D.}\ \bibnamefont
  {Somma}},\ }\bibfield  {title} {\bibinfo {title} {{Quantum eigenvalue
  estimation via time series analysis}},\ }\href
  {https://doi.org/10.1088/1367-2630/ab5c60} {\bibfield  {journal} {\bibinfo
  {journal} {New J. Phys.}\ }\textbf {\bibinfo {volume} {21}},\ \bibinfo
  {pages} {123025} (\bibinfo {year} {2019})},\ \Eprint
  {https://arxiv.org/abs/1907.11748} {arXiv:1907.11748} \BibitemShut {NoStop}%
\bibitem [{\citenamefont {Roggero}(2020)}]{aless2020spectral}%
  \BibitemOpen
  \bibfield  {author} {\bibinfo {author} {\bibfnamefont {A.}~\bibnamefont
  {Roggero}},\ }\bibfield  {title} {\bibinfo {title} {Spectral-density
  estimation with the gaussian integral transform},\ }\href
  {https://doi.org/10.1103/PhysRevA.102.022409} {\bibfield  {journal} {\bibinfo
   {journal} {Phys. Rev. A}\ }\textbf {\bibinfo {volume} {102}},\ \bibinfo
  {pages} {022409} (\bibinfo {year} {2020})}\BibitemShut {NoStop}%
\bibitem [{\citenamefont {Rall}(2020)}]{rall2020quantum}%
  \BibitemOpen
  \bibfield  {author} {\bibinfo {author} {\bibfnamefont {P.}~\bibnamefont
  {Rall}},\ }\bibfield  {title} {\bibinfo {title} {Quantum algorithms for
  estimating physical quantities using block encodings},\ }\href
  {https://doi.org/10.1103/PhysRevA.102.022408} {\bibfield  {journal} {\bibinfo
   {journal} {Phys. Rev. A}\ }\textbf {\bibinfo {volume} {102}},\ \bibinfo
  {pages} {022408} (\bibinfo {year} {2020})}\BibitemShut {NoStop}%
\bibitem [{\citenamefont {Troyer}\ and\ \citenamefont
  {Wiese}(2005)}]{WieseTroyer}%
  \BibitemOpen
  \bibfield  {author} {\bibinfo {author} {\bibfnamefont {M.}~\bibnamefont
  {Troyer}}\ and\ \bibinfo {author} {\bibfnamefont {U.-J.}\ \bibnamefont
  {Wiese}},\ }\bibfield  {title} {\bibinfo {title} {Computational complexity
  and fundamental limitations to fermionic quantum {Monte Carlo} simulations},\
  }\href {https://doi.org/10.1103/PhysRevLett.94.170201} {\bibfield  {journal}
  {\bibinfo  {journal} {Phys. Rev. Lett.}\ }\textbf {\bibinfo {volume} {94}},\
  \bibinfo {pages} {170201} (\bibinfo {year} {2005})}\BibitemShut {NoStop}%
\bibitem [{\citenamefont {Cerezo}\ \emph {et~al.}(2020)\citenamefont {Cerezo},
  \citenamefont {Arrasmith}, \citenamefont {Babbush}, \citenamefont {Benjamin},
  \citenamefont {Endo}, \citenamefont {Fujii}, \citenamefont {McClean},
  \citenamefont {Mitarai}, \citenamefont {Yuan}, \citenamefont {Cincio},\ and\
  \citenamefont {Coles}}]{VQAreview}%
  \BibitemOpen
  \bibfield  {author} {\bibinfo {author} {\bibfnamefont {M.}~\bibnamefont
  {Cerezo}}, \bibinfo {author} {\bibfnamefont {A.}~\bibnamefont {Arrasmith}},
  \bibinfo {author} {\bibfnamefont {R.}~\bibnamefont {Babbush}}, \bibinfo
  {author} {\bibfnamefont {S.~C.}\ \bibnamefont {Benjamin}}, \bibinfo {author}
  {\bibfnamefont {S.}~\bibnamefont {Endo}}, \bibinfo {author} {\bibfnamefont
  {K.}~\bibnamefont {Fujii}}, \bibinfo {author} {\bibfnamefont {J.~R.}\
  \bibnamefont {McClean}}, \bibinfo {author} {\bibfnamefont {K.}~\bibnamefont
  {Mitarai}}, \bibinfo {author} {\bibfnamefont {X.}~\bibnamefont {Yuan}},
  \bibinfo {author} {\bibfnamefont {L.}~\bibnamefont {Cincio}},\ and\ \bibinfo
  {author} {\bibfnamefont {P.~J.}\ \bibnamefont {Coles}},\ }\href@noop {}
  {\bibinfo {title} {Variational quantum algorithms}} (\bibinfo {year}
  {2020}),\ \Eprint {https://arxiv.org/abs/2012.09265} {arXiv:2012.09265
  [quant-ph]} \BibitemShut {NoStop}%
\bibitem [{\citenamefont {Stair}\ \emph {et~al.}(2020)\citenamefont {Stair},
  \citenamefont {Huang},\ and\ \citenamefont {Evangelista}}]{QuKrylov}%
  \BibitemOpen
  \bibfield  {author} {\bibinfo {author} {\bibfnamefont {N.~H.}\ \bibnamefont
  {Stair}}, \bibinfo {author} {\bibfnamefont {R.}~\bibnamefont {Huang}},\ and\
  \bibinfo {author} {\bibfnamefont {F.~A.}\ \bibnamefont {Evangelista}},\
  }\bibfield  {title} {\bibinfo {title} {A multireference quantum krylov
  algorithm for strongly correlated electrons},\ }\bibfield  {booktitle} {\emph
  {\bibinfo {booktitle} {Journal of Chemical Theory and Computation}},\ }\href
  {https://doi.org/10.1021/acs.jctc.9b01125} {\bibfield  {journal} {\bibinfo
  {journal} {Journal of Chemical Theory and Computation}\ }\textbf {\bibinfo
  {volume} {16}},\ \bibinfo {pages} {2236} (\bibinfo {year}
  {2020})}\BibitemShut {NoStop}%
\bibitem [{\citenamefont {McClean}\ \emph {et~al.}(2017)\citenamefont
  {McClean}, \citenamefont {Kimchi-Schwartz}, \citenamefont {Carter},\ and\
  \citenamefont {de~Jong}}]{mcclean2017hybrid}%
  \BibitemOpen
  \bibfield  {author} {\bibinfo {author} {\bibfnamefont {J.~R.}\ \bibnamefont
  {McClean}}, \bibinfo {author} {\bibfnamefont {M.~E.}\ \bibnamefont
  {Kimchi-Schwartz}}, \bibinfo {author} {\bibfnamefont {J.}~\bibnamefont
  {Carter}},\ and\ \bibinfo {author} {\bibfnamefont {W.~A.}\ \bibnamefont
  {de~Jong}},\ }\bibfield  {title} {\bibinfo {title} {Hybrid quantum-classical
  hierarchy for mitigation of decoherence and determination of excited
  states},\ }\href {https://doi.org/10.1103/PhysRevA.95.042308} {\bibfield
  {journal} {\bibinfo  {journal} {Phys. Rev. A}\ }\textbf {\bibinfo {volume}
  {95}},\ \bibinfo {pages} {042308} (\bibinfo {year} {2017})}\BibitemShut
  {NoStop}%
\bibitem [{\citenamefont {Parrish}\ and\ \citenamefont
  {McMahon}(2019)}]{parrish2019quantum}%
  \BibitemOpen
  \bibfield  {author} {\bibinfo {author} {\bibfnamefont {R.~M.}\ \bibnamefont
  {Parrish}}\ and\ \bibinfo {author} {\bibfnamefont {P.~L.}\ \bibnamefont
  {McMahon}},\ }\href@noop {} {\bibinfo {title} {Quantum filter
  diagonalization: Quantum eigendecomposition without full quantum phase
  estimation}} (\bibinfo {year} {2019}),\ \Eprint
  {https://arxiv.org/abs/1909.08925} {arXiv:1909.08925 [quant-ph]} \BibitemShut
  {NoStop}%
\bibitem [{\citenamefont {Parrish}\ \emph {et~al.}(2019)\citenamefont
  {Parrish}, \citenamefont {Hohenstein}, \citenamefont {McMahon},\ and\
  \citenamefont {Mart\'{\i}nez}}]{Parrish2019d}%
  \BibitemOpen
  \bibfield  {author} {\bibinfo {author} {\bibfnamefont {R.~M.}\ \bibnamefont
  {Parrish}}, \bibinfo {author} {\bibfnamefont {E.~G.}\ \bibnamefont
  {Hohenstein}}, \bibinfo {author} {\bibfnamefont {P.~L.}\ \bibnamefont
  {McMahon}},\ and\ \bibinfo {author} {\bibfnamefont {T.~J.}\ \bibnamefont
  {Mart\'{\i}nez}},\ }\bibfield  {title} {\bibinfo {title} {Quantum computation
  of electronic transitions using a variational quantum eigensolver},\ }\href
  {https://doi.org/10.1103/PhysRevLett.122.230401} {\bibfield  {journal}
  {\bibinfo  {journal} {Phys. Rev. Lett.}\ }\textbf {\bibinfo {volume} {122}},\
  \bibinfo {pages} {230401} (\bibinfo {year} {2019})}\BibitemShut {NoStop}%
\bibitem [{\citenamefont {Huggins}\ \emph {et~al.}(2020)\citenamefont
  {Huggins}, \citenamefont {Lee}, \citenamefont {Baek}, \citenamefont
  {O'Gorman},\ and\ \citenamefont {Whaley}}]{Huggins2019a}%
  \BibitemOpen
  \bibfield  {author} {\bibinfo {author} {\bibfnamefont {W.~J.}\ \bibnamefont
  {Huggins}}, \bibinfo {author} {\bibfnamefont {J.}~\bibnamefont {Lee}},
  \bibinfo {author} {\bibfnamefont {U.}~\bibnamefont {Baek}}, \bibinfo {author}
  {\bibfnamefont {B.}~\bibnamefont {O'Gorman}},\ and\ \bibinfo {author}
  {\bibfnamefont {K.~B.}\ \bibnamefont {Whaley}},\ }\bibfield  {title}
  {\bibinfo {title} {{A non-orthogonal variational quantum eigensolver}},\
  }\href {https://doi.org/10.1088/1367-2630/ab867b} {\bibfield  {journal}
  {\bibinfo  {journal} {New J. Phys.}\ }\textbf {\bibinfo {volume} {22}},\
  \bibinfo {pages} {073009} (\bibinfo {year} {2020})}\BibitemShut {NoStop}%
\bibitem [{\citenamefont {Peres}(1984)}]{Peres1984}%
  \BibitemOpen
  \bibfield  {author} {\bibinfo {author} {\bibfnamefont {A.}~\bibnamefont
  {Peres}},\ }\bibfield  {title} {\bibinfo {title} {{Stability of quantum
  motion in chaotic and regular systems}},\ }\href
  {https://doi.org/10.1103/PhysRevA.30.1610} {\bibfield  {journal} {\bibinfo
  {journal} {Phys. Rev. A}\ }\textbf {\bibinfo {volume} {30}},\ \bibinfo
  {pages} {1610} (\bibinfo {year} {1984})}\BibitemShut {NoStop}%
\bibitem [{\citenamefont {Wisniacki}(2012)}]{Wisniacki2012}%
  \BibitemOpen
  \bibfield  {author} {\bibinfo {author} {\bibfnamefont {A.}~\bibnamefont
  {Wisniacki}},\ }\bibfield  {title} {\bibinfo {title} {{Loschmidt echo}},\
  }\href {https://doi.org/10.4249/scholarpedia.11687} {\bibfield  {journal}
  {\bibinfo  {journal} {Scholarpedia}\ }\textbf {\bibinfo {volume} {7}},\
  \bibinfo {pages} {11687} (\bibinfo {year} {2012})}\BibitemShut {NoStop}%
\bibitem [{\citenamefont {Gardiner}\ \emph {et~al.}(1997)\citenamefont
  {Gardiner}, \citenamefont {Cirac},\ and\ \citenamefont
  {Zoller}}]{GardinerCiracZoller}%
  \BibitemOpen
  \bibfield  {author} {\bibinfo {author} {\bibfnamefont {S.~A.}\ \bibnamefont
  {Gardiner}}, \bibinfo {author} {\bibfnamefont {J.~I.}\ \bibnamefont
  {Cirac}},\ and\ \bibinfo {author} {\bibfnamefont {P.}~\bibnamefont
  {Zoller}},\ }\bibfield  {title} {\bibinfo {title} {{Quantum Chaos in an Ion
  Trap: The Delta-Kicked Harmonic Oscillator}},\ }\href
  {https://doi.org/10.1103/PhysRevLett.79.4790} {\bibfield  {journal} {\bibinfo
   {journal} {Phys. Rev. Lett.}\ }\textbf {\bibinfo {volume} {79}},\ \bibinfo
  {pages} {4790} (\bibinfo {year} {1997})}\BibitemShut {NoStop}%
\bibitem [{\citenamefont {Knap}\ \emph {et~al.}(2013)\citenamefont {Knap},
  \citenamefont {Kantian}, \citenamefont {Giamarchi}, \citenamefont {Bloch},
  \citenamefont {Lukin},\ and\ \citenamefont {Demler}}]{Knap2013}%
  \BibitemOpen
  \bibfield  {author} {\bibinfo {author} {\bibfnamefont {M.}~\bibnamefont
  {Knap}}, \bibinfo {author} {\bibfnamefont {A.}~\bibnamefont {Kantian}},
  \bibinfo {author} {\bibfnamefont {T.}~\bibnamefont {Giamarchi}}, \bibinfo
  {author} {\bibfnamefont {I.}~\bibnamefont {Bloch}}, \bibinfo {author}
  {\bibfnamefont {M.~D.}\ \bibnamefont {Lukin}},\ and\ \bibinfo {author}
  {\bibfnamefont {E.}~\bibnamefont {Demler}},\ }\bibfield  {title} {\bibinfo
  {title} {{Probing Real-Space and Time-Resolved Correlation Functions with
  Many-Body Ramsey Interferometry}},\ }\href
  {https://doi.org/10.1103/PhysRevLett.111.147205} {\bibfield  {journal}
  {\bibinfo  {journal} {Phys. Rev. Lett.}\ }\textbf {\bibinfo {volume} {111}},\
  \bibinfo {pages} {147205} (\bibinfo {year} {2013})}\BibitemShut {NoStop}%
\bibitem [{\citenamefont {Greenberger}\ \emph {et~al.}(1989)\citenamefont
  {Greenberger}, \citenamefont {Horne},\ and\ \citenamefont
  {Zeilinger}}]{GHZoriginalpaper}%
  \BibitemOpen
  \bibfield  {author} {\bibinfo {author} {\bibfnamefont {D.~M.}\ \bibnamefont
  {Greenberger}}, \bibinfo {author} {\bibfnamefont {M.~A.}\ \bibnamefont
  {Horne}},\ and\ \bibinfo {author} {\bibfnamefont {A.}~\bibnamefont
  {Zeilinger}},\ }\bibfield  {title} {\bibinfo {title} {{Going Beyond Bell's
  Theorem}},\ }in\ \href {https://doi.org/10.1007/978-94-017-0849-4_10} {\emph
  {\bibinfo {booktitle} {Bell's Theorem, Quantum Theory and Conceptions of the
  Universe}}},\ \bibinfo {series and number} {\bibinfo {number} {3}}\ (\bibinfo
   {publisher} {Springer Netherlands},\ \bibinfo {address} {Dordrecht},\
  \bibinfo {year} {1989})\ pp.\ \bibinfo {pages} {69--72}\BibitemShut {NoStop}%
\bibitem [{\citenamefont {Laflamme}\ \emph {et~al.}(1998)\citenamefont
  {Laflamme}, \citenamefont {Knill}, \citenamefont {Zurek}, \citenamefont
  {Catasti},\ and\ \citenamefont {Mariappan}}]{Laflamme1998NMRGHZ}%
  \BibitemOpen
  \bibfield  {author} {\bibinfo {author} {\bibfnamefont {R.}~\bibnamefont
  {Laflamme}}, \bibinfo {author} {\bibfnamefont {E.}~\bibnamefont {Knill}},
  \bibinfo {author} {\bibfnamefont {W.~H.}\ \bibnamefont {Zurek}}, \bibinfo
  {author} {\bibfnamefont {P.}~\bibnamefont {Catasti}},\ and\ \bibinfo {author}
  {\bibfnamefont {S.~V.~S.}\ \bibnamefont {Mariappan}},\ }\bibfield  {title}
  {\bibinfo {title} {{NMR} {Greenberger}–{Horne}–{Zeilinger} states},\
  }\href {https://doi.org/10.1098/rsta.1998.0257} {\bibfield  {journal}
  {\bibinfo  {journal} {Philos. Trans. R. Soc. London. Ser. A Math. Phys. Eng.
  Sci.}\ }\textbf {\bibinfo {volume} {356}},\ \bibinfo {pages} {1941} (\bibinfo
  {year} {1998})}\BibitemShut {NoStop}%
\bibitem [{\citenamefont {Neumann}\ \emph {et~al.}(2008)\citenamefont
  {Neumann}, \citenamefont {Mizuochi}, \citenamefont {Rempp}, \citenamefont
  {Hemmer}, \citenamefont {Watanabe}, \citenamefont {Yamasaki}, \citenamefont
  {Jacques}, \citenamefont {Gaebel}, \citenamefont {Jelezko},\ and\
  \citenamefont {Wrachtrup}}]{Neumann2009GHZNV}%
  \BibitemOpen
  \bibfield  {author} {\bibinfo {author} {\bibfnamefont {P.}~\bibnamefont
  {Neumann}}, \bibinfo {author} {\bibfnamefont {N.}~\bibnamefont {Mizuochi}},
  \bibinfo {author} {\bibfnamefont {F.}~\bibnamefont {Rempp}}, \bibinfo
  {author} {\bibfnamefont {P.}~\bibnamefont {Hemmer}}, \bibinfo {author}
  {\bibfnamefont {H.}~\bibnamefont {Watanabe}}, \bibinfo {author}
  {\bibfnamefont {S.}~\bibnamefont {Yamasaki}}, \bibinfo {author}
  {\bibfnamefont {V.}~\bibnamefont {Jacques}}, \bibinfo {author} {\bibfnamefont
  {T.}~\bibnamefont {Gaebel}}, \bibinfo {author} {\bibfnamefont
  {F.}~\bibnamefont {Jelezko}},\ and\ \bibinfo {author} {\bibfnamefont
  {J.}~\bibnamefont {Wrachtrup}},\ }\bibfield  {title} {\bibinfo {title}
  {{Multipartite Entanglement Among Single Spins in Diamond}},\ }\href
  {https://doi.org/10.1126/science.1157233} {\bibfield  {journal} {\bibinfo
  {journal} {Science}\ }\textbf {\bibinfo {volume} {320}},\ \bibinfo {pages}
  {1326} (\bibinfo {year} {2008})}\BibitemShut {NoStop}%
\bibitem [{\citenamefont {Leibfried}\ \emph {et~al.}(2005)\citenamefont
  {Leibfried}, \citenamefont {Knill}, \citenamefont {Seidelin}, \citenamefont
  {Britton}, \citenamefont {Blakestad}, \citenamefont {Chiaverini},
  \citenamefont {Hume}, \citenamefont {Itano}, \citenamefont {Jost},
  \citenamefont {Langer}, \citenamefont {Ozeri}, \citenamefont {Reichle},\ and\
  \citenamefont {Wineland}}]{Leibfried2005GHZ6}%
  \BibitemOpen
  \bibfield  {author} {\bibinfo {author} {\bibfnamefont {D.}~\bibnamefont
  {Leibfried}}, \bibinfo {author} {\bibfnamefont {E.}~\bibnamefont {Knill}},
  \bibinfo {author} {\bibfnamefont {S.}~\bibnamefont {Seidelin}}, \bibinfo
  {author} {\bibfnamefont {J.}~\bibnamefont {Britton}}, \bibinfo {author}
  {\bibfnamefont {R.~B.}\ \bibnamefont {Blakestad}}, \bibinfo {author}
  {\bibfnamefont {J.}~\bibnamefont {Chiaverini}}, \bibinfo {author}
  {\bibfnamefont {D.~B.}\ \bibnamefont {Hume}}, \bibinfo {author}
  {\bibfnamefont {W.~M.}\ \bibnamefont {Itano}}, \bibinfo {author}
  {\bibfnamefont {J.~D.}\ \bibnamefont {Jost}}, \bibinfo {author}
  {\bibfnamefont {C.}~\bibnamefont {Langer}}, \bibinfo {author} {\bibfnamefont
  {R.}~\bibnamefont {Ozeri}}, \bibinfo {author} {\bibfnamefont
  {R.}~\bibnamefont {Reichle}},\ and\ \bibinfo {author} {\bibfnamefont {D.~J.}\
  \bibnamefont {Wineland}},\ }\bibfield  {title} {\bibinfo {title} {{Creation
  of a six-atom ‘Schr{\"o}dinger cat’state}},\ }\href
  {https://doi.org/10.1038/nature04251} {\bibfield  {journal} {\bibinfo
  {journal} {Nature}\ }\textbf {\bibinfo {volume} {438}},\ \bibinfo {pages}
  {639} (\bibinfo {year} {2005})}\BibitemShut {NoStop}%
\bibitem [{\citenamefont {Monz}\ \emph {et~al.}(2011)\citenamefont {Monz},
  \citenamefont {Schindler}, \citenamefont {Barreiro}, \citenamefont {Chwalla},
  \citenamefont {Nigg}, \citenamefont {Coish}, \citenamefont {Harlander},
  \citenamefont {H{\"{a}}nsel}, \citenamefont {Hennrich},\ and\ \citenamefont
  {Blatt}}]{Monz2011GHZ14}%
  \BibitemOpen
  \bibfield  {author} {\bibinfo {author} {\bibfnamefont {T.}~\bibnamefont
  {Monz}}, \bibinfo {author} {\bibfnamefont {P.}~\bibnamefont {Schindler}},
  \bibinfo {author} {\bibfnamefont {J.~T.}\ \bibnamefont {Barreiro}}, \bibinfo
  {author} {\bibfnamefont {M.}~\bibnamefont {Chwalla}}, \bibinfo {author}
  {\bibfnamefont {D.}~\bibnamefont {Nigg}}, \bibinfo {author} {\bibfnamefont
  {W.~A.}\ \bibnamefont {Coish}}, \bibinfo {author} {\bibfnamefont
  {M.}~\bibnamefont {Harlander}}, \bibinfo {author} {\bibfnamefont
  {W.}~\bibnamefont {H{\"{a}}nsel}}, \bibinfo {author} {\bibfnamefont
  {M.}~\bibnamefont {Hennrich}},\ and\ \bibinfo {author} {\bibfnamefont
  {R.}~\bibnamefont {Blatt}},\ }\bibfield  {title} {\bibinfo {title} {{14-Qubit
  Entanglement: Creation and Coherence}},\ }\href
  {https://doi.org/10.1103/PhysRevLett.106.130506} {\bibfield  {journal}
  {\bibinfo  {journal} {Phys. Rev. Lett.}\ }\textbf {\bibinfo {volume} {106}},\
  \bibinfo {pages} {130506} (\bibinfo {year} {2011})}\BibitemShut {NoStop}%
\bibitem [{\citenamefont {DiCarlo}\ \emph {et~al.}(2010)\citenamefont
  {DiCarlo}, \citenamefont {Reed}, \citenamefont {Sun}, \citenamefont
  {Johnson}, \citenamefont {Chow}, \citenamefont {Gambetta}, \citenamefont
  {Frunzio}, \citenamefont {Girvin}, \citenamefont {Devoret},\ and\
  \citenamefont {Schoelkopf}}]{Dicarlo2010GHZ3}%
  \BibitemOpen
  \bibfield  {author} {\bibinfo {author} {\bibfnamefont {L.}~\bibnamefont
  {DiCarlo}}, \bibinfo {author} {\bibfnamefont {M.~D.}\ \bibnamefont {Reed}},
  \bibinfo {author} {\bibfnamefont {L.}~\bibnamefont {Sun}}, \bibinfo {author}
  {\bibfnamefont {B.~R.}\ \bibnamefont {Johnson}}, \bibinfo {author}
  {\bibfnamefont {J.~M.}\ \bibnamefont {Chow}}, \bibinfo {author}
  {\bibfnamefont {J.~M.}\ \bibnamefont {Gambetta}}, \bibinfo {author}
  {\bibfnamefont {L.}~\bibnamefont {Frunzio}}, \bibinfo {author} {\bibfnamefont
  {S.~M.}\ \bibnamefont {Girvin}}, \bibinfo {author} {\bibfnamefont {M.~H.}\
  \bibnamefont {Devoret}},\ and\ \bibinfo {author} {\bibfnamefont {R.~J.}\
  \bibnamefont {Schoelkopf}},\ }\bibfield  {title} {\bibinfo {title}
  {{Preparation and measurement of three-qubit entanglement in a
  superconducting circuit}},\ }\href {https://doi.org/10.1038/nature09416}
  {\bibfield  {journal} {\bibinfo  {journal} {Nature}\ }\textbf {\bibinfo
  {volume} {467}},\ \bibinfo {pages} {574} (\bibinfo {year}
  {2010})}\BibitemShut {NoStop}%
\bibitem [{\citenamefont {Song}\ \emph {et~al.}(2017)\citenamefont {Song},
  \citenamefont {Others}, \citenamefont {Xu}, \citenamefont {Liu},
  \citenamefont {Yang}, \citenamefont {Zheng}, \citenamefont {Deng},
  \citenamefont {Xie}, \citenamefont {Huang}, \citenamefont {Guo},
  \citenamefont {Zhang}, \citenamefont {Zhang}, \citenamefont {Xu},
  \citenamefont {Zheng}, \citenamefont {Zhu}, \citenamefont {Wang},
  \citenamefont {Chen}, \citenamefont {Lu}, \citenamefont {Han},\ and\
  \citenamefont {Pan}}]{Song2017GHZ10}%
  \BibitemOpen
  \bibfield  {author} {\bibinfo {author} {\bibfnamefont {C.}~\bibnamefont
  {Song}}, \bibinfo {author} {\bibnamefont {Others}}, \bibinfo {author}
  {\bibfnamefont {K.}~\bibnamefont {Xu}}, \bibinfo {author} {\bibfnamefont
  {W.}~\bibnamefont {Liu}}, \bibinfo {author} {\bibfnamefont {C.-p.}\
  \bibnamefont {Yang}}, \bibinfo {author} {\bibfnamefont {S.-B.}\ \bibnamefont
  {Zheng}}, \bibinfo {author} {\bibfnamefont {H.}~\bibnamefont {Deng}},
  \bibinfo {author} {\bibfnamefont {Q.}~\bibnamefont {Xie}}, \bibinfo {author}
  {\bibfnamefont {K.}~\bibnamefont {Huang}}, \bibinfo {author} {\bibfnamefont
  {Q.}~\bibnamefont {Guo}}, \bibinfo {author} {\bibfnamefont {L.}~\bibnamefont
  {Zhang}}, \bibinfo {author} {\bibfnamefont {P.}~\bibnamefont {Zhang}},
  \bibinfo {author} {\bibfnamefont {D.}~\bibnamefont {Xu}}, \bibinfo {author}
  {\bibfnamefont {D.}~\bibnamefont {Zheng}}, \bibinfo {author} {\bibfnamefont
  {X.}~\bibnamefont {Zhu}}, \bibinfo {author} {\bibfnamefont {H.}~\bibnamefont
  {Wang}}, \bibinfo {author} {\bibfnamefont {Y.-A.}\ \bibnamefont {Chen}},
  \bibinfo {author} {\bibfnamefont {C.-Y.}\ \bibnamefont {Lu}}, \bibinfo
  {author} {\bibfnamefont {S.}~\bibnamefont {Han}},\ and\ \bibinfo {author}
  {\bibfnamefont {J.-W.}\ \bibnamefont {Pan}},\ }\bibfield  {title} {\bibinfo
  {title} {{10-Qubit Entanglement and Parallel Logic Operations with a
  Superconducting Circuit}},\ }\href
  {https://doi.org/10.1103/PhysRevLett.119.180511} {\bibfield  {journal}
  {\bibinfo  {journal} {Phys. Rev. Lett.}\ }\textbf {\bibinfo {volume} {119}},\
  \bibinfo {pages} {180511} (\bibinfo {year} {2017})}\BibitemShut {NoStop}%
\bibitem [{\citenamefont {Wang}\ \emph {et~al.}(2018)\citenamefont {Wang},
  \citenamefont {Luo}, \citenamefont {Huang}, \citenamefont {Chen},
  \citenamefont {Su}, \citenamefont {Liu}, \citenamefont {Chen}, \citenamefont
  {Li}, \citenamefont {Fang}, \citenamefont {Jiang}, \citenamefont {Zhang},
  \citenamefont {Li}, \citenamefont {Liu}, \citenamefont {Lu},\ and\
  \citenamefont {Pan}}]{Wang2018GHZ18}%
  \BibitemOpen
  \bibfield  {author} {\bibinfo {author} {\bibfnamefont {X.-L.}\ \bibnamefont
  {Wang}}, \bibinfo {author} {\bibfnamefont {Y.-H.}\ \bibnamefont {Luo}},
  \bibinfo {author} {\bibfnamefont {H.-L.}\ \bibnamefont {Huang}}, \bibinfo
  {author} {\bibfnamefont {M.-C.}\ \bibnamefont {Chen}}, \bibinfo {author}
  {\bibfnamefont {Z.-E.}\ \bibnamefont {Su}}, \bibinfo {author} {\bibfnamefont
  {C.}~\bibnamefont {Liu}}, \bibinfo {author} {\bibfnamefont {C.}~\bibnamefont
  {Chen}}, \bibinfo {author} {\bibfnamefont {W.}~\bibnamefont {Li}}, \bibinfo
  {author} {\bibfnamefont {Y.-Q.}\ \bibnamefont {Fang}}, \bibinfo {author}
  {\bibfnamefont {X.}~\bibnamefont {Jiang}}, \bibinfo {author} {\bibfnamefont
  {J.}~\bibnamefont {Zhang}}, \bibinfo {author} {\bibfnamefont
  {L.}~\bibnamefont {Li}}, \bibinfo {author} {\bibfnamefont {N.-L.}\
  \bibnamefont {Liu}}, \bibinfo {author} {\bibfnamefont {C.-Y.}\ \bibnamefont
  {Lu}},\ and\ \bibinfo {author} {\bibfnamefont {J.-W.}\ \bibnamefont {Pan}},\
  }\bibfield  {title} {\bibinfo {title} {{18-Qubit Entanglement with Six
  Photons' Three Degrees of Freedom}},\ }\href
  {https://doi.org/10.1103/PhysRevLett.120.260502} {\bibfield  {journal}
  {\bibinfo  {journal} {Phys. Rev. Lett.}\ }\textbf {\bibinfo {volume} {120}},\
  \bibinfo {pages} {260502} (\bibinfo {year} {2018})}\BibitemShut {NoStop}%
\bibitem [{\citenamefont {Friis}\ \emph {et~al.}(2018)\citenamefont {Friis},
  \citenamefont {Marty}, \citenamefont {Maier}, \citenamefont {Hempel},
  \citenamefont {Holzapfel}, \citenamefont {Jurcevic}, \citenamefont {Plenio},
  \citenamefont {Huber}, \citenamefont {Roos}, \citenamefont {Blatt},\ and\
  \citenamefont {Lanyon}}]{Friis2018GHZ20}%
  \BibitemOpen
  \bibfield  {author} {\bibinfo {author} {\bibfnamefont {N.}~\bibnamefont
  {Friis}}, \bibinfo {author} {\bibfnamefont {O.}~\bibnamefont {Marty}},
  \bibinfo {author} {\bibfnamefont {C.}~\bibnamefont {Maier}}, \bibinfo
  {author} {\bibfnamefont {C.}~\bibnamefont {Hempel}}, \bibinfo {author}
  {\bibfnamefont {M.}~\bibnamefont {Holzapfel}}, \bibinfo {author}
  {\bibfnamefont {P.}~\bibnamefont {Jurcevic}}, \bibinfo {author}
  {\bibfnamefont {M.~B.}\ \bibnamefont {Plenio}}, \bibinfo {author}
  {\bibfnamefont {M.}~\bibnamefont {Huber}}, \bibinfo {author} {\bibfnamefont
  {C.}~\bibnamefont {Roos}}, \bibinfo {author} {\bibfnamefont {R.}~\bibnamefont
  {Blatt}},\ and\ \bibinfo {author} {\bibfnamefont {B.}~\bibnamefont
  {Lanyon}},\ }\bibfield  {title} {\bibinfo {title} {{Observation of Entangled
  States of a Fully Controlled 20-Qubit System}},\ }\href
  {https://doi.org/10.1103/PhysRevX.8.021012} {\bibfield  {journal} {\bibinfo
  {journal} {Phys. Rev. X}\ }\textbf {\bibinfo {volume} {8}},\ \bibinfo {pages}
  {21012} (\bibinfo {year} {2018})}\BibitemShut {NoStop}%
\bibitem [{\citenamefont {Omran}\ \emph {et~al.}(2019)\citenamefont {Omran},
  \citenamefont {Levine}, \citenamefont {Keesling}, \citenamefont {Semeghini},
  \citenamefont {Wang}, \citenamefont {Ebadi}, \citenamefont {Bernien},
  \citenamefont {Zibrov}, \citenamefont {Pichler}, \citenamefont {Choi},
  \citenamefont {Cui}, \citenamefont {Rossignolo}, \citenamefont {Rembold},
  \citenamefont {Montangero}, \citenamefont {Calarco}, \citenamefont {Endres},
  \citenamefont {Greiner}, \citenamefont {Vuleti{\'{c}}},\ and\ \citenamefont
  {Lukin}}]{LukinScienceSchCats}%
  \BibitemOpen
  \bibfield  {author} {\bibinfo {author} {\bibfnamefont {A.}~\bibnamefont
  {Omran}}, \bibinfo {author} {\bibfnamefont {H.}~\bibnamefont {Levine}},
  \bibinfo {author} {\bibfnamefont {A.}~\bibnamefont {Keesling}}, \bibinfo
  {author} {\bibfnamefont {G.}~\bibnamefont {Semeghini}}, \bibinfo {author}
  {\bibfnamefont {T.~T.}\ \bibnamefont {Wang}}, \bibinfo {author}
  {\bibfnamefont {S.}~\bibnamefont {Ebadi}}, \bibinfo {author} {\bibfnamefont
  {H.}~\bibnamefont {Bernien}}, \bibinfo {author} {\bibfnamefont {A.~S.}\
  \bibnamefont {Zibrov}}, \bibinfo {author} {\bibfnamefont {H.}~\bibnamefont
  {Pichler}}, \bibinfo {author} {\bibfnamefont {S.}~\bibnamefont {Choi}},
  \bibinfo {author} {\bibfnamefont {J.}~\bibnamefont {Cui}}, \bibinfo {author}
  {\bibfnamefont {M.}~\bibnamefont {Rossignolo}}, \bibinfo {author}
  {\bibfnamefont {P.}~\bibnamefont {Rembold}}, \bibinfo {author} {\bibfnamefont
  {S.}~\bibnamefont {Montangero}}, \bibinfo {author} {\bibfnamefont
  {T.}~\bibnamefont {Calarco}}, \bibinfo {author} {\bibfnamefont
  {M.}~\bibnamefont {Endres}}, \bibinfo {author} {\bibfnamefont
  {M.}~\bibnamefont {Greiner}}, \bibinfo {author} {\bibfnamefont
  {V.}~\bibnamefont {Vuleti{\'{c}}}},\ and\ \bibinfo {author} {\bibfnamefont
  {M.~D.}\ \bibnamefont {Lukin}},\ }\bibfield  {title} {\bibinfo {title}
  {{Generation and manipulation of Schr{\"{o}}dinger cat states in Rydberg atom
  arrays}},\ }\href {https://doi.org/10.1126/science.aax9743} {\bibfield
  {journal} {\bibinfo  {journal} {Science}\ }\textbf {\bibinfo {volume}
  {365}},\ \bibinfo {pages} {570} (\bibinfo {year} {2019})}\BibitemShut
  {NoStop}%
\bibitem [{\citenamefont {Ekert}\ \emph {et~al.}(2002)\citenamefont {Ekert},
  \citenamefont {Alves}, \citenamefont {Oi}, \citenamefont {Horodecki},
  \citenamefont {Horodecki},\ and\ \citenamefont {Kwek}}]{Ekert2002}%
  \BibitemOpen
  \bibfield  {author} {\bibinfo {author} {\bibfnamefont {A.~K.}\ \bibnamefont
  {Ekert}}, \bibinfo {author} {\bibfnamefont {C.~M.}\ \bibnamefont {Alves}},
  \bibinfo {author} {\bibfnamefont {D.~K.~L.}\ \bibnamefont {Oi}}, \bibinfo
  {author} {\bibfnamefont {M.}~\bibnamefont {Horodecki}}, \bibinfo {author}
  {\bibfnamefont {P.}~\bibnamefont {Horodecki}},\ and\ \bibinfo {author}
  {\bibfnamefont {L.~C.}\ \bibnamefont {Kwek}},\ }\bibfield  {title} {\bibinfo
  {title} {{Direct Estimations of Linear and Nonlinear Functionals of a Quantum
  State}},\ }\href {https://doi.org/10.1103/PhysRevLett.88.217901} {\bibfield
  {journal} {\bibinfo  {journal} {Phys. Rev. Lett.}\ }\textbf {\bibinfo
  {volume} {88}},\ \bibinfo {pages} {217901} (\bibinfo {year}
  {2002})}\BibitemShut {NoStop}%
\bibitem [{\citenamefont {Schuch}\ and\ \citenamefont
  {Cirac}(2010)}]{Schuch2010a}%
  \BibitemOpen
  \bibfield  {author} {\bibinfo {author} {\bibfnamefont {N.}~\bibnamefont
  {Schuch}}\ and\ \bibinfo {author} {\bibfnamefont {J.~I.}\ \bibnamefont
  {Cirac}},\ }\bibfield  {title} {\bibinfo {title} {{Matrix product state and
  mean-field solutions for one-dimensional systems can be found efficiently}},\
  }\href {https://doi.org/10.1103/PhysRevA.82.012314} {\bibfield  {journal}
  {\bibinfo  {journal} {Phys. Rev. A}\ }\textbf {\bibinfo {volume} {82}},\
  \bibinfo {pages} {12314} (\bibinfo {year} {2010})}\BibitemShut {NoStop}%
\bibitem [{\citenamefont {Verstraete}\ \emph {et~al.}(2008)\citenamefont
  {Verstraete}, \citenamefont {Murg},\ and\ \citenamefont
  {Cirac}}]{Verstraete2008a}%
  \BibitemOpen
  \bibfield  {author} {\bibinfo {author} {\bibfnamefont {F.}~\bibnamefont
  {Verstraete}}, \bibinfo {author} {\bibfnamefont {V.}~\bibnamefont {Murg}},\
  and\ \bibinfo {author} {\bibfnamefont {J.~I.}\ \bibnamefont {Cirac}},\
  }\bibfield  {title} {\bibinfo {title} {{Matrix product states, projected
  entangled pair states, and variational renormalization group methods for
  quantum spin systems}},\ }\href {https://doi.org/10.1080/14789940801912366}
  {\bibfield  {journal} {\bibinfo  {journal} {Adv. Phys.}\ }\textbf {\bibinfo
  {volume} {57}},\ \bibinfo {pages} {143} (\bibinfo {year} {2008})}\BibitemShut
  {NoStop}%
\bibitem [{\citenamefont {Schollw{\"{o}}ck}(2011)}]{schollwock2011density}%
  \BibitemOpen
  \bibfield  {author} {\bibinfo {author} {\bibfnamefont {U.}~\bibnamefont
  {Schollw{\"{o}}ck}},\ }\bibfield  {title} {\bibinfo {title} {{The
  density-matrix renormalization group in the age of matrix product states}},\
  }\href {https://doi.org/10.1016/j.aop.2010.09.012} {\bibfield  {journal}
  {\bibinfo  {journal} {Ann. Phys.}\ }\textbf {\bibinfo {volume} {326}},\
  \bibinfo {pages} {96} (\bibinfo {year} {2011})}\BibitemShut {NoStop}%
\bibitem [{\citenamefont {Schrodi}\ \emph {et~al.}(2017)\citenamefont
  {Schrodi}, \citenamefont {Silvi}, \citenamefont {Tschirsich}, \citenamefont
  {Fazio},\ and\ \citenamefont {Montangero}}]{Schrodi2017}%
  \BibitemOpen
  \bibfield  {author} {\bibinfo {author} {\bibfnamefont {F.}~\bibnamefont
  {Schrodi}}, \bibinfo {author} {\bibfnamefont {P.}~\bibnamefont {Silvi}},
  \bibinfo {author} {\bibfnamefont {F.}~\bibnamefont {Tschirsich}}, \bibinfo
  {author} {\bibfnamefont {R.}~\bibnamefont {Fazio}},\ and\ \bibinfo {author}
  {\bibfnamefont {S.}~\bibnamefont {Montangero}},\ }\bibfield  {title}
  {\bibinfo {title} {{Density of states of many-body quantum systems from
  tensor networks}},\ }\href {https://doi.org/10.1103/PhysRevB.96.094303}
  {\bibfield  {journal} {\bibinfo  {journal} {Phys. Rev. B}\ }\textbf {\bibinfo
  {volume} {96}},\ \bibinfo {pages} {094303} (\bibinfo {year}
  {2017})}\BibitemShut {NoStop}%
\bibitem [{\citenamefont {Srednicki}(1994)}]{Srednicki1994}%
  \BibitemOpen
  \bibfield  {author} {\bibinfo {author} {\bibfnamefont {M.}~\bibnamefont
  {Srednicki}},\ }\bibfield  {title} {\bibinfo {title} {{Chaos and quantum
  thermalization}},\ }\href {https://doi.org/10.1103/PhysRevE.50.888}
  {\bibfield  {journal} {\bibinfo  {journal} {Phys. Rev. E}\ }\textbf {\bibinfo
  {volume} {50}},\ \bibinfo {pages} {888} (\bibinfo {year} {1994})}\BibitemShut
  {NoStop}%
\bibitem [{\citenamefont {Deutsch}(2018)}]{reviewETH}%
  \BibitemOpen
  \bibfield  {author} {\bibinfo {author} {\bibfnamefont {J.~M.}\ \bibnamefont
  {Deutsch}},\ }\bibfield  {title} {\bibinfo {title} {{Eigenstate
  thermalization hypothesis}},\ }\href
  {https://doi.org/10.1088/1361-6633/aac9f1} {\bibfield  {journal} {\bibinfo
  {journal} {Reports Prog. Phys.}\ }\textbf {\bibinfo {volume} {81}},\ \bibinfo
  {pages} {082001} (\bibinfo {year} {2018})}\BibitemShut {NoStop}%
\bibitem [{\citenamefont {Dymarsky}\ and\ \citenamefont
  {Liu}(2019)}]{Dymarsky2017}%
  \BibitemOpen
  \bibfield  {author} {\bibinfo {author} {\bibfnamefont {A.}~\bibnamefont
  {Dymarsky}}\ and\ \bibinfo {author} {\bibfnamefont {H.}~\bibnamefont {Liu}},\
  }\bibfield  {title} {\bibinfo {title} {{New characteristic of quantum
  many-body chaotic systems}},\ }\href
  {https://doi.org/10.1103/PhysRevE.99.010102} {\bibfield  {journal} {\bibinfo
  {journal} {Phys. Rev. E}\ }\textbf {\bibinfo {volume} {99}},\ \bibinfo
  {pages} {010102} (\bibinfo {year} {2019})}\BibitemShut {NoStop}%
\bibitem [{\citenamefont {Yang}\ \emph {et~al.}(2020)\citenamefont {Yang},
  \citenamefont {Iblisdir}, \citenamefont {Cirac},\ and\ \citenamefont
  {Ba{\~{n}}uls}}]{Yilun}%
  \BibitemOpen
  \bibfield  {author} {\bibinfo {author} {\bibfnamefont {Y.}~\bibnamefont
  {Yang}}, \bibinfo {author} {\bibfnamefont {S.}~\bibnamefont {Iblisdir}},
  \bibinfo {author} {\bibfnamefont {J.~I.}\ \bibnamefont {Cirac}},\ and\
  \bibinfo {author} {\bibfnamefont {M.~C.}\ \bibnamefont {Ba{\~{n}}uls}},\
  }\bibfield  {title} {\bibinfo {title} {{Probing Thermalization through
  Spectral Analysis with Matrix Product Operators}},\ }\href
  {https://doi.org/10.1103/PhysRevLett.124.100602} {\bibfield  {journal}
  {\bibinfo  {journal} {Phys. Rev. Lett.}\ }\textbf {\bibinfo {volume} {124}},\
  \bibinfo {pages} {100602} (\bibinfo {year} {2020})}\BibitemShut {NoStop}%
\bibitem [{\citenamefont {Lukin}\ \emph {et~al.}(2001)\citenamefont {Lukin},
  \citenamefont {Fleischhauer}, \citenamefont {Cote}, \citenamefont {Duan},
  \citenamefont {Jaksch}, \citenamefont {Cirac},\ and\ \citenamefont
  {Zoller}}]{Lukin2001}%
  \BibitemOpen
  \bibfield  {author} {\bibinfo {author} {\bibfnamefont {M.~D.}\ \bibnamefont
  {Lukin}}, \bibinfo {author} {\bibfnamefont {M.}~\bibnamefont {Fleischhauer}},
  \bibinfo {author} {\bibfnamefont {R.}~\bibnamefont {Cote}}, \bibinfo {author}
  {\bibfnamefont {L.~M.}\ \bibnamefont {Duan}}, \bibinfo {author}
  {\bibfnamefont {D.}~\bibnamefont {Jaksch}}, \bibinfo {author} {\bibfnamefont
  {J.~I.}\ \bibnamefont {Cirac}},\ and\ \bibinfo {author} {\bibfnamefont
  {P.}~\bibnamefont {Zoller}},\ }\bibfield  {title} {\bibinfo {title} {{Dipole
  Blockade and Quantum Information Processing in Mesoscopic Atomic
  Ensembles}},\ }\href {https://doi.org/10.1103/PhysRevLett.87.037901}
  {\bibfield  {journal} {\bibinfo  {journal} {Phys. Rev. Lett.}\ }\textbf
  {\bibinfo {volume} {87}},\ \bibinfo {pages} {037901} (\bibinfo {year}
  {2001})}\BibitemShut {NoStop}%
\bibitem [{\citenamefont {Pupillo}\ \emph {et~al.}(2010)\citenamefont
  {Pupillo}, \citenamefont {Micheli}, \citenamefont {Boninsegni}, \citenamefont
  {Lesanovsky},\ and\ \citenamefont {Zoller}}]{Pupillo2010}%
  \BibitemOpen
  \bibfield  {author} {\bibinfo {author} {\bibfnamefont {G.}~\bibnamefont
  {Pupillo}}, \bibinfo {author} {\bibfnamefont {A.}~\bibnamefont {Micheli}},
  \bibinfo {author} {\bibfnamefont {M.}~\bibnamefont {Boninsegni}}, \bibinfo
  {author} {\bibfnamefont {I.}~\bibnamefont {Lesanovsky}},\ and\ \bibinfo
  {author} {\bibfnamefont {P.}~\bibnamefont {Zoller}},\ }\bibfield  {title}
  {\bibinfo {title} {{Strongly Correlated Gases of Rydberg-Dressed Atoms:
  Quantum and Classical Dynamics}},\ }\href
  {https://doi.org/10.1103/PhysRevLett.104.223002} {\bibfield  {journal}
  {\bibinfo  {journal} {Phys. Rev. Lett.}\ }\textbf {\bibinfo {volume} {104}},\
  \bibinfo {pages} {223002} (\bibinfo {year} {2010})}\BibitemShut {NoStop}%
\bibitem [{\citenamefont {Jau}\ \emph {et~al.}(2016)\citenamefont {Jau},
  \citenamefont {Hankin}, \citenamefont {Keating}, \citenamefont {Deutsch},\
  and\ \citenamefont {Biedermann}}]{Jau2016}%
  \BibitemOpen
  \bibfield  {author} {\bibinfo {author} {\bibfnamefont {Y.-Y.}\ \bibnamefont
  {Jau}}, \bibinfo {author} {\bibfnamefont {A.~M.}\ \bibnamefont {Hankin}},
  \bibinfo {author} {\bibfnamefont {T.}~\bibnamefont {Keating}}, \bibinfo
  {author} {\bibfnamefont {I.~H.}\ \bibnamefont {Deutsch}},\ and\ \bibinfo
  {author} {\bibfnamefont {G.~W.}\ \bibnamefont {Biedermann}},\ }\bibfield
  {title} {\bibinfo {title} {{Entangling atomic spins with a Rydberg-dressed
  spin-flip blockade}},\ }\href {https://doi.org/10.1038/nphys3487} {\bibfield
  {journal} {\bibinfo  {journal} {Nat. Phys.}\ }\textbf {\bibinfo {volume}
  {12}},\ \bibinfo {pages} {71} (\bibinfo {year} {2016})}\BibitemShut {NoStop}%
\bibitem [{\citenamefont {Glaetzle}\ \emph {et~al.}(2017)\citenamefont
  {Glaetzle}, \citenamefont {van Bijnen}, \citenamefont {Zoller},\ and\
  \citenamefont {Lechner}}]{Glaetzle2017}%
  \BibitemOpen
  \bibfield  {author} {\bibinfo {author} {\bibfnamefont {A.~W.}\ \bibnamefont
  {Glaetzle}}, \bibinfo {author} {\bibfnamefont {R.~M.~W.}\ \bibnamefont {van
  Bijnen}}, \bibinfo {author} {\bibfnamefont {P.}~\bibnamefont {Zoller}},\ and\
  \bibinfo {author} {\bibfnamefont {W.}~\bibnamefont {Lechner}},\ }\bibfield
  {title} {\bibinfo {title} {{A coherent quantum annealer with Rydberg
  atoms}},\ }\href {https://doi.org/10.1038/ncomms15813} {\bibfield  {journal}
  {\bibinfo  {journal} {Nat. Commun.}\ }\textbf {\bibinfo {volume} {8}},\
  \bibinfo {pages} {15813} (\bibinfo {year} {2017})}\BibitemShut {NoStop}%
\bibitem [{\citenamefont {Korsbakken}\ \emph {et~al.}(2007)\citenamefont
  {Korsbakken}, \citenamefont {Whaley}, \citenamefont {Dubois},\ and\
  \citenamefont {Cirac}}]{Korsbakken2007}%
  \BibitemOpen
  \bibfield  {author} {\bibinfo {author} {\bibfnamefont {J.~I.}\ \bibnamefont
  {Korsbakken}}, \bibinfo {author} {\bibfnamefont {K.~B.}\ \bibnamefont
  {Whaley}}, \bibinfo {author} {\bibfnamefont {J.}~\bibnamefont {Dubois}},\
  and\ \bibinfo {author} {\bibfnamefont {J.~I.}\ \bibnamefont {Cirac}},\
  }\bibfield  {title} {\bibinfo {title} {Measurement-based measure of the size
  of macroscopic quantum superpositions},\ }\href
  {https://doi.org/10.1103/PhysRevA.75.042106} {\bibfield  {journal} {\bibinfo
  {journal} {Phys. Rev. A}\ }\textbf {\bibinfo {volume} {75}},\ \bibinfo
  {pages} {042106} (\bibinfo {year} {2007})}\BibitemShut {NoStop}%
\bibitem [{\citenamefont {Lieb}(1973)}]{lieb1973}%
  \BibitemOpen
  \bibfield  {author} {\bibinfo {author} {\bibfnamefont {E.~H.}\ \bibnamefont
  {Lieb}},\ }\bibfield  {title} {\bibinfo {title} {The classical limit of
  quantum spin systems},\ }\href
  {https://projecteuclid.org:443/euclid.cmp/1103859040} {\bibfield  {journal}
  {\bibinfo  {journal} {Comm. Math. Phys.}\ }\textbf {\bibinfo {volume} {31}},\
  \bibinfo {pages} {327} (\bibinfo {year} {1973})}\BibitemShut {NoStop}%
\bibitem [{\citenamefont {Kim}\ and\ \citenamefont
  {Huse}(2013)}]{kim2013ballistic}%
  \BibitemOpen
  \bibfield  {author} {\bibinfo {author} {\bibfnamefont {H.}~\bibnamefont
  {Kim}}\ and\ \bibinfo {author} {\bibfnamefont {D.~A.}\ \bibnamefont {Huse}},\
  }\bibfield  {title} {\bibinfo {title} {Ballistic spreading of entanglement in
  a diffusive nonintegrable system},\ }\href
  {https://doi.org/10.1103/PhysRevLett.111.127205} {\bibfield  {journal}
  {\bibinfo  {journal} {Phys. Rev. Lett.}\ }\textbf {\bibinfo {volume} {111}},\
  \bibinfo {pages} {127205} (\bibinfo {year} {2013})}\BibitemShut {NoStop}%
\end{thebibliography}%

\end{document}